\documentclass[a4paper,reqno,10pt]{amsart}

\usepackage[
    top    = 2.0cm,
    bottom = 2.0cm,
    left   = 2.0cm,
    right  = 2.0cm]{geometry}
\usepackage[english]{babel}
\usepackage{amsmath}
\usepackage{graphicx}
\usepackage{amssymb}
\usepackage{bm} % bold math package

\usepackage{array}
\usepackage{setspace,cite,enumitem}
\usepackage[capitalise,noabbrev]{cleveref}
\usepackage[table]{xcolor}

\usepackage{tikz}
\usetikzlibrary{shapes}
\usetikzlibrary{arrows,positioning}

\tikzstyle{sv}=[circle, fill=black, inner sep=2pt, outer sep=0pt, minimum size=5pt]
\tikzstyle{osv}=[circle, draw=black, inner sep=2pt, outer sep=0pt, minimum size=5pt]

\numberwithin{equation}{section}

\newcolumntype{C}{>{$}c<{$}} % Defines math mode in tabular (array package)...
\newcolumntype{L}[1]{>{\raggedright}m{#1}}
\newcolumntype{D}[1]{>{\centering\arraybackslash\vspace{2mm}}m{#1}<{\vspace{2mm}}}
\newcolumntype{R}[1]{>{\raggedleft}m{#1}}

\newenvironment{amatrix}[1]{\left( \begin{array}{@{}#1@{}}}{\end{array} \right)}

\let\originalleft\left     % removes spurious spacing around \left and \right brackets
\let\originalright\right
\renewcommand{\left}{\mathopen{}\mathclose\bgroup\originalleft}
\renewcommand{\right}{\aftergroup\egroup\originalright}

\allowdisplaybreaks

\setitemize{leftmargin=*}   % Removes margin from itemized lists (enumitem package)
\setenumerate{leftmargin=*, label=\arabic*)} % Removes margin for enumerate lists (enumitem package)

\newcommand{\BKL}{\cellcolor[gray]{0.8}} % shade for boundary entries in Kac tables (xcolor package)
\newcommand{\IKL}{\cellcolor[gray]{0.6}} % shade for interior entries in Kac tables (xcolor package)
\newcommand{\CKL}{\cellcolor[gray]{0.4}} % shade for centre entries in Kac tables (xcolor package)

\newcommand{\alg}[1]{\mathfrak{#1}} % for algebras
\newcommand{\func}[2]{#1 \left( #2 \right)} % standard variations are for display
\newcommand{\tfunc}[2]{#1 \bigl( #2 \bigr)} % t-variations are for using in text

\newcommand{\brac}[1]{\left( #1 \right)}
\newcommand{\tbrac}[1]{\bigl( #1 \bigr)}
\newcommand{\sqbrac}[1]{\left[ #1 \right]}

\newcommand{\set}[1]{\left\{ #1 \right\}}

\newcommand{\bfset}[2]{\left\{ #1 \, \middle\vert \, #2 \right\}} % { bosons | fermions }

\newcommand{\st}{\mspace{5mu} : \mspace{5mu}} % "such that" in sets

\newcommand{\abs}[1]{\left\lvert #1 \right\rvert}

\newcommand{\bracket}[3]{\bigl\langle #1 \bigr\rvert #2 \bigl\lvert #3 \bigr\rangle} % bracket = < | | >

\newcommand{\ZZ}{\mathbb{Z}}

\newcommand{\QQ}{\mathbb{Q}}
\newcommand{\RR}{\mathbb{R}}

\newcommand{\CC}{\mathbb{C}}

\newcommand{\pd}{\partial}         % holomorphic partial d
\newcommand{\dd}{\mathrm{d}}   % d in derivatives and integrals
\newcommand{\ii}{\mathfrak{i}} % imaginary unit
\newcommand{\ee}{\mathsf{e}}   % ln e = 1

\newcommand{\wun}{\mathbf{1}}  % the unit of all sorts of things

\newcommand{\inner}[2]{\left\langle #1 , #2 \right\rangle} % scalar products
\newcommand{\normord}[1]{\mbox{${} : #1 : {}$}} % normal ordering ({} necessary to prevent := or =:)

\newcommand{\comm}[2]{\bigl[ #1 , #2 \bigr]}
\newcommand{\acomm}[2]{\bigl\{ #1 , #2 \bigr\}}

\newcommand{\ra}{\rightarrow}
\newcommand{\Ra}{\Rightarrow}
\newcommand{\lra}{\longrightarrow}

\newcommand{\MinMod}[2]{\mathsf{M} \bigl( #1 , #2 \bigr)}                   % Virasoro minimal models

\newcommand{\Mod}[1]{\mathcal{#1}}                 % arbitrary module
\newcommand{\Ver}[1]{\Mod{V}_{#1}}                 % Verma module
\newcommand{\NSVer}[1]{{}^{\text{NS}}\Ver{#1}}     % NS Verma module
\newcommand{\RVer}[1]{{}^{\text{R}}\Ver{#1}}       % R Verma module
\newcommand{\PreVer}[1]{\Mod{U}_{#1}}              % R pre-Verma module
\newcommand{\Irr}[1]{\Mod{L}_{#1}}                 % irreducible module
\newcommand{\NSIrr}[1]{{}^{\text{NS}}\Irr{#1}}     % NS irreducible module
\newcommand{\RIrr}[1]{{}^{\text{R}}\Irr{#1}}       % R irreducible module
\newcommand{\Fock}[1]{\Mod{F}_{#1}}                % Fock module
\newcommand{\NSFock}[1]{{}^{\text{NS}}\Fock{#1}}   % NS Fock module
\newcommand{\RFock}[1]{{}^{\text{R}}\Fock{#1}}     % R Fock module
\newcommand{\Orb}[1]{\Mod{B}_{#1}}                 % bosonic orbifold module
\newcommand{\NSOrb}[1]{{}^{\text{NS}}\Orb{#1}}     % NS bosonic orbifold module
\newcommand{\ROrb}[1]{{}^{\text{R}}\Orb{#1}}       % R bosonic orbifold module

\newcommand{\Kac}[1]{\mathcal{K}_{#1}}             % Kac module
\newcommand{\Stag}[2]{\mathcal{S}_{#1}^{#2}}       % staggered module

\newcommand{\spsub}[1]{#1^{\text{ss}}}             % special subspace of a module

\newcommand{\chmap}{\mathrm{ch}}
\newcommand{\schmap}{\mathrm{sch}}

\newcommand{\Gr}[1]{\bigl[ #1 \bigr]}              % element of a Grothendieck group/ring

\newcommand{\ch}[1]{\chmap \Gr{#1}}                % characters
\newcommand{\fch}[2]{\ch{#1} \bigl( #2 \bigr)}     % characters as functions of q and ...
\newcommand{\sch}[1]{\schmap \Gr{#1}}              % supercharacters
\newcommand{\fsch}[2]{\sch{#1} \bigl( #2 \bigr)}   % supercharacters as functions of q and ...

\newcommand{\jth}[1]{\vartheta_{#1}}               % Jacobi theta
\newcommand{\fjth}[2]{\jth{#1} \bigl( #2 \bigr)}   % Jacobi theta as a function of z, q

\newcommand{\modS}{\mathsf{S}} % modular S-matrix
\newcommand{\Sch}[1]{\modS \Bigl\{ \ch{#1} \Bigr\}}   % shorthand for applying S to a character
\newcommand{\Smat}[2]{\modS \bigl[ #1 \ra #2 \bigr]}  % S-matrix entry

\newcommand{\OrbmodS}{\mathsf{s}} % orbifold modular S-matrix
\newcommand{\OrbSmat}[2]{\OrbmodS \bigl[ #1 \ra #2 \bigr]}  % orbifold S-matrix entry

\newcommand{\fuse}{\mathbin{\times}}                                            % fusion
\newcommand{\Grfuse}{\mathbin{\boxtimes}}                                       % Grothendieck fusion
\newcommand{\fuscoeff}[3]{\mathsf{N}_{#1 \, #2}^{\hphantom{#1 \, #2} #3}}       % fusion coefficient
\newcommand{\orbfuscoeff}[3]{\mathsf{n}_{#1 \, #2}^{\hphantom{#1 \, #2} #3}}    % orbifold fusion coefficient

\newcommand{\coproductsymb}{\Delta}                                                % coproduct symbol
\newcommand{\coproduct}[1]{\coproductsymb \bigl( #1 \bigr)}                        % fusion coproduct
\newcommand{\Ncoproductsymb}[1]{\coproductsymb^{(#1)}}                             % alternative notation for coproduct symbol
\newcommand{\parNcoproductsymb}[2]{\Ncoproductsymb{#1}_{#2}}                       % alternative notation for parametrised coproduct symbol
\newcommand{\parcoproduct}[2]{\coproductsymb_{#1} \bigl( #2 \bigr)}                % fusion coproduct with general parameters
\newcommand{\parNcoproduct}[3]{\Ncoproductsymb{#1}_{#2} \bigl( #3 \bigr)}          % alternative fusion coproduct with general parameters

\newcommand{\dses}[5]{0 \lra #1 \overset{#2}{\lra} #3 \overset{#4}{\lra} #5 \lra 0} % displayed ses
\newcommand{\Res}[1]{#1 {\downarrow}} % restriction (the bracing {\arrows} prevents additional spacing
\newcommand{\Ind}[1]{#1 {\uparrow}}   % induction

\newcommand{\cft}{conformal field theory}
\newcommand{\cfts}{conformal field theories}
\newcommand{\voa}{vertex operator algebra}
\newcommand{\voas}{vertex operator algebras}
\newcommand{\vosa}{vertex operator superalgebra}
\newcommand{\vosas}{vertex operator superalgebras}
\newcommand{\voSa}{vertex operator (super)algebra}
\newcommand{\voSas}{vertex operator (super)algebras}
\newcommand{\uea}{universal enveloping algebra}
\newcommand{\lcft}{logarithmic conformal field theory}
\newcommand{\lcfts}{logarithmic conformal field theories}

\newcommand{\ope}{operator product expansion}
\newcommand{\opes}{operator product expansions}
\newcommand{\hw}{highest-weight}
\newcommand{\hwv}{\hw{} vector}
\newcommand{\hwvs}{\hw{} vectors}
\newcommand{\sv}{singular vector}
\newcommand{\svs}{singular vectors}
\newcommand{\ssv}{subsingular vector}
\newcommand{\ssvs}{subsingular vectors}
\newcommand{\hwm}{\hw{} module}
\newcommand{\hwms}{\hw{} modules}

\newcommand{\NGK}{Nahm-Gaberdiel-Kausch}
\newcommand{\PBW}{Poincar\'{e}-Birkhoff-Witt}
\newcommand{\lhs}{left-hand side}
\newcommand{\rhs}{right-hand side}
\newcommand{\ns}{Neveu-Schwarz}
\newcommand{\ram}{Ramond}

\newcommand{\eps}{\varepsilon}

\newcommand{\tS}{\widetilde{S}}
\newcommand{\tG}{\widetilde{G}}

\DeclareMathOperator{\vspn}{span}

\renewcommand{\ge}{\geqslant}
\renewcommand{\le}{\leqslant}

\makeatletter % stop amsthm environments from forcing what follows to be a new paragraph
\def\@endtheorem{\endtrivlist}%               NEW
\makeatother % ie can now ignore all those \noindent commands...

\theoremstyle{plain}

\begin{document}

\title[Fusion rules for the logarithmic $N=1$ minimal models]{Fusion rules for the logarithmic $\bm{N=1}$ superconformal \\ minimal models II:  including the Ramond sector}

\author[M Canagasabey]{Michael Canagasabey}

\address[Michael Canagasabey]{
Mathematical Sciences Institute \\
Australian National University \\
Acton, ACT 2601 \\
Australia
}

\email{nishan.canagasabey@anu.edu.au}

\author[D Ridout]{David Ridout}

\address[David Ridout]{
Department of Theoretical Physics \\
Research School of Physics and Engineering;
and
Mathematical Sciences Institute;
Australian National University \\
Acton, ACT 2601 \\
Australia
}

\email{david.ridout@anu.edu.au}

\thanks{\today}

\begin{abstract}
The Virasoro logarithmic minimal models were intensively studied by several groups over the last ten years with much attention paid to the fusion rules and the structures of the indecomposable representations that fusion generates.  The analogous study of the fusion rules of the $N=1$ superconformal logarithmic minimal models was initiated in \cite{CanFusI15} as a continuum counterpart to the lattice explorations of \cite{PeaLog14}.  These works restricted fusion considerations to Neveu-Schwarz representations.  Here, this is extended to include the Ramond sector.  Technical advances that make this possible include a fermionic Verlinde formula applicable to logarithmic conformal field theories and a twisted version of the fusion algorithm of Nahm and Gaberdiel-Kausch.  The results include the first construction and detailed analysis of logarithmic structures in the Ramond sector.
\end{abstract}

\maketitle

\onehalfspacing

\section{Introduction} \label{sec:Intro}

The last ten years have seen significant advances in the study of the so-called logarithmic conformal field theories \cite{RozQua92,GurLog93,GabLoc99}, making it clear that such theories are neither pathological nor intractable.  Rather, it is now recognised \cite{CreLog13,GaiLog13,QueSup13} that logarithmic theories successfully model non-local observables in statistical lattice models \cite{PeaLog06,ReaAss07,RidPer07,GaiAss12,RueLog13} and string theories with fermionic spacetime symmetries \cite{SalGL106,SalSU207,QueFre07,CreRel11}.  From a mathematical point of view, ``logarithmic'' means that the relevant category of modules over the \voa{} is non-semisimple.  More precisely, it means that the hamiltonian acts non-diagonalisably on the quantum state space.  This leads to many subtle mathematical questions and the field of logarithmic \voas{} is now being actively pursued by mathematicians, see \cite{HuaLog06,AdaTri08,AdaVer13,HuaTen13} for example.

In \cite{CanFusI15}, we instigated a detailed study of certain \lcfts{} with $N=1$ supersymmetry.  These are the $N=1$ \emph{logarithmic minimal models}, corresponding to the universal \voas{} associated with the \ns{} algebra.  Some abstract consequences of combining supersymmetry with logarithmic structures had already been studied in \cite{KhoLog98,MavNev03} and a detailed discussion of the $N=1$ triplet models may be found in \cite{AdaN=109}.  Our motivation, however, is a recent lattice-theoretic study, reported in \cite{PeaLog14}, in which certain fused loop models were conjectured to have the $N=1$ logarithmic minimal models as their continuum scaling limits.  We do not work with these loop models, instead preferring to study certain aspects of the $N=1$ logarithmic minimal models directly using field- and representation-theoretic methods.

In particular, we study the fusion rules of certain $N=1$ representations known as Kac modules.  Originally introduced non-constructively \cite{PeaLog06} for Virasoro logarithmic minimal models to describe conjectured limiting partition functions for boundary sectors of (non-fused) loop models, candidates for Virasoro Kac modules were proposed in \cite{RasCla11}, for certain models, then confirmed and generalised in \cite{MorKac15}.  In \cite{CanFusI15}, we introduced the $N=1$ Kac modules, following these papers and \cite{PeaLog14}, investigating them and their fusion rules in the \ns{} sector.  Here, we extend this investigation to include the Ramond sector, overcoming the significant technical difficulties that result from working with twisted representations.

The two main tools that we develop for this investigation are a fermionic analogue of the ``standard'' Verlinde formula of \cite{RidVer14} and a twisted version of the \NGK{} fusion algorithm \cite{NahQua94,GabInd96}.  The standard Verlinde formula is the centrepiece of the standard module formalism that is being developed to analyse the modular properties of \lcfts{} \cite{CreRel11,CreWAl11,CreMod12,BabTak12,CreMod13,RidMod13,RidBos14,MorKac15}.  Combining this formalism with simple current technology \cite{SchExt89,SchSimp90}, as was done for the rational Verlinde formula in \cite{EhoFus94}, we arrive at a Verlinde formula that gives the (super)character of a fusion product involving both \ns{} and Ramond modules.  On the other hand, the \NGK{} algorithm gives an algorithmic means of explicitly constructing fusion products and analysing the resulting structures.  Originally applying only to untwisted modules, a twisted generalisation was first discussed in \cite{GabFus97}.  We simplify this discussion significantly and detail the practical implementation of the algorithm, necessary for explicit fusion calculations with Ramond modules.

We begin, in \cref{sec:Back}, with a thorough review of the representation theory of the $N=1$ superconformal algebras, focusing on Verma modules and Fock spaces.  As the \ns{} sector was discussed in \cite{CanFusI15}, and is anyway very similar to Virasoro representation theory, we concentrate here on the Ramond sector.  In particular, we detail the unusual \ssv{} structures of the Verma modules and Fock spaces corresponding to the case where the conformal \hw{} $h$ and central charge $c$ satisfy $h = \frac{c}{24}$, referring to \cite{IohRepI03,IohRepII03,IohStr06} for a more complete treatment.  The section concludes with the definition of an $N=1$ Kac module.  These modules play a central role in what follows.

\cref{sec:CharMod} introduces the characters and supercharacters of the \ns{} and Ramond Fock spaces, these playing the role of the standard modules of the theory.  The S-matrix (here, the kernel of an integral transform similar to the Fourier transform) is computed and the results are generalised to the \ns{} and Ramond Kac modules.  We then introduce a fermionic version \eqref{eq:Verlinde} of the standard Verlinde formula and use it to compute the character and supercharacter of the fusion product of any two Kac modules.  This already settles several conjectures left unsolved in \cite{CanFusI15} concerning the relative parities of the direct summands of \ns{} Kac module fusion products.  A derivation of this fermionic Verlinde formula is presented in \cref{sec:FermVer}, followed by an explicit check of the formula, in \cref{app:VerRat}, applied to the free fermion.

Having determined the character and supercharacter of every Kac module fusion product, we now ask how to identify the indecomposable direct summands of such fusion products.  One means of exploring this question is the twisted \NGK{} fusion algorithm which we describe in detail in \cref{sec:TwFus}.  After introducing a convenient filtration of the fusion product, we present the twisted coproduct formulae that define the action of the superconformal modes on the fusion product.  These generalise the untwisted formulae of \cite{GabFus94,GabFus94b} and appear simpler than the twisted formulae of \cite{GabFus97}.  These formulae are derived in \cref{app:TwCoprod}, for completeness.  We use these new formulae to deduce the correct twisted versions of the special subspace and truncated subspace, generalising the untwisted results of \cite{NahQua94,GabInd96}.

These twisted results illustrate that working explicitly with Ramond modules is significantly more laborious than pure \ns{} calculations.  Nevertheless, we proceed to discuss two explicit fusion computations performed using the twisted \NGK{} algorithm.  We emphasise that the complete identification of these fusion products, and those that follow, is only possible because we can first determine their characters and supercharacters, hence their composition factors, from the fermionic Verlinde formula.

The first explicit computation, in \cref{sec:FusRR}, fuses two Ramond Kac modules and the result is found to be the direct sum of two \ns{} \emph{staggered modules} \cite{RohRed96,RidSta09,CanFusI15}, one being a parity-reversed copy of the other.  Such staggered modules are characterised by rank $2$ Jordan blocks for the action of the Virasoro zero mode $L_0$ and their identification, up to isomorphism, usually requires calculating a single auxiliary parameter, the logarithmic coupling $\beta \in \CC$.

\cref{sec:FusNSR} details the fusion of a Ramond Kac module with a \ns{} Kac module.  The result is a Ramond staggered module, the theory of such modules being outlined in \cref{app:RStag}.  To the best of our knowledge, this is the first time that such Ramond staggered modules have been constructed and their structure analysed.  We remark that the presence of rank $2$ Jordan blocks for $L_0$ automatically implies that the superpartner mode $G_0$ also acts with Jordan blocks (though making this manifest would destroy the natural splitting into bosonic and fermionic subspaces).  The Ramond staggered modules that we construct are usually characterised by a single logarithmic coupling.  However, the example detailed here exhibits a novelty in that two independent logarithmic couplings are required to completely fix its isomorphism class.

\cref{sec:Results} then outlines the further results that we have been able to obtain using the fermionic Verlinde formula and a computer implementation of the twisted \NGK{} fusion algorithm.  We present only Ramond by Ramond and Ramond by \ns{} results as our \ns{} by \ns{} results were already summarised in \cite{CanFusI15}.  After reporting a few conjectures that our results suggest, we turn to a brief discussion in \cref{sec:Conc}, putting our results in context and indicating future directions of research.

\section*{Acknowledgements}

MC is supported by an Australian Postgraduate Award from the Australian Government.  DR's research is supported by the Australian Research Council Discovery Project DP1093910.  The authors thank Scott Melville, J\o{}rgen Rasmussen and Simon Wood for helpful correspondence and discussions.  DR would also like to thank Kenji Iohara for numerous superconformal discussions during the trimestre ``Advanced Conformal Field Theory and Applications'' at the Institut Henri Poincar\'{e}, Paris, in 2011.

\section{$N=1$ representation theory} \label{sec:Back}

In this section, we review the $N=1$ superconformal algebras and certain aspects of their representation theories, thereby fixing our notation and conventions.  As is well known, much of this representation theory parallels that of the Virasoro algebra.  However, there are a few important differences that deserve emphasis, particularly as regards \sv{} multiplicities in the Ramond sector.

\subsection{$\bm{N=1}$ algebras} \label{sec:Alg}

The $N=1$ superconformal algebras are infinite-dimensional complex Lie superalgebras.  They may be defined as the vector superspaces spanned by even (bosonic) modes, $L_n$ and $C$, and odd (fermionic) modes $G_k$, equipped with the following brackets:
\begin{equation} \label{eq:CommN=1}
\begin{aligned}
\comm{L_m}{L_n} &= \brac{m-n} L_{m+n} + \frac{1}{12} \brac{m^3-m} \delta_{m+n=0} \, C, & \comm{L_m}{G_k} &= \brac{\frac{1}{2} m - k} G_{m+k}, \\
\acomm{G_j}{G_k} &= 2 L_{j+k} + \frac{1}{3} \brac{j^2-\frac{1}{4}} \delta_{j+k=0} \, C, & \comm{L_m}{C} &= \comm{G_j}{C} = 0.
\end{aligned}
\end{equation}
More precisely, there are two $N=1$ superconformal algebras which are distinguished by the values taken by the index $k$ of the fermionic modes $G_k$:  The \emph{\ns{} algebra} takes $k \in \ZZ + \frac{1}{2}$, whereas the \emph{Ramond algebra} takes $k \in \ZZ$.  Both algebras require the index $n$ of the bosonic modes $L_n$ to be an integer, hence the bosonic subalgebra of each is identified with the Virasoro algebra.

The algebraic structures of interest to us are the universal \vosas{} associated with the \ns{} algebra.  There are an infinite number of these, parametrised by the central charge $c \in \CC$, and they are realised \cite{KacVer94} on the \ns{} module generated by a \hwv{} $\Omega$ satisfying
\begin{equation}
L_0 \Omega = 0, \qquad C \, \Omega = c \, \Omega, \qquad G_{-1/2} \Omega = 0.
\end{equation}
In other words, each such universal \vosa{} is defined on the quotient of the \ns{} Verma module $\NSVer{0}$, of conformal weight $0$ and central charge $c$, by the submodule generated by the \sv{} of conformal weight $\frac{1}{2}$ (see \cref{sec:Verma} below).  We will refer to these universal \vosas{} as the \emph{$N=1$ algebras}, for short.

The category of modules over a given $N=1$ algebra is a full subcategory of the category of \ns{} modules consisting of the modules $\Mod{M}$ that satisfy the following conditions:  The central element $C$ acts on $\Mod{M}$ as $c$ times the identity operator and, for each $v \in \Mod{M}$, one has $L_n v = G_k v = 0$ for all sufficiently large $n$ and $k$.  The latter condition ensures that the orders of the poles in the \opes{} of $T(z)$ and $G(z)$ with $v(w)$, hence in those of every $N=1$ field with $v(w)$, are bounded above.  In what follows, we shall further restrict to modules that admit a $\ZZ_2$-grading compatible with that of the generators $L_n$ and $G_k$.  In other words, each $N=1$ module decomposes as a direct sum of two subspaces, one even and the other odd; each is preserved by the action of $L_n$ and they are swapped by the action of $G_k$.

For reasons of physical consistency, one is also led to consider the Ramond modules that satisfy the same conditions.  Mathematically, Ramond modules are \emph{twisted} modules over the \ns{} algebra, hence over the $N=1$ algebra, though we will usually drop this qualifier in what follows and use the term \emph{$N=1$ module} to include both \ns{} and Ramond modules.  We define the \emph{\ns{} sector} to consist of the $N=1$ modules that are \ns{} modules and the \emph{Ramond sector} to consist of the (twisted) $N=1$ modules that are Ramond modules.

Field-theoretically, each $N=1$ algebra extends the universal Virasoro \voa{} (of the same central charge) by a fermionic primary field $G(z)$ of conformal weight $\frac{3}{2}$.  With the mode decompositions
\begin{equation} \label{eq:TGFourier}
T(z) = \sum_{n \in \ZZ} L_n z^{-n-2}, \qquad 
G(z) = \sum_{k \in \ZZ + \eps} G_k z^{-k-3/2},
\end{equation}
where $\eps = \frac{1}{2}$ in the \ns{} sector and $\eps = 0$ in the Ramond sector, the Lie brackets \eqref{eq:CommN=1} are equivalent to the \opes{}
\begin{equation} \label{OPE:TTTGGG}
\begin{gathered}
T(z) T(w) \sim \frac{c/2}{\brac{z-w}^4} + \frac{2 \, T(w)}{\brac{z-w}^2} + \frac{\pd T(w)}{z-w}, \\
T(z) G(w) \sim \frac{\frac{3}{2} \, G(w)}{\brac{z-w}^2} + \frac{\pd G(w)}{z-w}, \qquad 
G(z) G(w) \sim \frac{2c/3}{\brac{z-w}^3} + \frac{2 \, T(w)}{z-w}.
\end{gathered}
\end{equation}
Note that the energy-momentum tensor and a Virasoro primary field are always mutually local in correlation functions (see \cite{RidLog07} for example):  $T(z) G(w) = G(w) T(z)$.  We emphasise that we have defined the $N=1$ algebra to be universal, meaning that the \opes{} \eqref{OPE:TTTGGG} generate a complete set of relations.  In particular, the $N=1$ algebra never coincides with an $N=1$ minimal model \vosa{}, even when $c$ is a minimal model central charge.

\subsection{Extended Kac tables} \label{sec:KacTables}

The standard parametrisation suggested by the $N=1$ analogues \cite{KacCon79,FriSup85,MeuHig86} of the Kac determinant formula is
\begin{equation}
c = \frac{15}{2} - 3 \brac{t+t^{-1}}, \qquad 
h_{r,s} = \frac{r^2-1}{8} t^{-1} - \frac{rs-1}{4} + \frac{s^2-1}{8} t + \frac{1}{16} \delta_{r \neq s \bmod{2}},
\end{equation}
where $r,s \in \ZZ$ and $t \in \CC \setminus \set{0}$.  We remark that in applications to representation theory, the conformal weight $h_{r,s}$ is associated to a module in the \ns{} sector, when $r=s \bmod{2}$, and to a module in the Ramond sector, when $r \neq s \bmod{2}$.  If $t$ is rational, then this parametrisation may be written in the form
\begin{equation} \label{eq:ParByt}
t = \frac{p}{p'}, \qquad 
c = \frac{3}{2} \brac{1 - \frac{2 \brac{p'-p}^2}{pp'}}, \qquad 
h_{r,s} = \frac{\brac{p'r-ps}^2 - \brac{p'-p}^2}{8pp'} + \frac{1}{16} \delta_{r \neq s \bmod{2}},
\end{equation}
where one customarily imposes the constraints $p=p' \bmod{2}$ and $\gcd \set{p, \frac{1}{2} \brac{p'-p}} = 1$.

The $N=1$ superconformal minimal models \cite{EicMin85,BerSup85,FriSup85} correspond to $p,p' \in \ZZ_{\ge 2}$ satisfying these constraints.  The indecomposable modules of the $N=1$ minimal model \vosa{} are precisely \cite{AdaRat97,MilCha07} the simple \hwms{} of conformal highest weight $h_{r,s}$, where $1 \le r \le p-1$ and $1 \le s \le p'-1$.  This range of $r$ and $s$ defines the ($N=1$) \emph{Kac table} in which the entries are the conformal weights $h_{r,s}$.

For studying the representation theory of the (universal) $N=1$ algebras, it is convenient to consider instead the \emph{extended Kac table} in which the entries $h_{r,s}$ are indexed by $r,s \in \ZZ_{>0}$.  This table is relevant for all values of $t$, hence all central charges, but we shall focus here exclusively on the case $t \in \QQ_{>0}$.  Defining $p,p' \in \ZZ_{>0}$ as above, we partition the entries of the extended Kac table into four disjoint subsets (some of which may be empty):
\begin{itemize}
\item If $p$ divides $r$ and $p'$ divides $s$, then we say that $(r,s)$ is of \emph{corner} type in the extended Kac table.
\item If $p$ divides $r$ or $p'$ divides $s$, but not both, then $(r,s)$ is said to be of \emph{boundary} type.
\item If $r = \frac{1}{2} p \bmod{p}$ and $s = \frac{1}{2} p' \bmod{p'}$, then $(r,s)$ is said to be of \emph{centre} type.
\item Otherwise, $(r,s)$ is said to be of \emph{interior} type.
\end{itemize}
We note the following facts:  If $p$ and $p'$ are odd, then there are no entries of centre type in the extended Kac table; if $p=1$ or $p'=1$, then there are no interior entries; if $p=p'=1$, then there are no boundary entries.  The extended Kac table for $t=1$, hence $(p,p') = (1,1)$ and $c=\frac{3}{2}$, therefore consists entirely of corner entries.  To illustrate the other possibilities, we present (parts of) four extended Kac tables in \cref{fig:KacTables}.

{
\renewcommand{\arraystretch}{1.1}
\begin{figure}
\begin{center}
\scalebox{0.85}{
\begin{tikzpicture}
\node (K13) at (0,0) {
\setlength{\extrarowheight}{4pt}
\begin{tabular}{|CC|C|CC|C|CC|C|CC|C|C}
\hline
\BKL 0 & \BKL -\frac{1}{16} & -\frac{1}{6} & \BKL -\frac{1}{16} & \BKL 0 & \frac{13}{48} & \BKL \frac{1}{2} & \BKL \frac{15}{16} & \frac{4}{3} & \BKL \frac{31}{16} & \BKL \frac{5}{2} & \frac{157}{48} & \BKL \cdots \\[1mm]
\hline
\BKL \frac{15}{16} & \BKL \frac{1}{2} & \frac{13}{48} & \BKL 0 & \BKL -\frac{1}{16} & -\frac{1}{6} & \BKL -\frac{1}{16} & \BKL 0 & \frac{13}{48} & \BKL \frac{1}{2} & \BKL \frac{15}{16} & \frac{4}{3} & \BKL \cdots \\[1mm]
\hline
\BKL \frac{5}{2} & \BKL \frac{31}{16} & \frac{4}{3} & \BKL \frac{15}{16} & \BKL \frac{1}{2} & \frac{13}{48} & \BKL 0 & \BKL -\frac{1}{16} & -\frac{1}{6} & \BKL -\frac{1}{16} & \BKL 0 & \frac{13}{48} & \BKL \cdots \\[1mm]
\hline
\BKL \frac{79}{16} & \BKL 4 & \frac{157}{48} & \BKL \frac{5}{2} & \BKL \frac{31}{16} & \frac{4}{3} & \BKL \frac{15}{16} & \BKL \frac{1}{2} & \frac{13}{48} & \BKL 0 & \BKL -\frac{1}{16} & -\frac{1}{6} & \BKL \cdots \\[1mm]
\hline
\BKL 8 & \BKL \frac{111}{16} & \frac{35}{6} & \BKL \frac{79}{16} & \BKL 4 & \frac{157}{48} & \BKL \frac{5}{2} & \BKL \frac{31}{16} & \frac{4}{3} & \BKL \frac{15}{16} & \BKL \frac{1}{2} & \frac{13}{48} & \BKL \cdots \\[1mm]
\hline
\BKL \frac{191}{16} & \BKL \frac{21}{2} & \frac{445}{48} & \BKL 8 & \BKL \frac{111}{16} & \frac{35}{6} & \BKL \frac{79}{16} & \BKL 4 & \frac{157}{48} & \BKL \frac{5}{2} & \BKL \frac{31}{16} & \frac{4}{3} & \BKL \cdots \\[1mm]
\hline
\BKL \vdots & \BKL \vdots & \vdots & \BKL \vdots & \BKL \vdots & \vdots & \BKL \vdots & \BKL \vdots & \vdots & \BKL \vdots & \BKL \vdots & \vdots & \BKL \ddots
\end{tabular}
};
\node [below=3mm of K13] {
$t = \dfrac{1}{3}, \qquad (p,p')=(1,3),\qquad c = -\dfrac{5}{2}.$
};
\node (K24) [below=20mm of K13] {
\setlength{\extrarowheight}{4pt}
\begin{tabular}{|CCC|C|CCC|C|CCC|C|C}
\hline
\IKL 0 & \CKL 0 & \IKL 0 & \BKL \frac{1}{4} & \IKL \frac{1}{2} & \CKL 1 & \IKL \frac{3}{2} & \BKL \frac{9}{4} & \IKL 3 & \CKL 4 & \IKL 5 & \BKL \frac{25}{4} & \IKL \cdots \\[1mm]
\hline
\BKL \frac{9}{16} & \BKL \frac{3}{16} & \BKL \frac{1}{16} & -\frac{1}{16} & \BKL \frac{1}{16} & \BKL \frac{3}{16} & \BKL \frac{9}{16} & \frac{15}{16} & \BKL \frac{25}{16} & \BKL \frac{35}{16} & \BKL \frac{49}{16} & \frac{63}{16} & \BKL \cdots \\[1mm]
\hline
\IKL \frac{3}{2} & \CKL 1 & \IKL \frac{1}{2} & \BKL \frac{1}{4} & \IKL 0 & \CKL 0 & \IKL 0 & \BKL \frac{1}{4} & \IKL \frac{1}{2} & \CKL 1 & \IKL \frac{3}{2} & \BKL \frac{9}{4} & \IKL \cdots \\[1mm]
\hline
\BKL \frac{49}{16} & \BKL \frac{35}{16} & \BKL \frac{25}{16} & \frac{15}{16} & \BKL \frac{9}{16} & \BKL \frac{3}{16} & \BKL \frac{1}{16} & -\frac{1}{16} & \BKL \frac{1}{16} & \BKL \frac{3}{16} & \BKL \frac{9}{16} & \frac{15}{16} & \BKL \cdots \\[1mm]
\hline
\IKL 5 & \CKL 4 & \IKL 3 & \BKL \frac{9}{4} & \IKL \frac{3}{2} & \CKL 1 & \IKL \frac{1}{2} & \BKL \frac{1}{4} & \IKL 0 & \CKL 0 & \IKL 0 & \BKL \frac{1}{4} & \IKL \cdots \\[1mm]
\hline
\BKL \frac{121}{16} & \BKL \frac{99}{16} & \BKL \frac{81}{16} & \frac{63}{16} & \BKL \frac{49}{16} & \BKL \frac{35}{16} & \BKL \frac{25}{16} & \frac{15}{16} & \BKL \frac{9}{16} & \BKL \frac{3}{16} & \BKL \frac{1}{16} & -\frac{1}{16} & \BKL \cdots \\[1mm]
\hline
\IKL \vdots & \CKL \vdots & \IKL \vdots & \BKL \vdots & \IKL \vdots & \CKL \vdots & \IKL \vdots & \BKL \vdots & \IKL \vdots & \CKL \vdots & \IKL \vdots & \BKL \vdots & \IKL \ddots
\end{tabular}
};
\node [below=3mm of K24] {
$t = \dfrac{1}{2}, \qquad (p,p')=(2,4),\qquad c = 0.$
};
\node (K35) [below=20mm of K24] {
\setlength{\extrarowheight}{4pt}
\begin{tabular}{|CCCC|C|CCCC|C|CCC}
\hline
\IKL 0 & \IKL \frac{3}{80} & \IKL \frac{1}{10} & \IKL \frac{7}{16} & \BKL \frac{4}{5} & \IKL \frac{23}{16} & \IKL \frac{21}{10} & \IKL \frac{243}{80} & \IKL 4 & \BKL \frac{419}{80} & \IKL \frac{13}{2} & \IKL \frac{643}{80} & \IKL \cdots \\[1mm]
\IKL \frac{7}{16} & \IKL \frac{1}{10} & \IKL \frac{3}{80} & \IKL 0 & \BKL \frac{19}{80} & \IKL \frac{1}{2} & \IKL \frac{83}{80} & \IKL \frac{8}{5} & \IKL \frac{39}{16} & \BKL \frac{33}{10} & \IKL \frac{71}{16} & \IKL \frac{28}{5} & \IKL \cdots \\[1mm]
\hline
\BKL \frac{7}{6} & \BKL \frac{169}{240} & \BKL \frac{4}{15} & \BKL \frac{5}{48} & -\frac{1}{30} & \BKL \frac{5}{48} & \BKL \frac{4}{15} & \BKL \frac{169}{240} & \BKL \frac{7}{6} & \frac{457}{240} & \BKL \frac{8}{3} & \BKL \frac{889}{240} & \BKL \cdots \\[1mm]
\hline
\IKL \frac{39}{16} & \IKL \frac{8}{5} & \IKL \frac{83}{80} & \IKL \frac{1}{2} & \BKL \frac{19}{80} & \IKL 0 & \IKL \frac{3}{80} & \IKL \frac{1}{10} & \IKL \frac{7}{16} & \BKL \frac{4}{5} & \IKL \frac{23}{16} & \IKL \frac{21}{10} & \IKL \cdots \\[1mm]
\IKL 4 & \IKL \frac{243}{80} & \IKL \frac{21}{10} & \IKL \frac{23}{16} & \BKL \frac{4}{5} & \IKL \frac{7}{16} & \IKL \frac{1}{10} & \IKL \frac{3}{80} & \IKL 0 & \BKL \frac{19}{80} & \IKL \frac{1}{2} & \IKL \frac{83}{80} & \IKL \cdots \\[1mm]
\hline
\BKL \frac{293}{48} & \BKL \frac{143}{30} & \BKL \frac{889}{240} & \BKL \frac{8}{3} & \frac{457}{240} & \BKL \frac{7}{6} & \BKL \frac{169}{240} & \BKL \frac{4}{15} & \BKL \frac{5}{48} & -\frac{1}{30} & \BKL \frac{5}{48} & \BKL \frac{4}{15} & \BKL \cdots \\[1mm]
\hline
\IKL \vdots & \IKL \vdots & \IKL \vdots & \IKL \vdots & \BKL \vdots & \IKL \vdots & \IKL \vdots & \IKL \vdots & \IKL \vdots & \BKL \vdots & \IKL \vdots & \IKL \vdots & \IKL \ddots
\end{tabular}
};
\node [below=3mm of K35] {
$t = \dfrac{3}{5}, \qquad (p,p')=(3,5),\qquad c = \dfrac{7}{10}.$
};
\node (K46) [below=20mm of K35] {
\setlength{\extrarowheight}{4pt}
\begin{tabular}{|CCCCC|C|CCCCC|C|C}
\hline
\IKL 0 & \IKL \frac{1}{16} & \IKL \frac{1}{6} & \IKL \frac{9}{16} & \IKL 1 & \BKL \frac{83}{48} & \IKL \frac{5}{2} & \IKL \frac{57}{16} & \IKL \frac{14}{3} & \IKL \frac{97}{16} & \IKL \frac{15}{2} & \BKL \frac{443}{48} & \IKL \cdots \\[1mm]
\IKL \frac{3}{8} & \IKL \frac{1}{16} & \CKL \frac{1}{24} & \IKL \frac{1}{16} & \IKL \frac{3}{8} & \BKL \frac{35}{48} & \IKL \frac{11}{8} & \IKL \frac{33}{16} & \CKL \frac{73}{24} & \IKL \frac{65}{16} & \IKL \frac{43}{8} & \BKL \frac{323}{48} & \IKL \cdots \\[1mm]
\IKL 1 & \IKL \frac{9}{16} & \IKL \frac{1}{6} & \IKL \frac{1}{16} & \IKL 0 & \BKL \frac{11}{48} & \IKL \frac{1}{2} & \IKL \frac{17}{16} & \IKL \frac{5}{3} & \IKL \frac{41}{16} & \IKL \frac{7}{2} & \BKL \frac{227}{48} & \IKL \cdots \\[1mm]
\hline
\BKL \frac{17}{8} & \BKL \frac{21}{16} & \BKL \frac{19}{24} & \BKL \frac{5}{16} & \BKL \frac{1}{8} & -\frac{1}{48} & \BKL \frac{1}{8} & \BKL \frac{5}{16} & \BKL \frac{19}{24} & \BKL \frac{21}{16} & \BKL \frac{17}{8} & \frac{143}{48} & \BKL \cdots \\[1mm]
\hline
\IKL \frac{7}{2} & \IKL \frac{41}{16} & \IKL \frac{5}{3} & \IKL \frac{17}{16} & \IKL \frac{1}{2} & \BKL \frac{11}{48} & \IKL 0 & \IKL \frac{1}{16} & \IKL \frac{1}{6} & \IKL \frac{9}{16} & \IKL 1 & \BKL \frac{83}{48} & \IKL \cdots \\[1mm]
\IKL \frac{43}{8} & \IKL \frac{65}{16} & \CKL \frac{73}{24} & \IKL \frac{33}{16} & \IKL \frac{11}{8} & \BKL \frac{35}{48} & \IKL \frac{3}{8} & \IKL \frac{1}{16} & \CKL \frac{1}{24} & \IKL \frac{1}{16} & \IKL \frac{3}{8} & \BKL \frac{35}{48} & \IKL \cdots \\[1mm]
\IKL \vdots & \IKL \vdots & \IKL \vdots & \IKL \vdots & \IKL \vdots & \BKL \vdots & \IKL \vdots & \IKL \vdots & \IKL \vdots & \IKL \vdots & \IKL \vdots & \BKL \vdots & \IKL \ddots
\end{tabular}
};
\node [below=3mm of K46] {
$t = \dfrac{2}{3}, \qquad (p,p')=(4,6),\qquad c = 1.$
};
\end{tikzpicture}
}
\caption{Parts of four of the extended $N=1$ Kac tables.  The rows of the tables are labelled by $r = 1, 2, 3, \ldots$ and the columns by $s = 1, 2, 3, \ldots$\,.  Centre entries are shaded dark grey, interior entries are grey, boundary entries are light grey, while corner entries are white.} \label{fig:KacTables}
\end{center}
\end{figure}
}

\subsection{Verma modules} \label{sec:Verma}

In the \ns{} sector, one obtains a \hw{} theory from the triangular decomposition that splits the \ns{} algebra into the spans of the positive modes $L_n$ and $G_k$, with $n,k>0$, the negative modes $L_n$ and $G_k$, with $n,k<0$, and the zero modes $L_0$ and $C$.  A \ns{} \hwv{} is therefore characterised by its $L_0$-eigenvalue $h$ (because $C = c \, \wun$ in the \vosa{}), also called its conformal weight.  We denote by $\NSVer{h}$ the \ns{} Verma module generated by a \hwv{} of conformal weight $h$.  Its (unique) simple quotient will be denoted by $\NSIrr{h}$.

The determinant formula \cite{KacCon79,MeuHig86} for \ns{} Verma modules indicates that a given Verma module $\NSVer{h}$ is simple, $\NSVer{h} = \NSIrr{h}$, unless $h = h_{r,s}$ for some $r,s \in \ZZ_{>0}$ with $r=s \bmod{2}$.  In this case, $\NSVer{h_{r,s}}$ possesses a \sv{} of \emph{depth} $\frac{1}{2} rs$, meaning that its conformal weight is $h_{r,s} + \frac{1}{2} rs$.  We will therefore denote the non-simple Verma module $\NSVer{h_{r,s}}$ by $\Ver{r,s}$, for clarity, implicitly understanding that it belongs to the \ns{} sector because $r=s \bmod{2}$.  Similarly, the simple quotient of $\Ver{r,s}$ will be denoted by $\Irr{r,s}$.

It is useful to note that \ns{} \hwms{} may be naturally $\ZZ_2$-graded because assigning a parity to a \hwv{} automatically results in a well-defined parity for each \PBW{} basis vector.  This generalises to other indecomposable \ns{} (generalised) weight modules if we replace ``\hwv{}'' by ``ground state'', meaning a vector of minimal conformal weight (all ground states must have the same parity).  Such a grading is required for many physical calculations, in particular for the fusion computations that we report here.  However, there are two choices of parity assignment for each \ns{} indecomposable:  either the ground states are bosonic or they are fermionic.

We will therefore affix a superscript $+$ or $-$ to indecomposable \ns{} modules according to the parity, bosonic or fermionic, respectively, of their ground states.\footnote{Actually, we shall sometimes omit this superscript on an $N=1$ module if its parity is not important for the discussion at hand.}  Thus, $\Ver{r,s}^+$ is generated by a bosonic \hwv{} whereas the \hwv{} generating $\Ver{r,s}^-$ is fermionic.  We remark that $\Mod{M}^+$ and $\Mod{M}^-$ are isomorphic as $N=1$ modules, but not as $\ZZ_2$-graded $N=1$ modules.  However, there is an obvious functor $\Pi$ that reverses the parity of each indecomposable \ns{} module.  Concretely, $\Pi$ amounts to tensoring with the one-dimensional trivial fermionic \ns{} module $\CC^-$ of central charge $0$.

In contrast, the concept of \hw{} theory is a little more subtle in the Ramond sector.  Because of $G_0$, the obvious splitting into positive, negative and zero modes no longer defines a triangular decomposition because the zero modes do not span an abelian Lie superalgebra:
\begin{equation} \label{eq:G_0^2}
G_0^2 = \frac{1}{2} \acomm{G_0}{G_0} = L_0 - \frac{C}{24}.
\end{equation}
Instead, one considers this splitting as defining a generalised triangular decomposition with respect to which (generalised) Verma modules are defined by inducing from an arbitrary simple module over the zero mode subalgebra (this notion is called a relaxed Verma module in \cite{RidRel15}).

The simple $\ZZ_2$-graded weight modules over $\vspn \set{L_0, C, G_0}$ are two-dimensional whenever $h \neq \frac{c}{24}$, because $v$ having conformal weight $h$ implies that $\vspn \set{v, G_0 v}$ is simple:
\begin{equation}
G_0 G_0 v = \brac{L_0 - \frac{C}{24}} v = \brac{h - \frac{c}{24}} v \neq 0.
\end{equation}
We remark that the $\ZZ_2$-grading requirement is necessary because $G_0$ has two linearly independent eigenvectors on this space, each of which spans a simple weight submodule that does not admit a $\ZZ_2$-grading (any $G_0$-eigenvector of non-zero eigenvalue cannot be consistently assigned a parity).\footnote{We mention that the literature does, from time to time, consider Ramond modules generated by simultaneous eigenvectors of $L_0$, $C$ and $G_0$.  As these modules cannot be $\ZZ_2$-graded, their physical relevance is unclear to us.}  If, however, $h = \frac{c}{24}$, then $G_0$ acts nilpotently and there is a unique simple $\ZZ_2$-graded weight module over $\vspn \set{L_0, C, G_0}$.  Its dimension is $1$.

We conclude that the Ramond Verma module $\RVer{h}$, generated by a \hwv{} of conformal weight $h$, has two independent ground states, $v$ and $G_0 v$ say, provided that $h \neq \frac{c}{24}$.  Since these ground states necessarily have opposite parities, it follows that the parity-reversing functor $\Pi$ fixes these Verma modules:  $\RVer{h} \cong \Pi \RVer{h}$ if $h \neq \frac{c}{24}$.  There is therefore no need to affix a superscript sign to these Verma modules.  The same is true for quotients of such Ramond Verma modules because their \svs{} always come in pairs of the same conformal weight, one bosonic and one fermionic \cite[Rem~3.2]{IohRepI03}.  When $h = \frac{c}{24}$, the Ramond Verma module $\RVer{c/24}$ has instead one independent ground state and $G_0$ acts on it as $0$.  This module is not fixed by $\Pi$, hence neither are its quotients, and so we shall affix a superscript sign to indicate the parity of the ground state.

The determinant formula \cite{FriSup85,MeuHig86} for Ramond Verma modules shows that $\RVer{h}$ is simple, unless $h = h_{r,s}$ for some $r,s \in \ZZ_{>0}$ with $r \neq s \bmod{2}$.  Again, $\RVer{h_{r,s}}$ has a singular vector at depth $\frac{1}{2} rs$ in this case.  We therefore define $\Ver{r,s} = \RVer{h_{r,s}}$ and $\Irr{r,s} = \RIrr{h_{r,s}}$, when $r,s \in \ZZ_{>0}$ and $r \neq s \bmod{2}$, complementing the \ns{} sector definition.  If $p$ and $p'$ are both even, so that centre type modules are defined, then the centre Verma modules $\Ver{p/2+mp,p'/2+np'}$ ($m,n \in \ZZ_{\ge 0}$) are all Ramond.  We note that the exceptional case with $h=\frac{c}{24}$ corresponds to the centre type module $\Ver{p/2,p'/2}$.  If $p$ and $p'$ are both odd, this exceptional case does not correspond to any entry in the extended Kac table.

We can now summarise the structure theory \cite{AstStr97,IohRepI03} of $N=1$ Verma modules, restricting to the case $t \in \QQ_{>0}$ and the modules $\Ver{r,s}$ that are of most relevance to this paper.  As with the structures of Virasoro Verma modules, it turns out that every non-zero submodule of an $N=1$ Verma module is generated by \svs{} \cite[Thm.~4.2]{IohRepI03}.  When $(r,s)$ is a corner or boundary entry in the extended Kac table, the \svs{} are arranged in an infinite chain pattern; when $(r,s)$ is an interior entry, the \svs{} form an infinite braid instead.  We illustrate these patterns in \cref{fig:VermaStructures} and refer to \cite{IohRepI03} for explicit formulae for the conformal weights of the \svs{}.  For \ns{} Verma modules, the multiplicity of a \sv{} in a given weight space ($L_0$-eigenspace) is either $1$ or $0$.  For non-centre Ramond Verma modules, this multiplicity is either $2$ or $0$ --- when a \sv{} exists, the weight space contains one of each parity.  For centre Ramond Verma modules, the \sv{} multiplicity can be $4$, $2$, $1$ or $0$.

\begin{figure}
\begin{tikzpicture}[->,>=stealth', node distance=1cm]
  \node[sv] (b) {};
  \node[] (bdry) [above of=b] {Corner/Boundary};
  \node[sv] (ba) [below of=b] {};
  \node[sv] (bb) [below of=ba] {};
  \node[sv] (bc) [below of=bb] {};
  \node[inner sep=2pt] (bd) [below of=bc] {$\vdots$};
  \path[]
  (b) edge node {} (ba)
  (ba) edge node {} (bb)
  (bb) edge node {} (bc)
  (bc) edge node {} (bd);
  \node[sv] (ia) [right = 2.5cm of ba] {};
  \node[sv] (i) [above right of=ia] {};
  \node[] (tmp) [right = 3.2cm of b] {};
  \node[] (int) [above of=tmp] {Interior};
  \node[sv] (ib) [below of=ia] {};
  \node[sv] (ic) [below of=ib] {};
  \node[inner sep=2pt] (id) [below of =ic] {$\vdots$};
  \node[sv] (ie) [below right of =i] {};
  \node[sv] (if) [below of =ie] {};
  \node[sv] (ig) [below of =if] {};
  \node[inner sep=2pt] (ih) [below of=ig] {$\vdots$};
  \path[]
  (i) edge node {} (ia)
  (ia) edge node {} (ib)
  (ia) edge node {} (if)
  (ib) edge node {} (ic)
  (ib) edge node {} (ig)
  (ic) edge node {} (id)
  (ic) edge node {} (ih)
  (i) edge node {} (ie)
  (ie) edge node {} (if)
  (ie) edge node {} (ib)
  (if) edge node {} (ig)
  (if) edge node {} (ic)
  (ig) edge node {} (ih)
  (ig) edge node {} (id);	
  \node[osv] (c) [right = 6.4cm of b] {};
  \node[sv] (ca) [below of=c] {};
  \node[sv] (cb) [below of=ca] {};
  \node[sv] (cc) [below of=cb] {};
  \node[inner sep=2pt] (cd) [below of=cc] {$\vdots$};
  \node[] (hc) [below of=cd] {$h = \frac{c}{24}$};
  \path[]
  (c) edge node {} (ca)
  (ca) edge node {} (cb)
  (cb) edge node {} (cc)
  (cc) edge node {} (cd);
  \node[] (tmp2) [right = 0.9cm of c] {};
  \node[] [above of=tmp2]{Centre};
  \node[sv] (d) [right = 2cm of c] {};
  \node[] (tmp3) [below of=d] {};
  \node[sv] (da) [left = -0.1cm of tmp3] {};
  \node[sv] (db) [right = -0.1cm of tmp3] {};
  \node[sv] (dc) [below of=da] {};
  \node[sv] (dd) [below of=db] {};
  \node[sv] (de) [below of=dc] {};
  \node[sv] (df) [below of=dd] {};
  \node[inner sep=2pt] (dg) [below of=de] {$\vphantom{\vdots}$};
  \node[inner sep=2pt] (dh) [below of=df] {$\vphantom{\vdots}$};
  \node[] at (cd-|d) {$\vdots$};
  \node[] at (hc-|d) {$h \neq \frac{c}{24}$};
  \path[]
  (d) edge node {} (da)
  (d) edge node {} (db)
  (da) edge node {} (dc)
  (db) edge node {} (dd)
  (dc) edge node {} (de)
  (dd) edge node {} (df)
  (de) edge node {} (dg)
  (df) edge node {} (dh);
\end{tikzpicture}
\caption{The \sv{} structure of the $N=1$ Verma modules $\Ver{r,s}$ with $t \in \QQ_{>0}$.  Each black circle corresponds to a \sv{} in the \ns{} sector and a pair of \svs{}, one bosonic and one fermionic, in the Ramond sector.  The white circle indicates a Ramond \sv{} whose multiplicity is one and the double circles indicate Ramond \svs{} of multiplicity four.  Arrows from one \sv{} (pair of \svs{}) to another indicate that the latter may be obtained from the former by acting with a suitable polynomial in the $L_n$ and $G_j$.} \label{fig:VermaStructures}
\end{figure}

Aside from the doubling of the \sv{} multiplicities in the Ramond sector, due to $G_0$, the structures of the non-centre $N=1$ Verma modules are analogous to those of the Virasoro algebra.  The new features are exhibited in the centre modules.  Despite the chain-like depiction of the \sv{} structures in \cref{fig:VermaStructures}, the centre Verma modules may be thought of as degenerations of interior modules in which the conformal weights of the \svs{} at the same horizontal level coincide (whence the multiplicity $4$ \svs{}).  However, the braided pattern of the interior modules is absent in this degeneration.  For $h \neq \frac{c}{24}$, each \sv{} space instead splits uniformly in two \cite{IohRepI03}, leading to the double-chain pattern of \cref{fig:VermaStructures}.

An example helps to clarify the peculiarities of the centre Verma modules.  For $(p,p') = (2,4)$, the Verma module $\Ver{1,2}$ is of centre type (the extended Kac table is given in \cref{fig:KacTables}).  As $h_{1,2} = 0 = \frac{c}{24}$, its ground state space is one-dimensional and it has two-dimensional \sv{} spaces of conformal weights $1, 4, 9, \ldots$\,.  The Verma module $\Ver{1,6} = \Ver{3,2}$ is also of centre type, with $h_{1,6} = h_{3,2} = 1$, and it has \sv{} spaces of weights $4, 9, 16, \ldots$ as well, but these are four-dimensional.  It follows that a module homomorphism from $\Ver{1,6}$ to $\Ver{1,2}$ cannot be an inclusion, a fact reinforced by consideration of their characters (see \cref{sec:CharMod} below):
\begin{equation}
\ch{\Ver{1,2}} = 1 + 2q + 4q^2 + 8q^3 + 14q^4 + \cdots, \qquad \ch{\Ver{1,6}} = 2q + 4q^2 + 8q^3 + 16q^4 + \cdots.
\end{equation}
Indeed, such a (non-zero) homomorphism maps one chain of \svs{} of $\Ver{1,6}$ onto those of $\Ver{1,2}$ and the other chain to $0$.  In other words, the submodule of $\Ver{1,2}$ generated by the \svs{} of weight $1$ is not isomorphic to a Verma module, despite the fact \cite{AubZer85} that the \uea{} of the Ramond algebra has no zero divisors.\footnote{The loophole is that Ramond Verma modules are not obtained from the free action of a \uea{} on a \hwv{} because $G_0$ does not act freely.}

We conclude this survey by noting that when $h = \frac{c}{24}$, one can further relax the definition of a Ramond Verma module to allow inducing indecomposable modules over the zero mode subalgebra.  Then, one may induce the two-dimensional module spanned by the weight vectors $v$ and $G_0 v$ to obtain the module $\PreVer{c/24}$ that is called a \emph{pre-Verma module} in \cite{IohRepI03,IohStr06}.  This module is again not fixed by parity-reversal and we accordingly attach a superscript $\pm$ to match the parity of $v$.  It is, in fact, a non-split extension of the corresponding Verma module by its parity-reversed counterpart:
\begin{equation} \label{es:PreVerma}
\dses{\RVer{c/24}^{\mp}}{}{\PreVer{c/24}^{\pm}}{}{\RVer{c/24}^{\pm}}.
\end{equation}
Unlike the case of Ramond Verma modules, there are submodules of pre-Verma modules that are not generated by \svs{}.  Instead, one has to introduce \emph{\ssvs{}} which are vectors that become singular upon taking an appropriate quotient.  The structure of the pre-Verma modules was determined in \cite{IohStr06} and we indicate this structure, for $t \in \QQ_{>0}$ hence $(r,s) = (\frac{p}{2}, \frac{p'}{2})$, in \cref{fig:PreVermaStructures}.

\begin{figure}
\begin{tikzpicture}[->,>=stealth', node distance=1cm]
  \node[osv, label=above right:{$v$}] (c) {};
  \node[osv, label=above left:{$G_0 v$}] (gc) [left = 2cm of c] {};
  \node[sv] (ca) [below of=c] {};
  \node[sv] (cb) [below of=ca] {};
  \node[sv] (cc) [below of=cb] {};
  \node[inner sep=2pt] (cd) [below of=cc] {$\vdots$};
  \node[sv] (gca) [below of=gc] {};
  \node[sv] (gcb) [below of=gca] {};
  \node[sv] (gcc) [below of=gcb] {};
  \node[inner sep=2pt] (gcd) [below of=gcc] {$\vdots$};
  \path[]
  (c) edge node {} (ca)
  (ca) edge node {} (cb)
  (cb) edge node {} (cc)
  (cc) edge node {} (cd)
  (gc) edge node {} (gca)
  (gca) edge node {} (gcb)
  (gcb) edge node {} (gcc)
  (gcc) edge node {} (gcd)
  (c) edge node [above] {$G_0$} (gc)
  (ca) edge node {} (gca)
  (cb) edge node {} (gcb)
  (cc) edge node {} (gcc)
  (ca) edge node {} (gc)
  (cb) edge node {} (gca)
  (cc) edge node {} (gcb)
  (cd) edge node {} (gcc);
\end{tikzpicture}
\caption{The \ssv{} structure of the $N=1$ pre-Verma module $\PreVer{c/24}$ for $t \in \QQ_{>0}$.  The white circles indicate \svs{} of multiplicity $1$, the black circles on the left indicate \svs{} of multiplicity $2$, and those on the right correspond to \ssvs{} of multiplicity $2$.  Arrows between (sub)\svs{} have the same meaning as in \cref{fig:VermaStructures}.  Note that (sub)\svs{} at the same horizontal level have the same conformal weights.  The separation is intended to emphasise subsingularity and accords with \eqref{es:PreVerma}.} \label{fig:PreVermaStructures}
\end{figure}

\subsection{Fock spaces} \label{sec:Fock}

The $N=1$ superconformal algebras have a free field realisation as a subalgebra of the tensor product of the free boson and free fermion \vosas{}.  In particular, the $N=1$ algebra acts on the tensor product of any Fock space over the Heisenberg algebra with either the \ns{} or Ramond fermionic Fock space.  We shall refer to such tensor products as \emph{$N=1$ Fock spaces} for brevity.

The free boson and free fermion \vosas{} are generated by fields $\func{a}{z} = \sum_n a_n z^{-n-1}$ and $\func{b}{z} = \sum_j b_j z^{-j-1/2}$, respectively, that satisfy
\begin{equation}
\func{a}{z} \func{a}{w} \sim \frac{1}{\brac{z-w}^2}, \qquad \func{b}{z} \func{b}{w} \sim \frac{1}{z-w}.
\end{equation}
The energy-momentum tensor and its superpartner, the generators of the $N=1$ algebra, are then given by
\begin{equation} \label{eq:DefTG}
\func{T}{z} = \frac{1}{2} \normord{\func{a}{z} \func{a}{z}} + \frac{1}{2} Q \func{\pd a}{z} + \frac{1}{2} \normord{\func{\pd b}{z} \func{b}{z}}, \qquad 
\func{G}{z} = \func{a}{z} \func{b}{z} + Q \func{\pd b}{z},
\end{equation}
where $\normord{\cdots}$ denotes normal ordering and we omit the tensor product symbols for brevity.  The resulting central charge is $c = \frac{3}{2} - 3Q^2$ which matches the $N=1$ parametrisation \eqref{eq:ParByt} if we set
\begin{equation}
\alpha = \sqrt{\frac{p \vphantom{p'}}{4p'}}, \qquad \alpha' = \sqrt{\frac{p'}{4p}}, \qquad
Q = 2 \brac{\alpha' - \alpha} = \frac{p'-p}{\sqrt{pp'}}.
\end{equation}

In the \ns{} sector, the $N=1$ Fock space $\NSFock{\lambda}$ is defined to be the tensor product of the free boson Verma module of $a_0$-eigenvalue $\lambda$ with the free fermion vacuum Verma module.  It therefore has a one-dimensional space of ground states and the conformal weight of these ground states is
\begin{subequations}
\begin{equation}
h_{\lambda} = \frac{1}{2} \lambda \brac{\lambda - Q} = \frac{4pp' \brac{\lambda - Q/2}^2 - \brac{p'-p}^2}{8pp'}.
\end{equation}
\ns{} Fock spaces inherit a choice of parity for the ground state from that of the free fermion vacuum module; as before, we indicate this choice by a superscript $\pm$.  In the Ramond sector, an $N=1$ Fock space $\RFock{\lambda}$ is the tensor product of the free boson Verma module of $a_0$-eigenvalue $\lambda$ with the free fermion Ramond Verma module.  It therefore has a two-dimensional space of ground states whose common conformal weight is
\begin{equation}
h_{\lambda} = \frac{1}{2} \lambda \brac{\lambda - Q} + \frac{1}{16} = \frac{4pp' \brac{\lambda - Q/2}^2 - \brac{p'-p}^2}{8pp'} + \frac{1}{16}.
\end{equation}
\end{subequations}
Ramond Fock spaces are preserved by the parity-reversing functor $\Pi$, even when the conformal weight of the ground states satisfies $h_{\lambda} = \frac{c}{24}$.

The contribution to the conformal weight of the ground states from the free fermion Ramond module accords perfectly with the $N=1$ parametrisation \eqref{eq:ParByt}.  Indeed, in both sectors, we have $h_{\lambda} = h_{r,s}$ when
\begin{equation} \label{eq:DefLambda}
\lambda = \lambda_{r,s} \equiv -\alpha' \brac{r-1} + \alpha \brac{s-1},
\end{equation}
with $r=s \bmod{2}$ in the \ns{} sector and $r \neq s \bmod{2}$ in the Ramond sector.  We note the following symmetries for later use:
\begin{equation} \label{eq:FFSymm}
\lambda_{r+p,s} = \lambda_{r,s} - \frac{1}{2} \sqrt{pp'}, \quad 
\lambda_{r,s+p'} = \lambda_{r,s} + \frac{1}{2} \sqrt{pp'} \qquad \Ra \qquad 
\lambda_{r+p,s+p'} = \lambda_{r,s}.
\end{equation}

We therefore define $\Fock{r,s}$ to be $\NSFock{\lambda_{r,s}}$ or $\RFock{\lambda_{r,s}}$ depending on whether $r-s$ is even or odd, respectively.  The $\Fock{r,s}$, with $r,s \in \ZZ$, exhaust the non-simple Fock spaces \cite{IohRepII03}:  A \ns{} Fock space $\NSFock{\lambda}$ is simple, unless $\lambda = \lambda_{r,s}$ with $r=s \bmod{2}$, and a Ramond Fock space $\RFock{\lambda}$ is simple, unless $\lambda = \lambda_{r,s}$ with $r \neq s \bmod{2}$.  As was the case for Verma modules, the centre Fock spaces of the form $\Fock{r,s}$ are all Ramond and only exist when $p$ and $p'$ are even.  For $t \in \QQ_{>0}$, we depict the submodule structure of the $\Fock{r,s}$ in \cref{fig:FockStructures}.  Unlike the case of $N=1$ Verma modules, submodules of Fock spaces are generated by \ssvs{} in general.

\begin{figure}
\begin{tikzpicture}[->,>=stealth', node distance=1cm]
  \node[sv] (a) {};
  \node[] (cor) [above of=a] {Corner};
  \node[sv] (aa) [below of=a] {};
  \node[sv] (ab) [below of=aa] {};
  \node[sv] (ac) [below of=ab] {};
  \node[inner sep=2pt] (ad) [below of=ac] {$\vdots$};
  \node[sv] (b) [right = 1.5cm of a] {};
  \node[sv] (ba) [below of=b] {};
  \node[sv] (bb) [below of=ba] {};
  \node[sv] (bc) [below of=bb] {};
  \node[inner sep=2pt] (bd) [below of=bc] {$\vdots$};
  \path[]
  (b) edge node {} (ba)
  (bb) edge node {} (ba)
  (bb) edge node {} (bc)
  (bd) edge node {} (bc);
  \node[] (ugh) [right = 0.4cm of b] {};
  \node[] at (ugh|-cor) {Boundary};
  \node[sv] (B) [right = 1cm of b] {};
  \node[sv] (Ba) [below of=B] {};
  \node[sv] (Bb) [below of=Ba] {};
  \node[sv] (Bc) [below of=Bb] {};
  \node[inner sep=2pt] (Bd) [below of=Bc] {$\vdots$};
  \path[]
  (Ba) edge node {} (B)
  (Ba) edge node {} (Bb)
  (Bc) edge node {} (Bb)
  (Bc) edge node {} (Bd);
  \node[sv] (ia) [right = 1.5cm of Ba] {};
  \node[sv] (i) [above right of=ia] {};
  \node[sv] (ib) [below of=ia] {};
  \node[sv] (ic) [below of=ib] {};
  \node[inner sep=2pt] (id) [below of =ic] {$\vdots$};
  \node[sv] (ie) [below right of =i] {};
  \node[sv] (if) [below of =ie] {};
  \node[sv] (ig) [below of =if] {};
  \node[inner sep=2pt] (ih) [below of=ig] {$\vdots$};
  \path[]
  (i) edge node {} (ia)
  (ib) edge node {} (ia)
  (ib) edge node {} (ic)
  (id) edge node {} (ic)
  (ie) edge node {} (i)
  (ie) edge node {} (ib)
  (ie) edge node {} (if)
  (if) edge node {} (ia)
  (if) edge node {} (ic)
  (ig) edge node {} (ib)
  (ig) edge node {} (id)
  (ig) edge node {} (if)
  (ig) edge node {} (ih)
  (ih) edge node {} (ic);
  \node[] (tmp) [right = 3.3cm of B] {};
  \node[] (int) [above of=tmp] {Interior};
  \node[sv] (Ia) [right = 2cm of ia] {};
  \node[sv] (I) [above right of=Ia] {};
  \node[sv] (Ib) [below of=Ia] {};
  \node[sv] (Ic) [below of=Ib] {};
  \node[inner sep=2pt] (Id) [below of =Ic] {$\vdots$};
  \node[sv] (Ie) [below right of =I] {};
  \node[sv] (If) [below of =Ie] {};
  \node[sv] (Ig) [below of =If] {};
  \node[inner sep=2pt] (Ih) [below of=Ig] {$\vdots$};
  \path[]
  (Ia) edge node {} (I)
  (Ia) edge node {} (Ib)
  (Ic) edge node {} (Ib)
  (Ic) edge node {} (Id)
  (I) edge node {} (Ie)
  (Ib) edge node {} (Ie)
  (If) edge node {} (Ie)
  (Ia) edge node {} (If)
  (Ic) edge node {} (If)
  (Ib) edge node {} (Ig)
  (Id) edge node {} (Ig)
  (If) edge node {} (Ig)
  (Ih) edge node {} (Ig)
  (Ic) edge node {} (Ih);	
  \node[sv] (ca) [right = 1.5cm of Ie] {};
  \node[osv] (c1) [above of=ca] {};
  \node[sv] (cb) [below of=ca] {};
  \node[sv] (cc) [below of=cb] {};
  \node[inner sep=2pt] (cd) [below of=cc] {$\vdots$};
  \node[] (ick) [right = 0.45cm of cd] {};
  \node[] (hc) [below of=ick] {$h = \frac{c}{24}$};
  \node[] (duh) [above right of=ca] {};
  \node[sv] (ce) [below right of=duh] {};
  \node[osv] (c2) [above of=ce] {};
  \node[sv] (cf) [below of=ce] {};
  \node[sv] (cg) [below of=cf] {};
  \node[inner sep=2pt] (ch) [below of=cg] {$\vdots$};
  \path[]
  (c1) edge node {} (ca)
  (cb) edge node {} (ca)
  (cb) edge node {} (cc)
  (cd) edge node {} (cc)
  (c2) edge node {} (ca)
  (ce) edge node {} (c1)
  (ce) edge node {} (cb)
  (ce) edge node {} (c2)
  (ce) edge node {} (cf)
  (cf) edge node {} (ca)
  (cf) edge node {} (cc)
  (cg) edge node {} (cb)
  (cg) edge node {} (cd)
  (cg) edge node {} (cf)
  (cg) edge node {} (ch)
  (ch) edge node {} (cc);
  \node[] (tmp2) [right = 0.2cm of c2] {};
  \node[] [above of=tmp2]{Centre};
  \node[sv] (Ca) [right = 2cm of ca] {};
  \node[sv] (C) [above right of=Ca] {};
  \node[sv] (Cb) [below of=Ca] {};
  \node[sv] (Cc) [below of=Cb] {};
  \node[inner sep=2pt] (Cd) [below of =Cc] {$\vdots$};
  \node[sv] (Ce) [below right of =C] {};
  \node[sv] (Cf) [below of =Ce] {};
  \node[sv] (Cg) [below of =Cf] {};
  \node[inner sep=2pt] (Ch) [below of=Cg] {$\vdots$};
  \node[] at (hc-|C) {$h \neq \frac{c}{24}$};
  \path[]
  (C) edge node {} (Ca)
  (Cb) edge node {} (Ca)
  (Cb) edge node {} (Cc)
  (Cd) edge node {} (Cc)
  (Ce) edge node {} (C)
  (Ce) edge node {} (Cb)
  (Ce) edge node {} (Cf)
  (Cf) edge node {} (Ca)
  (Cf) edge node {} (Cc)
  (Cg) edge node {} (Cb)
  (Cg) edge node {} (Cd)
  (Cg) edge node {} (Cf)
  (Cg) edge node {} (Ch)
  (Ch) edge node {} (Cc);
\end{tikzpicture}
\caption{The \ssv{} structure of the $N=1$ Fock spaces $\Fock{r,s}$ when $t \in \QQ_{>0}$.  Each black circle corresponds to a \ssv{} in the \ns{} sector and a pair of \ssvs{}, one bosonic and one fermionic, in the Ramond sector.  The white circles indicate a Ramond \sv{} whose multiplicity is one.  Arrows from one \ssv{} (pair of \ssvs{}) to another have the same meaning as in \cref{fig:VermaStructures,fig:PreVermaStructures}.  The two structures for interior Fock spaces are mirror images, the repetition serving to remind us that these Fock spaces are not self-contragredient.} \label{fig:FockStructures}
\end{figure}

We remark that there are two possible structures for boundary and interior Fock spaces $\Fock{r,s}$, corresponding to the fact that these modules are not isomorphic to their contragredient duals $\Fock{Q-\lambda_{r,s}} = \Fock{-r,-s}$.  There is no ambiguity for corner and centre Fock spaces as they are self-contragredient.  For $r,s \in \ZZ_{>0}$, the conformal weights of the \ssvs{} of the Fock space $\Fock{r,s}$ (and its contragredient $\Fock{-r,-s}$) coincide with those of the \svs{} of the Verma module $\Ver{r,s}$.  Both $\Fock{r,s}$ and $\Fock{-r,-s}$ therefore have \ssvs{} of depth $\frac{1}{2} rs$.  For $r,s \in \ZZ_{>0}$, the depth $\frac{1}{2} rs$ \ssvs{} of $\Fock{r,s}$ are always associated to the head of the Fock space (its circle in \cref{fig:FockStructures} has all arrows pointing away from it).

This fixes the ambiguity in the structure of a boundary Fock space $\Fock{r,s}$:  One uses the symmetries \eqref{eq:FFSymm} to shift $r$ and $s$ to positive integers, thereby identifying the depth $\frac{1}{2} rs$ \ssvs{} as elements associated to the head.  Because the \ssvs{} of lesser depth are easily determined, this is sufficient to distinguish between the two possibilities in \cref{fig:FockStructures}.  For an interior Fock space $\Fock{r,s}$, this information should be supplemented by the following fact.  First, note that every second horizontal level in the interior structures of \cref{fig:FockStructures} indicates \svs{} (associated to the socle of $\Fock{r,s}$) and \ssvs{} (associated to the head).  The relevant fact is that if the depth of the \svs{} is greater than that of the \ssvs{}, at some given horizontal level, then it will also be greater at the other horizontal levels (and vice versa).  One may then check which has greater depth in a given module because one knows that the depth $\frac{1}{2} rs$ \ssvs{} are associated with the head, for $r,s \in \ZZ_{>0}$.

It remains to discuss the centre Fock spaces.  The space of ground states is two-dimensional and the structure, when the conformal weight $h$ of the ground states is not $\frac{c}{24}$, is similar to the structure of the interior Fock spaces.  The only difference is that the \ssvs{} appearing at the same horizontal levels in \cref{fig:FockStructures} now have the same conformal weight (this never happens for interior Fock spaces).  The case where $h = \frac{c}{24}$ differs further in that the ground states do not define a simple module over the zero mode subalgebra $\vspn \set{L_0, C, G_0}$.  Instead, the ground states decompose as a direct sum of two simple modules upon which $G_0$ acts as the zero operator.  This is easy to check as the ground states have the form $v \otimes w$ and $v \otimes b_0 w$, where $v$ is a Heisenberg \hwv{} with $a_0$-eigenvalue
\begin{equation}
\lambda_{p/2, p'/2} = -\alpha' \frac{p-2}{2} + \alpha \frac{p'-2}{2} = \alpha' - \alpha = \frac{Q}{2}
\end{equation}
and $w$ is a Ramond \hwv{} for the free fermion algebra.  Using \eqref{eq:DefTG}, we verify that
\begin{equation}
G_0 \brac{v \otimes w} = a_0 v \otimes b_0 w - \frac{Q}{2} v \otimes b_0 w = \brac{\lambda_{p/2,p'/2} - \frac{Q}{2}} v \otimes w = 0
\end{equation}
and, similarly, that $G_0 \brac{v \otimes b_0 w} = 0$.

\subsection{Kac modules} \label{sec:Kac}

In what follows, we will be interested in the fusion rules of certain modules $\Kac{r,s}$, indexed by $r,s \in \ZZ_{>0}$, that we shall refer to as $N=1$ Kac modules.  Analogues of these modules over the Virasoro algebra were introduced non-constructively in \cite{PeaLog06} in order to describe the quantum state space for a class of boundary sectors in the scaling limit of certain integrable lattice models.  Their characters were determined in many examples, but a concrete proposal for the identities of the corresponding Virasoro modules was only made recently, for corner and boundary entries $(r,s)$, as submodules of the corresponding Fock spaces \cite{RasCla11}.  More recent work \cite{MorKac15} has extended this proposal to interior entries of the extended Virasoro Kac table and has also provided a significant amount of additional evidence for its correctness.

$N=1$ Kac modules in the \ns{} sector have been recently considered from the lattice \cite{PeaLog14} and continuum \cite{CanFusI15} points of view.  The lattice analysis only studied the action of $L_0$ on a few examples, thereby obtaining a limited amount of structural information such as the non-diagonalisability of $L_0$ on several fusion products.  The continuum analysis built upon this by describing explicit fusion calculations that confirmed these non-semisimple actions and, moreover, detailed a series of conjectures for the structures of certain \ns{} Kac module fusion products.  One of the aims of this work is to extend these calculations to the Ramond sector in order to test the hypothesis that $N=1$ Kac modules are submodules of Fock spaces.  A second aim is to explore the structural features exhibited by fusion products involving Ramond Kac modules.

\begin{figure}
\scalebox{0.5}{
\begin{tabular}{|D{3.5cm}|D{0.5cm}|D{5.5cm}|D{0.5cm}|D{5.5cm}|D{0.5cm}|D{5.5cm}|D{0.5cm}|D{5.5cm}}
\hline
% row 1
\IKL 
%%%%%%%%%%%%%%%%%%%%%%%%%%%%%%%%% TWO CHAIN %%%%%%%%%%%%%%%%%%%%%%%%%%%%%%%%%%%
\begin{tikzpicture} 
[->,thick,-stealth', transform shape, node distance=1.1cm]
\node[sv] (1) [] {};
\node[sv] (3) [below of=1] {};
\node[sv] (5) [right=1.5cm of 3] {};
\node[osv] (4) [above left of=5] {};
\node[osv] (6) [above right of=5] {};
\path[]
   (1) edge node {} (3)
   (4) edge node {} (5)
   (6) edge node {} (5);
\end{tikzpicture}
%%%%%%%%%%%%%%%%%%%%%%%%%%%%%%%%%%%%%%%%%%%%%%%%%%%%%%%%%%%%%%%%%%%%%%%%%%%%%%
& \BKL 
\begin{tikzpicture} 
\node[sv] (1) [] {};
\end{tikzpicture}
& \IKL 
\begin{tikzpicture} 
[->,thick,-stealth', transform shape, node distance=1.1cm]
\node[sv] (1) [] {};
\node[sv] (3) [below of=1] {};
\path[]
   (1) edge node {} (3);
\end{tikzpicture}
& \BKL 
\begin{tikzpicture} 
\node[sv] (1) [] {};
\end{tikzpicture}
& \IKL 
\begin{tikzpicture} 
[->,thick,-stealth', transform shape, node distance=1.1cm]
\node[sv] (1) [] {};
\node[sv] (3) [below of=1] {};
\path[]
   (1) edge node {} (3);
\end{tikzpicture}
& \BKL 
\begin{tikzpicture} 
\node[sv] (1) [] {};
\end{tikzpicture}
& \IKL 
%%%%%%%%%%%%%%%%%%%%%%%%%%%%%%%% TWO CHAIN %%%%%%%%%%%%%%%%%%%%%%%%%%%%%%%%%%%
\begin{tikzpicture} 
[->,thick,-stealth', transform shape, node distance=1.1cm]
\node[sv] (1) [] {};
\node[sv] (3) [below of=1] {};
\path[]
   (1) edge node {} (3);
\end{tikzpicture}
%%%%%%%%%%%%%%%%%%%%%%%%%%%%%%%%%%%%%%%%%%%%%%%%%%%%%%%%%%%%%%%%%%%%%%%%%%%%%%
& \BKL 
\begin{tikzpicture} 
\node[sv] (1) [] {};
\end{tikzpicture}
& \IKL 
%%%%%%%%%%%%%%%%%%%%%%%%%%%%%%%%% TWO CHAIN %%%%%%%%%%%%%%%%%%%%%%%%%%%%%%%%%%%
\begin{tikzpicture} 
[->,thick,-stealth', transform shape, node distance=1.1cm]
\node[sv] (1) [] {};
\node[sv] (3) [below of=1] {};
\path[]
   (1) edge node {} (3);
\end{tikzpicture}
%%%%%%%%%%%%%%%%%%%%%%%%%%%%%%%%%%%%%%%%%%%%%%%%%%%%%%%%%%%%%%%%%%%%%%%%%%%%%%
\\
\hline
% row 2
\BKL
\begin{tikzpicture}
\node[sv] (1) [] {};
\end{tikzpicture}
& 
\begin{tikzpicture}
\node[sv] (1) [] {};
\end{tikzpicture}
& \BKL 
\begin{tikzpicture} 
[->,thick,-stealth', transform shape, node distance=1.1cm]
\node[sv] (1) [] {};
\node[sv] (3) [below of=1] {};
\path[]
   (1) edge node {} (3);
\end{tikzpicture}
& 
\begin{tikzpicture}
\node[sv] (1) [] {};
\end{tikzpicture}
& \BKL
\begin{tikzpicture}
 [->,thick,-stealth', transform shape, node distance=1.1cm]
 \node[sv] (E6) [] {};
 \node[sv] (E6a) [below of=E6] {};
 \path[]
   (E6) edge node {} (E6a);
\end{tikzpicture}
& 
\begin{tikzpicture}
 \node[sv] (1) [] {};
\end{tikzpicture}
&\BKL
%%%%%%%%%%%%%%%%%%%%%%%%%%%%%%%%% TWO CHAIN %%%%%%%%%%%%%%%%%%%%%%%%%%%%%%%%%%%
\begin{tikzpicture} 
[->,thick,-stealth', transform shape, node distance=1.1cm]
\node[sv] (1) [] {};
\node[sv] (3) [below of=1] {};
\path[]
   (1) edge node {} (3);
\end{tikzpicture}
%%%%%%%%%%%%%%%%%%%%%%%%%%%%%%%%%%%%%%%%%%%%%%%%%%%%%%%%%%%%%%%%%%%%%%%%%%%%%%
& 
\begin{tikzpicture}
 \node[sv] (1) [] {};
\end{tikzpicture}
&\BKL
%%%%%%%%%%%%%%%%%%%%%%%%%%%%%%%%% TWO CHAIN %%%%%%%%%%%%%%%%%%%%%%%%%%%%%%%%%%%
\begin{tikzpicture} 
[->,thick,-stealth', transform shape, node distance=1.1cm]
\node[sv] (1) [] {};
\node[sv] (3) [below of=1] {};
\path[]
   (1) edge node {} (3);
\end{tikzpicture}
%%%%%%%%%%%%%%%%%%%%%%%%%%%%%%%%%%%%%%%%%%%%%%%%%%%%%%%%%%%%%%%%%%%%%%%%%%%%%%
\\
\hline
% row 3
\IKL 
\begin{tikzpicture} 
[->,thick,-stealth', transform shape, node distance=1.1cm]
\node[sv] (1) [] {};
\node[sv] (3) [below of=1] {};
\path[]
   (1) edge node {} (3);
\end{tikzpicture}
& \BKL 
\begin{tikzpicture} 
[->,thick,-stealth', transform shape, node distance=1.1cm]
\node[sv] (1) [] {};
\node[sv] (3) [below of=1] {};
\path[]
   (1) edge node {} (3);
\end{tikzpicture}
& \IKL 
%%%%%%%%%%%%%%%%%%%%%%%%% HEXAGON %%%%%%%%%%%%%%%%%%%%%%%%%%%%
\begin{tikzpicture} 
[->,thick,-stealth', transform shape, node distance=1.1cm]
 \node[sv] (1f) [] {};
 \node[sv] (2f) [below left of=1f] {};
 \node[sv] (3f) [below right of=1f] {};
 \node[sv] (4f) [below of=2f] {};
 \node[sv] (5f) [below of=3f] {};
 \node[sv] (6f) [below left of=5f] {};
 \node[sv] (6g) [right=2.5cm of 6f] {};
 \node[sv] (5g) [above right of=6g] {};
 \node[sv] (4g) [above left of=6g] {};
 \node[sv] (3g) [above of=5g] {};
 \node[sv] (2g) [above of=4g] {};
 \node[osv] (1g) [above of=3g] {};
 \node[osv] (0g) [above of=2g] {};
 \path[]
   (2f) edge node {} (1f)
   (1f) edge node {} (3f)
   (2f) edge node {} (4f)
   (2f) edge node {} (5f)
   (4f) edge node {} (3f)
   (5f) edge node {} (3f)
   (5f) edge node {} (6f)
   (4f) edge node {} (6f)
   (0g) edge node {} (2g)
   (1g) edge node {} (2g)
   (3g) edge node {} (0g)
   (3g) edge node {} (1g)
   (3g) edge node {} (4g)
   (3g) edge node {} (5g)
   (4g) edge node {} (2g)
   (4g) edge node {} (6g)
   (5g) edge node {} (2g)
   (5g) edge node {} (6g);
\end{tikzpicture}
%%%%%%%%%%%%%%%%%%%%%%%%%%%%%%%%%%%%%%%%%%%%%%%%%%%%%%%%%%%%%%
& \BKL 
\begin{tikzpicture}
 [->,thick,-stealth', transform shape, node distance=1.1cm]
 \node[sv] (E6) [] {};
 \node[sv] (E6a) [below of=E6] {};
 \node[sv] (E6b) [below of=E6a] {};
 \path[]
   (E6a) edge node {} (E6)
   (E6a) edge node {} (E6b);
\end{tikzpicture}
& \IKL
%%%%%%%%%%%%%%%%%%%%%%%%% HEXAGON %%%%%%%%%%%%%%%%%%%%%%%%%%%%
\begin{tikzpicture} 
[->,thick,-stealth', transform shape, node distance=1.1cm]
 \node[sv] (1f) [] {};
 \node[sv] (2f) [below left of=1f] {};
 \node[sv] (3f) [below right of=1f] {};
 \node[sv] (4f) [below of=2f] {};
 \node[sv] (5f) [below of=3f] {};
 \node[sv] (6f) [below left of=5f] {};
 \path[]
   (2f) edge node {} (1f)
   (1f) edge node {} (3f)
   (2f) edge node {} (4f)
   (2f) edge node {} (5f)
   (4f) edge node {} (3f)
   (5f) edge node {} (3f)
   (5f) edge node {} (6f)
   (4f) edge node {} (6f);
\end{tikzpicture}
%%%%%%%%%%%%%%%%%%%%%%%%%%%%%%%%%%%%%%%%%%%%%%%%%%%%%%%%%%%%%%
& \BKL 
\begin{tikzpicture}
 [->,thick,-stealth', transform shape, node distance=1.1cm]
 \node[sv] (E6) [] {};
 \node[sv] (E6a) [below of=E6] {};
 \node[sv] (E6b) [below of=E6a] {};
 \path[]
   (E6a) edge node {} (E6)
   (E6a) edge node {} (E6b);
\end{tikzpicture}
&\IKL
%%%%%%%%%%%%%%%%%%%%%%%%% HEXAGON %%%%%%%%%%%%%%%%%%%%%%%%%%%%
\begin{tikzpicture} 
[->,thick,-stealth', transform shape, node distance=1.1cm]
 \node[sv] (1f) [] {};
 \node[sv] (2f) [below left of=1f] {};
 \node[sv] (3f) [below right of=1f] {};
 \node[sv] (4f) [below of=2f] {};
 \node[sv] (5f) [below of=3f] {};
 \node[sv] (6f) [below left of=5f] {};
 \path[]
   (2f) edge node {} (1f)
   (1f) edge node {} (3f)
   (2f) edge node {} (4f)
   (2f) edge node {} (5f)
   (4f) edge node {} (3f)
   (5f) edge node {} (3f)
   (5f) edge node {} (6f)
   (4f) edge node {} (6f);
\end{tikzpicture}
%%%%%%%%%%%%%%%%%%%%%%%%%%%%%%%%%%%%%%%%%%%%%%%%%%%%%%%%%%%%%%
& \BKL 
\begin{tikzpicture}
 [->,thick,-stealth', transform shape, node distance=1.1cm]
 \node[sv] (E6) [] {};
 \node[sv] (E6a) [below of=E6] {};
 \node[sv] (E6b) [below of=E6a] {};
 \path[]
   (E6a) edge node {} (E6)
   (E6a) edge node {} (E6b);
\end{tikzpicture}
&\IKL
%%%%%%%%%%%%%%%%%%%%%%%%% HEXAGON %%%%%%%%%%%%%%%%%%%%%%%%%%%%
\begin{tikzpicture} 
[->,thick,-stealth', transform shape, node distance=1.1cm]
 \node[sv] (1f) [] {};
 \node[sv] (2f) [below left of=1f] {};
 \node[sv] (3f) [below right of=1f] {};
 \node[sv] (4f) [below of=2f] {};
 \node[sv] (5f) [below of=3f] {};
 \node[sv] (6f) [below left of=5f] {};
 \path[]
   (2f) edge node {} (1f)
   (1f) edge node {} (3f)
   (2f) edge node {} (4f)
   (2f) edge node {} (5f)
   (4f) edge node {} (3f)
   (5f) edge node {} (3f)
   (5f) edge node {} (6f)
   (4f) edge node {} (6f);
\end{tikzpicture}
%%%%%%%%%%%%%%%%%%%%%%%%%%%%%%%%%%%%%%%%%%%%%%%%%%%%%%%%%%%%%%
\\
\hline
% row 4
\BKL 
\begin{tikzpicture}
\node[sv] (1) [] {};
\end{tikzpicture}
& 
\begin{tikzpicture}
\node[sv] (1) [] {};
\end{tikzpicture}
& \BKL 
\begin{tikzpicture}
 [->,thick,-stealth', transform shape, node distance=1.1cm]
 \node[sv] (E6) [] {};
 \node[sv] (E6a) [below of=E6] {};
 \node[sv] (E6b) [below of=E6a] {};
 \path[]
   (E6a) edge node {} (E6)
   (E6a) edge node {} (E6b);
\end{tikzpicture}
& 
\begin{tikzpicture}
 [->,thick,-stealth', transform shape, node distance=1.1cm]
 \node[sv] (E6) [] {};
 \node[sv] (E6a) [below of=E6] {};
\end{tikzpicture}
& \BKL 
%%%%%%%%%%%%%%%%%%%%%%%%%%%%%%%% FOUR CHAIN %%%%%%%%%%%%%%%%%%%%%%%%%%%%%%%
\begin{tikzpicture}
 [->,thick,-stealth', transform shape, node distance=1.1cm]
 \node[sv] (E6) [] {};
 \node[sv] (E7) [above of=E6] {};
 \node[sv] (E6a) [below of=E6] {};
 \node[sv] (E6b) [below of=E6a] {};
 \path[]
   (E7) edge node {} (E6)
   (E6a) edge node {} (E6)
   (E6a) edge node {} (E6b);
\end{tikzpicture}
%%%%%%%%%%%%%%%%%%%%%%%%%%%%%%%%%%%%%%%%%%%%%%%%%%%%%%%%%%%%%%%%%%%%%%%%%%
& 
\begin{tikzpicture}
 [->,thick,-stealth', transform shape, node distance=1.1cm]
 \node[sv] (E6) [] {};
 \node[sv] (E6a) [below of=E6] {};
\end{tikzpicture}
&\BKL
%%%%%%%%%%%%%%%%%%%%%%%%%%%%%%%% FOUR CHAIN %%%%%%%%%%%%%%%%%%%%%%%%%%%%%%%
\begin{tikzpicture}
 [->,thick,-stealth', transform shape, node distance=1.1cm]
 \node[sv] (E6) [] {};
 \node[sv] (E7) [above of=E6] {};
 \node[sv] (E6a) [below of=E6] {};
 \node[sv] (E6b) [below of=E6a] {};
 \path[]
   (E7) edge node {} (E6)
   (E6a) edge node {} (E6)
   (E6a) edge node {} (E6b);
\end{tikzpicture}
%%%%%%%%%%%%%%%%%%%%%%%%%%%%%%%%%%%%%%%%%%%%%%%%%%%%%%%%%%%%%%%%%%%%%%%%%%
& 
%%%%%%%%%%%%%%%%%%%%%%%%%%%%%%% TWO CHAIN %%%%%%%%%%%%%%%%%%%%%%%%%%%%%%%%%
\begin{tikzpicture}
 [->,thick,-stealth', transform shape, node distance=1.1cm]
 \node[sv] (E6) [] {};
 \node[sv] (E6a) [below of=E6] {};
\end{tikzpicture}
%%%%%%%%%%%%%%%%%%%%%%%%%%%%%%%%%%%%%%%%%%%%%%%%%%%%%%%%%%%%%%%%%%%%%%%%%%%%
&\BKL
%%%%%%%%%%%%%%%%%%%%%%%%%%%%%%%% FOUR CHAIN %%%%%%%%%%%%%%%%%%%%%%%%%%%%%%%
\begin{tikzpicture}
 [->,thick,-stealth', transform shape, node distance=1.1cm]
 \node[sv] (E6) [] {};
 \node[sv] (E7) [above of=E6] {};
 \node[sv] (E6a) [below of=E6] {};
 \node[sv] (E6b) [below of=E6a] {};
 \path[]
   (E7) edge node {} (E6)
   (E6a) edge node {} (E6)
   (E6a) edge node {} (E6b);
\end{tikzpicture}
%%%%%%%%%%%%%%%%%%%%%%%%%%%%%%%%%%%%%%%%%%%%%%%%%%%%%%%%%%%%%%%%%%%%%%%%%%
\\
\hline
% row 5
\IKL 
\begin{tikzpicture} 
[->,thick,-stealth', transform shape, node distance=1.1cm]
\node[sv] (1) [] {};
\node[sv] (3) [below of=1] {};
\path[]
   (1) edge node {} (3);
\end{tikzpicture}
& \BKL 
\begin{tikzpicture} 
[->,thick,-stealth', transform shape, node distance=1.1cm]
\node[sv] (1) [] {};
\node[sv] (3) [below of=1] {};
\path[]
   (1) edge node {} (3);
\end{tikzpicture}
& \IKL 
\begin{tikzpicture} 
[->,thick,-stealth', transform shape, node distance=1.1cm]
 \node[sv] (1f) [] {};
 \node[sv] (2f) [below left of=1f] {};
 \node[sv] (3f) [below right of=1f] {};
 \node[sv] (4f) [below of=2f] {};
 \node[sv] (5f) [below of=3f] {};
 \node[sv] (6f) [below left of=5f] {};
 \path[]
   (2f) edge node {} (1f)
   (1f) edge node {} (3f)
   (2f) edge node {} (4f)
   (2f) edge node {} (5f)
   (4f) edge node {} (3f)
   (5f) edge node {} (3f)
   (5f) edge node {} (6f)
   (4f) edge node {} (6f);
\end{tikzpicture}
& \BKL 
\begin{tikzpicture}
 [->,thick,-stealth', transform shape, node distance=1.1cm]
 \node[sv] (E6) [] {};
 \node[sv] (E7) [above of=E6] {};
 \node[sv] (E6a) [below of=E6] {};
 \node[sv] (E6b) [below of=E6a] {};
 \path[]
   (E7) edge node {} (E6)
   (E6a) edge node {} (E6)
   (E6a) edge node {} (E6b);
\end{tikzpicture}
& \IKL
%%%%%%%%%%%%%%%%%%%%%%%%%%%%%%%%%%% DECAGON %%%%%%%%%%%%%%%%%%%%%%%%%%%%%%%%%%%%%%%
\begin{tikzpicture} 
[->,thick,-stealth', transform shape, node distance=1.1cm]
 \node[sv] (1f) [] {};
 \node[sv] (2f) [below left of=1f] {};
 \node[sv] (3f) [below right of=1f] {};
 \node[sv] (4f) [below of=2f] {};
 \node[sv] (5f) [below of=3f] {};
 \node[sv] (6f) [below of=4f] {};
 \node[sv] (7f) [below of=5f] {};
 \node[sv] (8f) [below of =6f] {};
 \node[sv] (9f) [below of =7f] {};
 \node[sv] (10f) [below left of=9f] {};
 \node[sv] (10g) [right=2.5cm of 10f] {};
 \node[sv] (9g) [above right of=10g] {};
 \node[sv] (8g) [above left of=10g] {};
 \node[sv] (7g) [above of=9g] {};
 \node[sv] (6g) [above of=8g] {};
 \node[sv] (5g) [above of=7g] {};
 \node[sv] (4g) [above of=6g] {};
 \node[sv] (3g) [above of=5g] {};
 \node[sv] (2g) [above of=4g] {};
 \node[osv] (1g) [above of=3g] {};
 \node[osv] (0g) [above of=2g] {};
 \path[]
   (2f) edge node {} (1f)
   (1f) edge node {} (3f)
   (2f) edge node {} (4f)
   (2f) edge node {} (5f)
   (4f) edge node {} (3f)
   (5f) edge node {} (3f)
   (6f) edge node {} (4f)
   (4f) edge node {} (7f)
   (6f) edge node {} (5f)
   (5f) edge node {} (7f)
   (6f) edge node {} (8f)
   (6f) edge node {} (9f)
   (8f) edge node {} (7f)
   (9f) edge node {} (7f)
   (8f) edge node {} (10f)
   (9f) edge node {} (10f)
   (0g) edge node {} (2g)
   (1g) edge node {} (2g)
   (3g) edge node {} (0g)
   (3g) edge node {} (1g)
   (3g) edge node {} (4g)
   (3g) edge node {} (5g)
   (4g) edge node {} (2g)
   (4g) edge node {} (6g)
   (5g) edge node {} (2g)
   (5g) edge node {} (6g)
   (7g) edge node {} (4g)
   (7g) edge node {} (5g)
   (7g) edge node {} (8g)
   (7g) edge node {} (9g)
   (8g) edge node {} (6g)
   (8g) edge node {} (10g)
   (9g) edge node {} (6g)
   (9g) edge node {} (10g);
\end{tikzpicture}
%%%%%%%%%%%%%%%%%%%%%%%%%%%%%%%%%%%%%%%%%%%%%%%%%%%%%%%%%%%%%%%%%%%%%%%%%%%%%%%%%%%%%%% 
& \BKL 
%%%%%%%%%%%%%%%%%%%%%%%%%%%%%%%%%%%%%%%%%% FIVE CHAIN %%%%%%%%%%%%%%%%%%%%%%%%%%%%
\begin{tikzpicture}
 [->,thick,-stealth', transform shape, node distance=1.1cm]
 \node[sv] (E6) [] {};
 \node[sv] (E7) [above of=E6] {};
 \node[sv] (E8) [above of=E7] {};
 \node[sv] (E6a) [below of=E6] {};
 \node[sv] (E6b) [below of=E6a] {};
 \path[]
   (E7) edge node {} (E8)
   (E7) edge node {} (E6)
   (E6a) edge node {} (E6)
   (E6a) edge node {} (E6b);
\end{tikzpicture}
%%%%%%%%%%%%%%%%%%%%%%%%%%%%%%%%%%%%%%%%%%%%%%%%%%%%%%%%%%%%%%%%%%%%%%%%%%%%%%%%%
& \IKL
%%%%%%%%%%%%%%%%%%%%%%%%%%%%%%%%%%% DECAGON %%%%%%%%%%%%%%%%%%%%%%%%%%%%%%%%%%%%%%%
\begin{tikzpicture} 
[->,thick,-stealth', transform shape, node distance=1.1cm]
 \node[sv] (1f) [] {};
 \node[sv] (2f) [below left of=1f] {};
 \node[sv] (3f) [below right of=1f] {};
 \node[sv] (4f) [below of=2f] {};
 \node[sv] (5f) [below of=3f] {};
 \node[sv] (6f) [below of=4f] {};
 \node[sv] (7f) [below of=5f] {};
 \node[sv] (8f) [below of =6f] {};
 \node[sv] (9f) [below of =7f] {};
 \node[sv] (10f) [below left of=9f] {};
 \path[]
   (2f) edge node {} (1f)
   (1f) edge node {} (3f)
   (2f) edge node {} (4f)
   (2f) edge node {} (5f)
   (4f) edge node {} (3f)
   (5f) edge node {} (3f)
   (6f) edge node {} (4f)
   (4f) edge node {} (7f)
   (6f) edge node {} (5f)
   (5f) edge node {} (7f)
   (6f) edge node {} (8f)
   (6f) edge node {} (9f)
   (8f) edge node {} (7f)
   (9f) edge node {} (7f)
   (8f) edge node {} (10f)
   (9f) edge node {} (10f);
\end{tikzpicture}
%%%%%%%%%%%%%%%%%%%%%%%%%%%%%%%%%%%%%%%%%%%%%%%%%%%%%%%%%%%%%%%%%%%%%%%%%%%%%%%%%%%%%%% 
& \BKL 
%%%%%%%%%%%%%%%%%%%%%%%%%%%%%%%%%%%%%%%%%% FIVE CHAIN %%%%%%%%%%%%%%%%%%%%%%%%%%%%
\begin{tikzpicture}
 [->,thick,-stealth', transform shape, node distance=1.1cm]
 \node[sv] (E6) [] {};
 \node[sv] (E7) [above of=E6] {};
 \node[sv] (E8) [above of=E7] {};
 \node[sv] (E6a) [below of=E6] {};
 \node[sv] (E6b) [below of=E6a] {};
 \path[]
   (E7) edge node {} (E8)
   (E7) edge node {} (E6)
   (E6a) edge node {} (E6)
   (E6a) edge node {} (E6b);
\end{tikzpicture}
%%%%%%%%%%%%%%%%%%%%%%%%%%%%%%%%%%%%%%%%%%%%%%%%%%%%%%%%%%%%%%%%%%%%%%%%%%%%%%%%%
& \IKL
%%%%%%%%%%%%%%%%%%%%%%%%%%%%%%%%%%% DECAGON %%%%%%%%%%%%%%%%%%%%%%%%%%%%%%%%%%%%%%%
\begin{tikzpicture} 
[->,thick,-stealth', transform shape, node distance=1.1cm]
 \node[sv] (1f) [] {};
 \node[sv] (2f) [below left of=1f] {};
 \node[sv] (3f) [below right of=1f] {};
 \node[sv] (4f) [below of=2f] {};
 \node[sv] (5f) [below of=3f] {};
 \node[sv] (6f) [below of=4f] {};
 \node[sv] (7f) [below of=5f] {};
 \node[sv] (8f) [below of =6f] {};
 \node[sv] (9f) [below of =7f] {};
 \node[sv] (10f) [below left of=9f] {};
 \path[]
   (2f) edge node {} (1f)
   (1f) edge node {} (3f)
   (2f) edge node {} (4f)
   (2f) edge node {} (5f)
   (4f) edge node {} (3f)
   (5f) edge node {} (3f)
   (6f) edge node {} (4f)
   (4f) edge node {} (7f)
   (6f) edge node {} (5f)
   (5f) edge node {} (7f)
   (6f) edge node {} (8f)
   (6f) edge node {} (9f)
   (8f) edge node {} (7f)
   (9f) edge node {} (7f)
   (8f) edge node {} (10f)
   (9f) edge node {} (10f);
\end{tikzpicture}
%%%%%%%%%%%%%%%%%%%%%%%%%%%%%%%%%%%%%%%%%%%%%%%%%%%%%%%%%%%%%%%%%%%%%%%%%%%%%%%%%
\\
\hline
% row 6
\BKL 
\begin{tikzpicture}
\node[sv] (1) [] {};
\end{tikzpicture}
& 
\begin{tikzpicture}
\node[sv] (1) [] {};
\end{tikzpicture}
& \BKL 
\begin{tikzpicture}
 [->,thick,-stealth', transform shape, node distance=1.1cm]
 \node[sv] (E6) [] {};
 \node[sv] (E6a) [below of=E6] {};
 \node[sv] (E6b) [below of=E6a] {};
 \path[]
   (E6a) edge node {} (E6)
   (E6a) edge node {} (E6b);
\end{tikzpicture}
& 
\begin{tikzpicture} 
[->,thick,-stealth', transform shape, node distance=1.1cm]
\node[sv] (1) [] {};
\node[sv] (3) [below of=1] {};
\end{tikzpicture}
& \BKL 
%%%%%%%%%%%%%%%%%%%%%%%%%%%%%%%%%%%%%%%%%% FIVE CHAIN %%%%%%%%%%%%%%%%%%%%%%%%%%%%
\begin{tikzpicture}
 [->,thick,-stealth', transform shape, node distance=1.1cm]
 \node[sv] (E6) [] {};
 \node[sv] (E7) [above of=E6] {};
 \node[sv] (E8) [above of=E7] {};
 \node[sv] (E6a) [below of=E6] {};
 \node[sv] (E6b) [below of=E6a] {};
 \path[]
   (E7) edge node {} (E8)
   (E7) edge node {} (E6)
   (E6a) edge node {} (E6)
   (E6a) edge node {} (E6b);
\end{tikzpicture}
%%%%%%%%%%%%%%%%%%%%%%%%%%%%%%%%%%%%%%%%%%%%%%%%%%%%%%%%%%%%%%%%%%%%%%%%%%%%%%%%%
& 
\begin{tikzpicture}
 [->,thick,-stealth', transform shape, node distance=1.1cm]
 \node[sv] (E6) [] {};
 \node[sv] (E6a) [below of=E6] {};
 \node[sv] (E6b) [below of=E6a] {};
\end{tikzpicture}
&\BKL
%%%%%%%%%%%%%%%%%%%%%%%%%%%%%%%%%%%%%%%%%% SIX CHAIN %%%%%%%%%%%%%%%%%%%%%%%%%%%%
\begin{tikzpicture}
 [->,thick,-stealth', transform shape, node distance=1.1cm]
 \node[sv] (E6) [] {};
 \node[sv] (E7) [above of=E6] {};
 \node[sv] (E8) [above of=E7] {};
 \node[sv] (E9) [above of=E8] {};
 \node[sv] (E6a) [below of=E6] {};
 \node[sv] (E6b) [below of=E6a] {};
 \path[]
   (E9) edge node {} (E8)
   (E7) edge node {} (E8)
   (E7) edge node {} (E6)
   (E6a) edge node {} (E6)
   (E6a) edge node {} (E6b);
\end{tikzpicture}
%%%%%%%%%%%%%%%%%%%%%%%%%%%%%%%%%%%%%%%%%%%%%%%%%%%%%%%%%%%%%%%%%%%%%%%%%%%%%%%%%
& 
\begin{tikzpicture}
 [->,thick,-stealth', transform shape, node distance=1.1cm]
 \node[sv] (E6) [] {};
 \node[sv] (E6a) [below of=E6] {};
 \node[sv] (E6b) [below of=E6a] {};
\end{tikzpicture}
&\BKL
%%%%%%%%%%%%%%%%%%%%%%%%%%%%%%%%%%%%%%%%%% SIX CHAIN %%%%%%%%%%%%%%%%%%%%%%%%%%%%
\begin{tikzpicture}
 [->,thick,-stealth', transform shape, node distance=1.1cm]
 \node[sv] (E6) [] {};
 \node[sv] (E7) [above of=E6] {};
 \node[sv] (E8) [above of=E7] {};
 \node[sv] (E9) [above of=E8] {};
 \node[sv] (E6a) [below of=E6] {};
 \node[sv] (E6b) [below of=E6a] {};
 \path[]
   (E9) edge node {} (E8)
   (E7) edge node {} (E8)
   (E7) edge node {} (E6)
   (E6a) edge node {} (E6)
   (E6a) edge node {} (E6b);
\end{tikzpicture}
%%%%%%%%%%%%%%%%%%%%%%%%%%%%%%%%%%%%%%%%%%%%%%%%%%%%%%%%%%%%%%%%%%%%%%%%%%%%%%%%%
\end{tabular}
}
\caption{A depiction of the structures of the Kac modules $\Kac{r,s}$ as $(r,s)$ varies over (a part of) the extended Kac table.  The genuine Kac table, bounded by $1 \le r \le p-1$ and $1 \le s \le p'-1$, is represented by the dark grey rectangle in the upper-left corner.  Dark grey corresponds to interior and centre entries of the extended Kac table and light grey and white correspond to boundary and corner entries, respectively, as in \cref{fig:KacTables}.  When a dark grey cell contains two structures, the rightmost indicates that of a centre entry with $h = \frac{c}{24}$.  If $p=1$ or $p'=1$ (or both), then one should remove the rows or columns (or both) that contain interior labels.} \label{fig:KacStructures}
\end{figure}
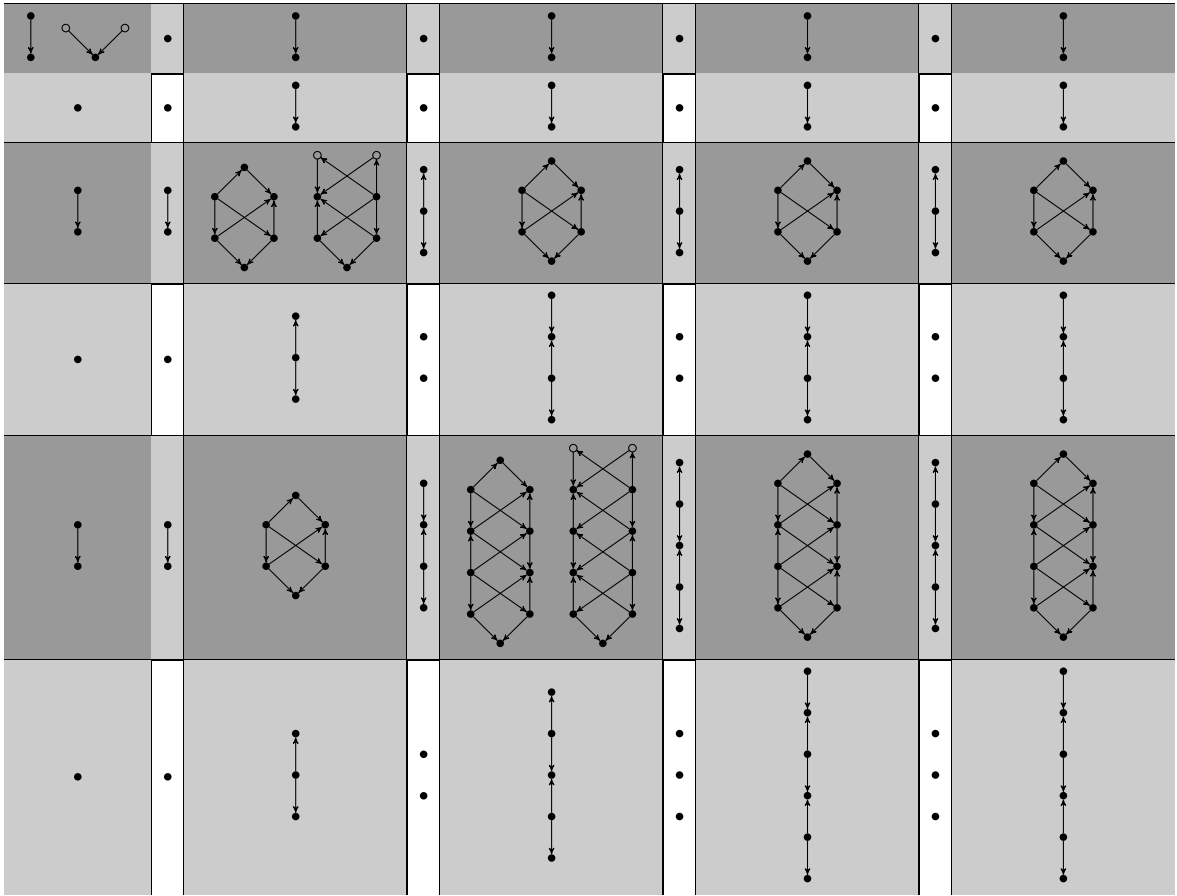

We will therefore define, for the purposes of this paper, the \emph{$N=1$ Kac module} $\Kac{r,s}$, with $r,s \in \ZZ_{>0}$, to be the submodule of the Fock space $\Fock{r,s}$ that is generated by the \ssvs{} of depths strictly less than $\frac{1}{2} rs$.  This generalises the definition proposed in \cite{MorKac15} for the Virasoro algebra and \cite{CanFusI15} for the \ns{} algebra.  We note that this definition does not preclude $\Kac{r,s}$ from having \svs{} of depth $\frac{1}{2} rs$ or greater.  A selection of Kac module structures is illustrated in \cref{fig:KacStructures}.

Inspection shows that the parity-reversing functor $\Pi$ fixes each Ramond Kac module $\Kac{r,s}$ ($r+s$ odd) and maps each \ns{} Kac module ($r+s$ even) to an inequivalent counterpart.  We will therefore affix a superscript $\pm$ to indicate the ground state parity in the \ns{} sector.  Note that $\Kac{1,1}^+$ is always a \hwm{} with an even conformal weight $0$ ground state and at most $2$ composition factors.  It plays the role of the \emph{vacuum module}, meaning that it carries the structure of the universal \vosa{} (the $N=1$ algebra).\footnote{We recall that the axioms of \vosas{} invariably require that the vacuum itself be bosonic.  This convention gives fields and their corresponding states the same parity.  Indeed, if we had instead declared that the vacuum module was $\Kac{1,1}^-$, so that the vacuum $\Omega$ was fermionic, then the fermionic field $G(z)$ would correspond to the bosonic state $G_{-3/2} \Omega$.}

\section{Characters, modular transforms and the Verlinde formula} \label{sec:CharMod}

We report here the characters and supercharacters for the \ns{} and \ram{} Fock spaces, as well as those of the Kac modules, before turning to their behaviour under modular transformations.  The block form of the resulting S-matrix is then used to formulate a fermionic Verlinde formula from which the Grothendieck fusion rules of the Kac modules are easily obtained.  As characters and supercharacters are blind to the difference between a module and the direct sum of its composition factors, the fermionic Verlinde formula only allows one to deduce the multiplicities of the composition factors of a fusion product, not the module structure of the fusion product itself.  We will address questions of structure in later sections.

\subsection{Modular transformations} \label{sec:Mod}

The characters and supercharacters of the Fock spaces are easily determined from those of the free boson and free fermion.  With $q = \ee^{2 \pi \ii \tau}$ as usual, we have
\begin{equation} \label{ch:Fock}
\begin{aligned}
\fch{\NSFock{\lambda}^{\pm}}{\tau} &= %q^{h_{\lambda} - c/24} \prod_{j=1}^{\infty} \frac{1+q^{j-1/2}}{1-q^j} =
\frac{q^{\brac{\lambda - Q/2}^2 / 2}}{\func{\eta}{q}} \sqrt{\frac{\fjth{3}{1;q}}{\func{\eta}{q}}}, &
\fch{\RFock{\lambda}}{\tau} &= %q^{h_{\lambda} - c/24} \prod_{j=1}^{\infty} \frac{1+q^j}{1-q^j} =
\frac{q^{\brac{\lambda - Q/2}^2 / 2}}{\func{\eta}{q}} \sqrt{\frac{2\,\fjth{2}{1;q}}{\func{\eta}{q}}}, \\
\fsch{\NSFock{\lambda}^{\pm}}{\tau} &= %q^{h_{\lambda} - c/24} \prod_{j=1}^{\infty} \frac{1-q^{j-1/2}}{1-q^j} =
\pm \frac{q^{\brac{\lambda - Q/2}^2 / 2}}{\func{\eta}{q}} \sqrt{\frac{\fjth{4}{1;q}}{\func{\eta}{q}}}, &
\fsch{\RFock{\lambda}}{\tau} &= 0,
\end{aligned}
\end{equation}
where we refer to \cite[App.~B]{RidSL208} for our conventions regarding Jacobi theta functions.  We note that the parity-reversing functor $\Pi$ has no effect on characters, but it negates supercharacters.  As every Ramond Fock space is fixed by $\Pi$, their supercharacters vanish identically.
%With the eta function and theta identities as follows
%\begin{equation}
%\func{\eta}{q} = q^{1/24} \prod_{j=1}^{\infty} \brac{1-q^j}, \qquad 
%\fjth{2}{z;q} = z^{1/2}q^{1/8}\prod_{j=0}^{\infty} \brac{1+zq^{j+1}} \brac{1-q^{j+1}} \brac{1+z^{-1}q^j}
%\end{equation}
%\begin{equation}
%\fjth{3}{z;q} = \prod_{j=1}^{\infty} \brac{1+zq^{j-1/2}} \brac{1-q^j} \brac{1+z^{-1}q^{j-1/2}} \qquad 
%\fjth{4}{z;q} = \prod_{j=1}^{\infty} \brac{1-zq^{j-1/2}} \brac{1-q^j} \brac{1-z^{-1}q^{j-1/2}} 
%\end{equation}

Because the \ram{} Fock space supercharacters are trivial, there are only three modular S-transforms to compute.  These follow from the transforms of the theta functions and the evaluation of a gaussian integral:
\begin{equation} \label{eq:FockSMat}
\begin{aligned}
\fch{\NSFock{\lambda}^+}{-\frac{1}{\tau}} &= \int_{-\infty}^{\infty} \Smat{\NSFock{\lambda}^+}{\NSFock{\mu}^+} \fch{\NSFock{\mu}^+}{\tau} \: \dd \mu, \quad \Smat{\NSFock{\lambda}^+}{\NSFock{\mu}^+} = \cos \sqbrac{2 \pi (\lambda - \tfrac{Q}{2}) (\mu - \tfrac{Q}{2})},\\
\fch{\RFock{\lambda}}{-\frac{1}{\tau}} &= \int_{-\infty}^{\infty} \Smat{\RFock{\lambda}}{\overline{\NSFock{\mu}^+}} \fsch{\NSFock{\mu}^+}{\tau} \: \dd \mu, \quad \Smat{\RFock{\lambda}}{\overline{\NSFock{\mu}^+}} = \sqrt{2} \cos \sqbrac{2 \pi (\lambda - \tfrac{Q}{2}) (\mu - \tfrac{Q}{2})},\\
\fsch{\NSFock{\lambda}^+}{-\frac{1}{\tau}} &= \int_{-\infty}^{\infty} \Smat{\overline{\NSFock{\lambda}^+}}{\RFock{\mu}} \fch{\RFock{\mu}}{\tau} \: \dd \mu, \quad \Smat{\overline{\NSFock{\lambda}^+}}{\RFock{\mu}} = \frac{1}{\sqrt{2}} \cos \sqbrac{2 \pi (\lambda - \tfrac{Q}{2}) (\mu - \tfrac{Q}{2})}.
\end{aligned}
\end{equation}
Here, we have indicated S-transforms involving a supercharacter, instead of a character, by a bar.  We have also assumed that the parity of each \ns{} Fock space is positive for simplicity.  S-matrix entries involving negative parities follow immediately from $\ch{\NSFock{\lambda}^-} = \ch{\NSFock{\lambda}^+}$ and $\sch{\NSFock{\lambda}^-} = -\sch{\NSFock{\lambda}^+}$.  Finally, we have followed \cite{CanFusI15} in extending the natural integration range from $[\frac{Q}{2}, \infty)$ to $(-\infty,\infty)$.  This convenience is allowed because $\Fock{\lambda}^{\pm}$ and its contragredient dual $\Fock{Q-\lambda}^{\pm}$ have the same (super)character.

With respect to the block-ordering $\set{\ch{\NSFock{}}, \ch{\RFock{}}, \sch{\NSFock{}}}$ of characters and supercharacters, the Fock space S-matrix may be summarised as
\begin{equation}
\begin{pmatrix}
\Smat{\NSFock{\lambda}}{\NSFock{\mu}} & 0 & 0 \\
0 & 0 & \Smat{\overline{\NSFock{\lambda}}}{\RFock{\mu}} \\
0 &\Smat{\RFock{\lambda}}{\overline{\NSFock{\mu}}} & 0
\end{pmatrix}
.
\end{equation}
We note that this S-matrix is not quite symmetric in this basis, but it is easily checked to be unitary and to square to the conjugation permutation.

The Fock spaces constitute a set of \emph{standard modules} \cite{CreLog13,RidVer14} for the $N=1$ algebra.  This means, among other things, that their characters form a (topological) basis for the space spanned by the characters of all the $N=1$ modules (in the module category of interest).  In particular, the Kac module characters must be expressible in terms of Fock space characters.  Recall that $\Kac{r,s}$ is a submodule of $\Fock{r,s}$, by definition.  Inspection shows that the quotient $\Fock{r,s} / \Kac{r,s}$ is not isomorphic to another Fock space or Kac module, in general, but that the character of the quotient matches that of a Fock space.  More precisely, we have the identity
\begin{equation} \label{eq:KacChar}
\ch{\Kac{r,s}} = \ch{\Fock{r,s}} - \ch{\Fock{-r,s}} = \ch{\Fock{r,s}} - \ch{\Fock{r,-s}}.
\end{equation}
The analogous identity for supercharacters is a little more complicated.  Since $h_{-r,s} = h_{r,s} + \frac{1}{2} rs$, it follows that $\Fock{r,s}$ and (the submodule whose character matches that of) $\Fock{-r,s}$ have opposite parity in the \ns{} sector, if $r$ and $s$ are both odd, and the same parity if $r$ and $s$ are both even.  In the Ramond sector, the supercharacters all vanish, hence
\begin{equation} \label{eq:KacSChar}
\begin{aligned}
\sch{\Kac{r,s}^{\pm}} &= \sch{\Fock{r,s}^{\pm}} - (-1)^r \sch{\Fock{-r,s}^{\pm}} & &\text{(\(r+s\) even),} \\
\sch{\Kac{r,s}} &= 0 & &\text{(\(r+s\) odd).}
\end{aligned}
\end{equation}

It is worth noting at this point that \eqref{eq:KacChar} and \eqref{eq:KacSChar} allow one to formally extend the Kac characters and supercharacters from $r,s \in \ZZ_{>0}$ to all $r,s \in \ZZ$.  Upon doing this, one arrives at the relations
\begin{equation} \label{eq:ChSchRels}
\begin{gathered}
\ch{\Kac{-r,s}} = -\ch{\Kac{r,s}} = \ch{\Kac{r,-s}}, \quad \ch{\Kac{r,0}} = \ch{\Kac{0,s}} = 0, \quad \ch{\Kac{-r,-s}} = \ch{\Kac{r,s}}, \\
\sch{\Kac{r,s}^{\pm}} = (-1)^{r-1} \sch{\Kac{-r,s}^{\pm}} = (-1)^{s-1} \sch{\Kac{r,-s}^{\pm}} = \sch{\Kac{-r,-s}^{\pm}}.
\end{gathered}
\end{equation}
These relations will be important for interpreting the Verlinde formula calculations that follow.

By combining \eqref{eq:FockSMat} with \eqref{eq:KacChar} and \eqref{eq:KacSChar}, we obtain the S-matrix entries for the Kac module characters and supercharacters as differences of Fock space S-matrix entries:
\begin{equation}
\begin{aligned}
\Smat{\Kac{r,s}^+}{\NSFock{\mu}^+} &= 2 \sin \sqbrac{2 \pi \alpha' r (\mu - \tfrac{Q}{2})} \sin \sqbrac{2 \pi \alpha s (\mu - \tfrac{Q}{2})} \mspace{50mu} \text{(\(r+s\) even),} \\
\Smat{\Kac{r,s}}{\overline{\NSFock{\mu}^+}} &= 2 \sqrt{2} \sin \sqbrac{2 \pi \alpha' r (\mu - \tfrac{Q}{2})} \sin \sqbrac{2 \pi \alpha s (\mu - \tfrac{Q}{2})} \mspace{30mu} \text{(\(r+s\) odd),} \\
\Smat{\overline{\Kac{r,s}^+}}{\RFock{\mu}} &=
\begin{cases}
\sqrt{2} \cos \sqbrac{2 \pi \alpha' r (\mu - \tfrac{Q}{2})} \cos \sqbrac{2 \pi \alpha s (\mu - \tfrac{Q}{2})} & \text{(\(r\), \(s\) odd),} \\[1ex]
\sqrt{2} \sin \sqbrac{2 \pi \alpha' r (\mu - \tfrac{Q}{2})} \sin \sqbrac{2 \pi \alpha s (\mu - \tfrac{Q}{2})} & \text{(\(r\), \(s\) even).}
\end{cases}
\end{aligned}
\end{equation}
Again, we have assumed positive parity ground states in the \ns{} sector for simplicity.  We remark that S-matrix entries of the form $\Smat{\Kac{r,s}}{\Kac{r',s'}}$ are not defined in this setup.

\subsection{Grothendieck fusion products} \label{sec:VerProd}

We are interested in the fusion rules of the Kac modules $\Kac{r,s}$, for $r,s \in \ZZ_{>0}$.  Consider therefore the category of $N=1$ modules that is generated by the Kac modules under finite iterated fusion products.  Because Kac modules are believed to define boundary sectors of the logarithmic $N=1$ superconformal minimal models \cite{PeaLog14} and because fusing with a module defining a boundary sector is believed to define an exact endofunctor on the module category relevant to the \cft{} \cite{GabFus09}, we will assume that fusing with a Kac module defines an exact functor on our category.  If we further assume that fusion defines a tensor structure on our module category, then fusing with any module from this category defines an exact functor \cite{EtiTen15}.  The fusion product $\fuse$ then descends to a commutative associative product $\Grfuse$ on the Grothendieck group of the category.  We call the resulting ring the \emph{Grothendieck fusion ring} and call $\Grfuse$ the \emph{Grothendieck fusion product}.

In bosonic \cft{}, one is accustomed to identifying the Grothendieck fusion ring with the ring generated by the characters of the simple modules equipped with $\Grfuse$, checking first that these characters are linearly independent.  In the fermionic case, one cannot do this because $\ch{\Kac{1,1}^+} = \ch{\Kac{1,1}^-}$ (for example).  Instead, one equips the characters and, separately, the supercharacters with $\Grfuse$, noting that knowledge of an identity of characters and the corresponding identity of supercharacters allows one to reconstruct the identity in the Grothendieck fusion ring.  We denote the image of an $N=1$ module $\Mod{M}$ in the Grothendieck fusion ring by $\Gr{\Mod{M}}$ so that its character $\ch{\Mod{M}}$ and supercharacter $\sch{\Mod{M}}$ are obtained by applying the (formal) linear operators $\chmap$ and $\schmap$, respectively.

Although the $N=1$ Fock spaces $\NSFock{\lambda}$ and $\RFock{\lambda}$ are not in the category that we are considering, the (super)characters of the Kac modules may be expressed as linear combinations of Fock space (super)characters.  Indeed, the standard module formalism of \cite{CreLog13,RidVer14} requires that we use the Fock space (super)characters as a canonical basis in all modular computations.  It follows that if $\Mod{M}$ and $\Mod{N}$ are $N=1$ Kac modules, then we may decompose the (super)character of their fusion product into a linear combination of Fock space (super)characters:
\begin{equation} \label{eq:VerChSch}
\begin{aligned}
\ch{\Mod{M} \fuse \Mod{N}} &= \ch{\Mod{M}} \Grfuse \ch{\Mod{N}} = \int_{-\infty}^{\infty} \sqbrac{\fuscoeff{\Mod{M}}{\Mod{N}}{\NSFock{\nu}^+} \ch{\NSFock{\nu}^+} + \fuscoeff{\Mod{M}}{\Mod{N}}{\RFock{\nu}} \ch{\RFock{\nu}}} \: \dd \nu, \\
\sch{\Mod{M} \fuse \Mod{N}} &= \sch{\Mod{M}} \Grfuse \sch{\Mod{N}} = \int_{-\infty}^{\infty} \fuscoeff{\overline{\Mod{M}}}{\overline{\Mod{N}}}{\overline{\NSFock{\nu}^+}} \sch{\NSFock{\nu}^+} \: \dd \nu.
\end{aligned}
\end{equation}
Here, bars indicate supercharacters, as above, and we have recalled that the $\sch{\RFock{\nu}}$ all vanish.  The multiplicities $\fuscoeff{\Mod{M}}{\Mod{N}}{\NSFock{\nu}^+}$, $\fuscoeff{\Mod{M}}{\Mod{N}}{\RFock{\nu}}$ and $\fuscoeff{\overline{\Mod{M}}}{\overline{\Mod{N}}}{\overline{\NSFock{\nu}^+}}$ will be referred to as the \emph{Verlinde coefficients} because we conjecture that they may be computed, in terms of the S-matrix entries, by the following version of the Verlinde formula:
\begin{equation} \label{eq:Verlinde}
\fuscoeff{\Mod{M}}{\Mod{N}}{\Fock{\nu}} = A_{\Mod{M} \Mod{N}} \int_{-\infty}^{\infty} \frac{\Smat{\Mod{M}}{\Fock{\rho}} \Smat{\Mod{N}}{\Fock{\rho}} \Smat{\Fock{\nu}}{\Fock{\rho}}^*}{\Smat{\Kac{1,1}^+}{\Fock{\rho}}} \: \dd \rho.
\end{equation}
This formula covers all Verlinde coefficients if interpreted as follows:  First, the $\Fock{\rho}$ run over both the \ns{} and Ramond Fock spaces, in principle, but in practice, only one sector contributes.  Second, whenever a module in one of the S-matrix entries on the \rhs{} is Ramond, then the other module is understood to be barred.  Finally, the constant $A_{\Mod{M} \Mod{N}}$ is unity unless $\Mod{M}$ and $\Mod{N}$ are in different sectors, in which case $A_{\Mod{M} \Mod{N}} = \frac{1}{2}$, or they are barred, in which case $A_{\Mod{M} \Mod{N}} \equiv A_{\overline{\Mod{M}} \overline{\Mod{N}}} = 2$.  We call \eqref{eq:Verlinde} the \emph{$N=1$ Verlinde formula}.  It is derived, assuming the standard Verlinde formula for the bosonic orbifold of the $N=1$ algebra, in \cref{sec:FermVer}.

We illustrate the use of this Verlinde formula by computing the Grothendieck fusion rules involving $\Kac{2,1}$.  Fusing first with the Ramond Kac module $\Kac{r,s}$ (so $r+s$ is odd), the $N=1$ Verlinde formula \eqref{eq:Verlinde} becomes
\begin{align}
\fuscoeff{\Kac{2,1}}{\Kac{r,s}}{\NSFock{\nu}^+} &= \int_{-\infty}^{\infty} \frac{\Smat{\Kac{2,1}}{\overline{\NSFock{\rho}^+}} \Smat{\Kac{r,s}}{\overline{\NSFock{\rho}^+}} \Smat{\NSFock{\nu}^+}{\NSFock{\rho}^+}^*}{\Smat{\Kac{1,1}^+}{\NSFock{\rho}^+}} \: \dd \rho \notag \\
&= 8 \int_{-\infty}^{\infty} \cos \sqbrac{2 \pi \alpha' \rho} \sin \sqbrac{2 \pi \alpha' r \rho} \sin \sqbrac{2 \pi \alpha s \rho} \cos \sqbrac{2 \pi (\nu - \tfrac{Q}{2}) \rho} \: \dd \rho \notag \\
&= \func{\delta}{\nu = \lambda_{r-1,s}} - \func{\delta}{\nu = \lambda_{-(r-1),s}} - \func{\delta}{\nu = \lambda_{r-1,-s}} + \func{\delta}{\nu = \lambda_{-(r-1),-s}} \notag \\
&\mspace{20mu} + \func{\delta}{\nu = \lambda_{r+1,s}} - \func{\delta}{\nu = \lambda_{-(r+1),s}} - \func{\delta}{\nu = \lambda_{r+1,-s}} + \func{\delta}{\nu = \lambda_{-(r+1),-s}}
\end{align}
and $\fuscoeff{\Kac{2,1}}{\Kac{r,s}}{\RFock{\nu}} = 0$.  Substituting into \eqref{eq:VerChSch}, while remembering \eqref{eq:KacChar} and \eqref{eq:ChSchRels}, we obtain
\begin{align}
\ch{\Kac{2,1}} \Grfuse \ch{\Kac{r,s}} &= 2 \Bigl( \ch{\Fock{r-1,s}} - \ch{\Fock{-(r-1),s}} + \ch{\Fock{r+1,s}} - \ch{\Fock{-(r+1),s}} \Bigr) \notag \\
&= 2 \Bigl( \ch{\Kac{r-1,s}} + \ch{\Kac{r+1,s}} \Bigr).
\end{align}
As $\sch{\Kac{2,1}} \Grfuse \sch{\Kac{r,s}} = 0$, the overall multiplicity of $2$ appearing in this character must correspond to each (\ns{}) factor appearing once with positive parity and once with negative parity.  This lets us deduce the following Grothendieck fusion rule:
\begin{equation} \label{GrFR:K21xKrsR}
\Gr{\Kac{2,1}} \Grfuse \Gr{\Kac{r,s}} = \Gr{\Kac{r-1,s}^+} + \Gr{\Kac{r-1,s}^-} + \Gr{\Kac{r+1,s}^+} + \Gr{\Kac{r+1,s}^-} \qquad \text{(\(r+s\) odd).}
\end{equation}
Applying $\chmap$ and $\schmap$ then recovers the corresponding character and supercharacter identities, respectively.

Fusing $\Kac{2,1}$ with a \ns{} Kac module $\Kac{r,s}^{\pm}$ (so $r+s$ is even), we note that $A_{\Kac{2,1} \Kac{r,s}^{\pm}} = \frac{1}{2}$ and thus
\begin{align}
\fuscoeff{\Kac{2,1}}{\Kac{r,s}^{\pm}}{\RFock{\nu}} &= \frac{1}{2} \int_{-\infty}^{\infty} \frac{\Smat{\Kac{2,1}}{\overline{\NSFock{\rho}^+}} \Smat{\Kac{r,s}^{\pm}}{\NSFock{\rho}^+} \Smat{\RFock{\nu}}{\overline{\NSFock{\rho}^+}}^*}{\Smat{\Kac{1,1}^+}{\NSFock{\rho}^+}} \: \dd \rho \notag \\
&= 4 \int_{-\infty}^{\infty} \cos \sqbrac{2 \pi \alpha' \rho} \sin \sqbrac{2 \pi \alpha' r \rho} \sin \sqbrac{2 \pi \alpha s \rho} \cos \sqbrac{2 \pi (\nu - \tfrac{Q}{2}) \rho} \: \dd \rho.
\end{align}
The result is therefore half that of the previous calculation:
\begin{gather}
\ch{\Kac{2,1}} \Grfuse \ch{\Kac{r,s}^{\pm}} = \ch{\Kac{r-1,s}} + \ch{\Kac{r+1,s}}, \qquad \sch{\Kac{2,1}} \Grfuse \sch{\Kac{r,s}^{\pm}} = 0 \notag \\
\Ra \qquad \Gr{\Kac{2,1}} \Grfuse \Gr{\Kac{r,s}^{\pm}} = \Gr{\Kac{r-1,s}} + \Gr{\Kac{r+1,s}} \qquad \text{(\(r+s\) even).} \label{GrFR:K21xKrsNS}
\end{gather}
Similar calculations with $\Kac{2,1}$ replaced by $\Kac{1,2}$ lead to results analogous to \eqref{GrFR:K21xKrsR} and \eqref{GrFR:K21xKrsNS}:
\begin{equation}
\begin{aligned}
\Gr{\Kac{1,2}} \Grfuse \Gr{\Kac{r,s}^{\pm}} &= \Gr{\Kac{r,s-1}} + \Gr{\Kac{r,s+1}} & &\text{(\(r+s\) even),} \\
\Gr{\Kac{1,2}} \Grfuse \Gr{\Kac{r,s}} &= \Gr{\Kac{r,s-1}^+} + \Gr{\Kac{r,s-1}^-} + \Gr{\Kac{r,s+1}^+} + \Gr{\Kac{r,s+1}^-} & &\text{(\(r+s\) odd).}
\end{aligned}
\end{equation}

One can now use associativity to explore the Grothendieck fusion rules of general Kac modules.  However, because fusing two Ramond Kac modules gives back \ns{} Kac modules of both parities, associativity does not completely determine the Grothendieck fusion rules of the \ns{} Kac modules.  Rather, it only fixes these rules up to parity.  To determine the missing information, we apply the $N=1$ Verlinde formula to Grothendieck fusion products involving $\Kac{3,1}^+$, $\Kac{2,2}^+$ and $\Kac{1,3}^+$.  Applying associativity to these results will then determine the Grothendieck fusion rule parities for all Kac modules.

In \cite{CanFusI15}, we used the standard Verlinde formula (that applies to \ns{} characters) to deduce that
\begin{equation}
\ch{\Kac{3,1}^+} \Grfuse \ch{\Kac{r,s}^+} = \ch{\Kac{r-2,s}^+} + \ch{\Kac{r,s}^+} + \ch{\Kac{r+2,s}^+} \qquad \text{(\(r+s\) even),}
\end{equation}
interpreting the \rhs{} using \eqref{eq:ChSchRels} if necessary.  The supercharacter version of this now follows from \eqref{eq:KacSChar} and the $N=1$ Verlinde formula \eqref{eq:Verlinde}:
\begin{gather}
\begin{aligned}
\fuscoeff{\overline{\Kac{3,1}^+}}{\overline{\Kac{r,s}^+}}{\overline{\NSFock{\nu}^+}} &= 2 \int_{-\infty}^{\infty} \frac{\Smat{\overline{\Kac{3,1}^+}}{\RFock{\rho}} \Smat{\overline{\Kac{r,s}^+}}{\RFock{\rho}} \Smat{\overline{\NSFock{\nu}^+}}{\RFock{\rho}}^*}{\Smat{\overline{\Kac{1,1}^+}}{\RFock{\rho}}} \: \dd \rho \\
&=
\begin{cases}
\displaystyle 2 \int_{-\infty}^{\infty} \Bigl( 2 \, \cos \sqbrac{4 \pi \alpha' \rho} - 1 \Bigr) \cos \sqbrac{2 \pi \alpha' r \rho} \cos \sqbrac{2 \pi \alpha s \rho} \cos \sqbrac{2 \pi (\nu - \tfrac{Q}{2}) \rho} \: \dd \rho & \text{(\(r\), \(s\) odd),} \\[2ex]
\displaystyle 2 \int_{-\infty}^{\infty} \Bigl( 2 \, \cos \sqbrac{4 \pi \alpha' \rho} - 1 \Bigr) \sin \sqbrac{2 \pi \alpha' r \rho} \sin \sqbrac{2 \pi \alpha s \rho} \cos \sqbrac{2 \pi (\nu - \tfrac{Q}{2}) \rho} \: \dd \rho & \text{(\(r\), \(s\) even)}
\end{cases}
\end{aligned}
\notag \\
\Ra \qquad \sch{\Kac{3,1}^+} \Grfuse \sch{\Kac{r,s}^+} = \sch{\Kac{r-2,s}^+} - \sch{\Kac{r,s}^+} + \sch{\Kac{r+2,s}^+} \qquad \text{(\(r+s\) even).}
\end{gather}
It therefore follows that the Grothendieck fusion rule is
\begin{subequations}
\begin{align}
\Gr{\Kac{3,1}^+} \Grfuse \Gr{\Kac{r,s}^+} &= \Gr{\Kac{r-2,s}^+} + \Gr{\Kac{r,s}^-} + \Gr{\Kac{r+2,s}^+} & &\text{(\(r+s\) even),}
\intertext{consistent with the explicit \ns{} fusion calculations reported in \cite{CanFusI15}.  We similarly obtain}
\Gr{\Kac{1,3}^+} \Grfuse \Gr{\Kac{r,s}^+} &= \Gr{\Kac{r,s-2}^+} + \Gr{\Kac{r,s}^-} + \Gr{\Kac{r,s+2}^+} & &\text{(\(r+s\) even),} \\
\Gr{\Kac{2,2}^+} \Grfuse \Gr{\Kac{r,s}^+} &= \Gr{\Kac{r-1,s-1}^+} + \Gr{\Kac{r-1,s+1}^-} + \Gr{\Kac{r+1,s-1}^-} + \Gr{\Kac{r+1,s+1}^+} & &\text{(\(r+s\) even).}
\end{align}
\end{subequations}
It is clear that Grothendieck fusion respects parities in the sense that changing the parity of one of the modules being fused results in a global change of parity of the fusion product.

Associativity now completely determines the Grothendieck fusion rules of the $N=1$ Kac modules.  The simplest are the mixed fusion rules involving a \ns{} and a Ramond module:
\begin{subequations} \label{GrFR:KxK}
\begin{align}
\Gr{\Kac{r,s}^{\pm}} \Grfuse \Gr{\Kac{r',s'}} &= \sideset{}{'} \sum_{r''=\abs{r-r'}+1}^{r+r'-1} \ \sideset{}{'} \sum_{s'' = \abs{s-s'}+1}^{s+s'-1} \Gr{\Kac{r'',s''}} & &\text{(\(r+s\) even, \(r'+s'\) odd).} \label{GrFR:NSxR}
\intertext{Here, a primed summation indicates that the summation variable increases in steps of two.  The Ramond-Ramond fusion rules are similar, but the result decomposes into \ns{} modules of both parities:}
\Gr{\Kac{r,s}} \Grfuse \Gr{\Kac{r',s'}} &= \sideset{}{'} \sum_{r''=\abs{r-r'}+1}^{r+r'-1} \ \sideset{}{'} \sum_{s'' = \abs{s-s'}+1}^{s+s'-1} \brac{\Gr{\Kac{r'',s''}^+} + \Gr{\Kac{r'',s''}^-}} & &\text{(\(r+s\), \(r'+s'\) odd).} \label{GrFR:RxR}
\intertext{Finally, fusing a \ns{} module with another \ns{} module results in}
\Gr{\Kac{r,s}^+} \Grfuse \Gr{\Kac{r',s'}^+} &= \sideset{}{'} \sum_{r''=\abs{r-r'}+1}^{r+r'-1} \ \sideset{}{'} \sum_{s'' = \abs{s-s'}+1}^{s+s'-1} \Gr{\Kac{r'',s''}^{\bullet}} & &\text{(\(r+s\), \(r'+s'\) even).} \label{GrFR:NSxNS}
\end{align}
\end{subequations}
The parity $\bullet$ is $+$ if $\frac{1}{2} (r+s + r'+s' + r''+s'')$ is odd and is $-$ otherwise.  Alternatively, $\bullet$ is $+$ for $(r'',s'') = (r+r'-1, s+s'-1)$ and it changes sign every time $r''$ or $s''$ decreases by $2$.

\section{Fusing twisted modules} \label{sec:TwFus}

Whilst the Verlinde formula \eqref{eq:Verlinde} allows one to determine the character and supercharacter of a fusion product, it does not reveal any structural details beyond identifying the composition factors.  To determine the structure, one can try to construct the fusion product explicitly using the \NGK{} fusion algorithm of \cite{NahQua94,GabInd96}.  This algorithm applies directly to untwisted modules, so our task in this section is to develop a twisted version of this algorithm that can be applied to both \ns{} and Ramond modules.  Such a twisted fusion algorithm was first outlined in \cite{GabFus97}, where coproduct formulae were derived for the action of a \vosa{} on a fusion product.  However, the implementation there was limited to the depth zero truncation of certain generic fusion products, where indecomposable structure was ignored.  Here, we extend the algorithm to all depths while significantly simplifying the coproduct formulae.  A derivation of these formulae is detailed in \cref{app:TwCoprod}, for completeness, as is a definition \eqref{eq:DefFusion} of the fusion product $\Mod{M} \fuse \Mod{N}$ of two (twisted) modules $\Mod{M}$ and $\Mod{N}$.

A key feature of fusion products, as far as the (twisted) \NGK{} algorithm is concerned, is that they admit consistent truncations from which one can (hopefully) determine the full structure unambiguously.  In many cases, including those considered here, it is enough to consider finite-dimensional truncations which are easily encoded in a computer algebra system.  For examples in which infinite-dimensional truncations are unavoidable, see \cite{GabFus01,RidFus10,RidBos14}.  

The truncations that we will compute in what follows are labelled by a non-negative integer $d$ (the depth) and correspond to quotienting the fusion product by the subspace generated by the action of monomials in the modes whose total weight is greater than $d$.  More precisely, define the following subalgebra of the mode algebra:
\begin{equation}
\alg{U}^d = \vspn \set{S^{(k_1)}_{n_1} \cdots S^{(k_r)}_{n_r} \st r \in \ZZ_{\ge 0}, \ n_1 + \cdots + n_r < -d}.
\end{equation}
Here, the indices $k_1, \ldots, k_r$ serve to distinguish the generating fields of the \vosa{}, though we shall often drop them in what follows to lighten the notation.  The \emph{depth $d$ truncation} of a module $\Mod{M}$ is then
\begin{equation}
\Mod{M}^d = \frac{\Mod{M}}{\alg{U}^d \Mod{M}}.
\end{equation}
This defines truncations of fusion products $\Mod{M} \fuse \Mod{N}$ wherein the action of $\alg{U}^d$ is obtained as (a quotient of) the action defined by the twisted coproduct formulae on $\Mod{M} \otimes_{\CC} \Mod{N}$, see \eqref{eq:DefFusion}.

In the untwisted case, the key fact upon which the \NGK{} fusion algorithm rests is the (vector space) inclusion \cite{GabInd96}
\begin{equation} \label{eq:OldKey}
\brac{\Mod{M} \fuse \Mod{N}}^d \subseteq \spsub{\Mod{M}} \otimes_{\CC} \Mod{N}^d,
\end{equation}
where $\spsub{\Mod{M}}$ is the so-called \emph{special subspace} of $\Mod{M}$.  This realises each truncated fusion product inside a tensor product space where the action of the modes may be explicitly computed using the untwisted coproduct formulae.  Our primary aim in this section is to generalise this inclusion to truncations of fusion products of twisted modules.  

For this, it is convenient to review the twisted coproduct formulae \eqref{eq:TwCoprods} which we write in the form of three master equations:
\begin{subequations} \label{eq:Master}
\begin{align}
\coproduct{\tS_n} &= \sum_{m=-h-\eps_1+1}^{\infty} \binom{n+h+\eps_1-1}{m+h+\eps_1-1} w^{n-m} \brac{S_m \otimes \wun} & &\text{(\(n \ge -h-\eps+1\))} \notag \\
&\mspace{50mu} + \mu_1 \sum_{j=0}^{\infty} \binom{-\eps_1}{j} \brac{-w}^{-\eps_1-j} \brac{\wun \otimes S_{n+\eps_1+j}}, \label{eq:Master1} \\
\coproduct{\tS_{-n}} &= \sum_{m=-h-\eps_1+1}^{\infty} \binom{m+n-1}{m+h+\eps_1-1} \brac{-1}^{m+h+\eps_1-1} w^{-m-n} \brac{S_m \otimes \wun} & &\text{(\(n \ge h+\eps\))} \notag \\
&\mspace{50mu} + \mu_1 \sum_{j=0}^{\infty} \binom{-\eps_1}{j} \brac{-w}^{-\eps_1-j} \brac{\wun \otimes S_{-n+\eps_1+j}}, \label{eq:Master2} \\
\sum_{j=0}^{\infty} \binom{-\eps_2}{j} &w^{-\eps_2-j} \brac{S_{-n+\eps_2+j} \otimes \wun} = \sum_{j,k=0}^{\infty} \brac{-1}^j \binom{\eps_1}{j} \binom{n-h-\eps_2+k}{k} w^{j+k} \coproduct{\tS_{-n-j-k}} & &\text{(\(n \ge h+\eps\))} \notag \\
&\mspace{50mu} + \mu_1 \sum_{m=-h-\eps_2+1}^{\infty} \binom{m+n-1}{m+h+\eps_2-1} \brac{-1}^{m+h+\eps_2} \brac{-w}^{-m-n} \brac{\wun \otimes S_m}. \label{eq:Master3}
\end{align}
\end{subequations}
Here, we have lightened the notation somewhat, as compared with \cref{app:TwCoprod}, by writing $\coproductsymb$ for $\parNcoproductsymb{2}{w,0}$ and $\tS_n$ for $\tS_n^{w,0}$ (see \eqref{eq:STilde} for the definition of the tilde modes).  Note that we have kept $w$ and $-w$ as formal indeterminates, instead of evaluating them at $w=1$ (say), because they may appear with non-integer exponents.  These master equations are to be understood as acting on a tensor product state $\psi_1 \otimes \psi_2$.  Then, $\eps_i$ is the twist parameter for $\psi_i(w)$, with respect to $\func{S}{z}$, see \eqref{eq:TwOPE}, $\eps = \eps_1 + \eps_2$, and $\mu_1$ is the mutual locality parameter for $\func{S}{z}$ and $\func{\psi_1}{w}$, see \eqref{eq:Locality}.\footnote{Strictly speaking, the factors of $\mu_1$ appearing in \eqref{eq:Master} should not be present because acting with $\wun \otimes S_m$ on $\psi_1 \otimes \psi_2$ will reproduce $\mu_1$ from the parities of $S_m$ and $\psi_1$, since $\otimes$ is a graded tensor product.  However, we have decided to keep the $\mu_1$ factors as an explicit reminder of parity and to be consistent with the conventions of \cite{GabFus94b,GabFus97}.}  In particular, for $S=G$, $\mu_1 = +1$ if $\psi_1$ is bosonic and $\mu_1 = -1$ if $\psi_1$ is fermionic.

We remark that \eqref{eq:Master3} is obtained by combining the coproduct formula \eqref{eq:TwCoprods2} with the translation formula \eqref{eq:Translation} to eliminate the alternative coproduct $\parNcoproductsymb{1}{0,-w}$.  Imposing this relation captures the definition \eqref{eq:DefFusion} of the fusion product as the largest quotient of the tensor product that is consistent with locality.  We also mention that the twist parameters $\eps_i$ are only defined modulo $1$, in principle.  However, it is clear that the coproduct formulae \eqref{eq:Master}, and hence the actual implementation of the twisted \NGK{} algorithm, are not invariant under shifting the $\eps_i$ by an integer.  Whilst the structure of a fusion product cannot depend on the choice of twist parameters used, we shall see that investigating this structure algorithmically becomes hopelessly impractical for all but a small number of choices.

To determine the appropriate generalisation of \eqref{eq:OldKey} for twisted modules, the idea is to apply the master equations \eqref{eq:Master}, along with $\coproduct{\alg{U}^d} = 0$, to elements $\psi_1 \otimes \psi_2 \in \Mod{M} \otimes_{\CC} \Mod{N}$.  For the twisted special subspace $\spsub{\Mod{M}}$, we substitute \eqref{eq:Master2} into \eqref{eq:Master3}, assuming that $n \ge h + \eps$ and suppressing all coefficients for brevity:
\begin{align}
\sum_{j=0}^{\infty} \brac{S_{-n+\eps_2+j} \otimes \wun} &\sim \sum_{j,k=0}^{\infty} \coproduct{\tS_{-n-j-k}} + \sum_{m=-h-\eps_2+1}^{\infty} \brac{\wun \otimes S_m} \notag \\
&\sim \sum_{j,k=0}^{\infty} \sqbrac{\sum_{m=-h-\eps_1+1}^{\infty} \brac{S_m \otimes \wun} + \sum_{\ell=0}^{\infty} \brac{\wun \otimes S_{-n-j-k+\eps_1+\ell}}} + \sum_{m=-h-\eps_2+1}^{\infty} \brac{\wun \otimes S_m}.
\end{align}
We interpret this as saying that $S_{-n+\eps_2} \psi_1 \otimes \psi_2$ may always be written as a linear combination of terms of the form $S_k \psi_1 \otimes \psi_2$ and $\psi_1 \otimes S_{\ell} \psi_2$, where $k \ge \min \set{-n+\eps_2+1, -h-\eps_1+1}$.  By iteration, it follows that any $S_n \psi_1 \otimes \psi_2$ with $n \le -h-\eps_1$ may be written as a linear combination of similar terms with $n > -h-\eps_1$ and terms of the form $\psi_1 \otimes S_m \psi_2$.  This suggests the following definition for the \emph{twisted special subspace} of $\Mod{M}$:
\begin{equation}
\spsub{\Mod{M}} = \frac{\Mod{M}}{\spsub{\alg{U}} \Mod{M}}, \qquad 
\spsub{\alg{U}} = \left\langle S^{(k)}_n \st n \le -h^{(k)}-\eps_1^{(k)} \right\rangle.
\end{equation}
The twisted special subspace $\spsub{\Mod{M}}$ therefore depends upon the twist parameters $\eps_1^{(k)}$, defined with respect to each (generating) field $\func{S^{(k)}}{z}$, of the fields of $\Mod{M}$.

We illustrate this definition for the $N=1$ algebra.  The generating fields are $\func{T}{z}$ and $\func{G}{z}$ and we may assume that all twist parameters with respect to $\func{T}{z}$ are $0$.  In the \ns{} sector, we may also assume that the twist parameters with respect to $\func{G}{z}$ are $0$, hence we obtain
\begin{equation}
\spsub{\alg{U}} = \left\langle L_m, \ G_n \st m \le -2, \ n \le -\tfrac{3}{2} \right\rangle.
\end{equation}
In particular, a \ns{} Verma module $\NSVer{}$ generated by a \hwv{} $\psi_1$ has special subspace
\begin{equation}
\spsub{\NSVer{}} = \vspn \set{L_{-1}^j G_{-1/2}^k \psi_1 \st j,k \in \ZZ_{\ge 0}}.
\end{equation}
In the Ramond sector, one might choose the twist parameter $\eps_1$ for $\func{G}{z}$ to be $+\frac{1}{2}$ or $-\frac{1}{2}$ (or another element in $\ZZ + \frac{1}{2}$).  However, the resulting (generic) Verma module special subspaces are quite different:
\begin{equation} \label{eq:VermaSS}
\begin{aligned}
\eps_1 &= -\frac{1}{2}: & \spsub{\alg{U}} &= \left\langle L_m, \ G_n \st m \le -2, \ n \le -1 \right\rangle, & \spsub{\Ver{}} &= \vspn \set{L_{-1}^j G_0^k \psi_1 \st j \in \ZZ_{\ge 0}, \ k=0,1}. \\
\eps_1 &= +\frac{1}{2}: & \spsub{\alg{U}} &= \left\langle L_m, \ G_n \st m,n \le -2 \right\rangle, & \spsub{\Ver{}} &= \vspn \set{L_{-1}^j G_{-1}^k G_0^{\ell} \psi_1 \st j \in \ZZ_{\ge 0}, \ k,\ell=0,1}.
\end{aligned}
\end{equation}
One therefore has some freedom in choosing the twist parameter so as to optimise the role of the special subspace in fusion computations.  However, we shall see that this has to be balanced against other considerations.\footnote{Observe that the twisted special subspace is \emph{empty} for $\eps_1 \le -\frac{3}{2}$, indicating that the twisted \NGK{} algorithm, as presented here, cannot be employed to determine the structure of the fusion product with this choice of twist parameter.  Whilst it may be possible to modify the algorithm so as to overcome this problem, see \cite[Sec.~7]{RidBos14} for a similar situation, we shall avoid it entirely by simply not choosing these values for $\eps_1$.}

In particular, we also need to determine the twisted generalisation of the truncated subspace $\Mod{N}^d$ appearing on the \rhs{} of \eqref{eq:OldKey}.  For this, we may assume that the master equations have already been utilised to convert $\psi_1 \otimes \psi_2 \in \Mod{M} \otimes_{\CC} \Mod{N}$ into a linear combination of similar terms in which each $\psi_1 \in \spsub{\Mod{M}}$.  Because we are truncating $\Mod{M} \fuse \Mod{N}$ to depth $d$, we may assert that $\coproduct{S_n} = 0$, for all $n<-d$.  It follows immediately from \eqref{eq:STilde} that $\coproduct{\tS_n} = 0$, for all $n<-d$, as well.  Substituting this into \eqref{eq:Master1} and \eqref{eq:Master2} then results in
\begin{equation}
\sum_{j=0}^{\infty} \brac{\wun \otimes S_{n+\eps_1+j}} \sim \sum_{m=-h-\eps_1+1}^{\infty} \brac{S_m \otimes \wun},
\end{equation}
where we again suppress all constants.  Thus, a term of the form $\psi_1 \otimes S_{n+\eps_1} \psi_2$, where $\psi_1 \in \spsub{\Mod{M}}$ and $n<-d$, may be written as a linear combination of such terms with $n \ge -d$ and terms of the form $S_m \psi_1 \otimes \psi_2$ with $m>-h-\eps_1$.  Note that $S_m \psi_1$ is (usually) an element of $\spsub{\Mod{M}}$ under these conditions; we will return to this point shortly.

Repeating these manipulations for $\coproduct{S_{n_1}^{(k_1)} \cdots S_{n_r}^{(k_r)}} = 0$, which holds whenever $n_1 + \cdots + n_r < -d$, thereby motivates the definition of the \emph{twisted truncated subspace} of $\Mod{N}$:
\begin{equation} \label{eq:DefTwTruncSubSp}
\Mod{N}^{(d)} = \frac{\Mod{N}}{\alg{U}^{(d)} \Mod{N}}, \qquad \alg{U}^{(d)} = \vspn \set{S^{(k_1)}_{n_1} \cdots S^{(k_r)}_{n_r} \st r \in \ZZ_{>0}, \ n_1 + \cdots + n_r < -d + \eps_1^{(k_1)} + \cdots + \eps_1^{(k_r)}}.
\end{equation}
These twisted truncations of $\Mod{N}$ therefore also depend upon the twist parameters $\eps_1^{(k)}$ of the fields of $\Mod{M}$.  This may seem surprising, but recall that this notion of truncation is chosen to facilitate the fusion of $\Mod{M}$ with $\Mod{N}$, so perhaps it should have been more surprising that $\spsub{\Mod{M}}$ did not depend upon $\Mod{N}$.  Indeed, the relative asymmetry that we have just observed between these two definitions is a consequence of the fact that we have chosen to present the master equations \eqref{eq:Master} in terms of $\coproductsymb = \parNcoproductsymb{2}{w,0}$ instead of $\parNcoproductsymb{1}{0,-w}$.

We also illustrate examples of twisted truncated subspaces for the $N=1$ algebra.  If $\Mod{M}$ belongs to the \ns{} sector, then we may choose $\eps_1 = 0$ so that the twisted and untwisted truncated subspaces coincide, $\Mod{N}^{(d)} = \Mod{N}^d$ for all $d$.  When $\Mod{M}$ is Ramond, we tabulate the $d=0, \frac{1}{2}, 1$ and $\eps_1 = \pm \frac{1}{2}$ truncations of a generic Verma module $\Ver{}$, generated by a \hwv{} $\psi_2$, in \cref{fig:Truncations}.  Comparing with \eqref{eq:VermaSS}, we see that there is a tradeoff between the sizes of the special subspace and the truncated subspaces as we vary $\eps_1$.

{
\renewcommand{\arraystretch}{1.1}
\setlength{\extrarowheight}{4pt}
\begin{figure}
\begin{center}
\begin{tabular}{C|CC}
\NSVer{}^{(d)} & \eps_1 = -\frac{1}{2} & \eps_1 = +\frac{1}{2} \\
\hline
d=0 & \vspn \set{\psi_2, G_{-1/2} \psi_2} & \vspn \set{\psi_2} \\
d=\frac{1}{2} & \vspn \set{\psi_2, G_{-1/2} \psi_2} & \vspn \set{\psi_2} \\
d=1 & \vspn \set{\psi_2, L_{-1} \psi_2, G_{-1/2} \psi_2, L_{-1} G_{-1/2} \psi_2, G_{-3/2} \psi_2, G_{-3/2} G_{-1/2} \psi_2} & \vspn \set{\psi_2, L_{-1} \psi_2, G_{-1/2} \psi_2}
\end{tabular}

\vspace{5mm}

\begin{tabular}{C|CC}
\RVer{}^{(d)} & \eps_1 = -\frac{1}{2} & \eps_1 = +\frac{1}{2} \\
\hline
d=0 & \vspn \set{\psi_2, G_0 \psi_2, G_{-1} G_0 \psi_2} & \vspn \set{\psi_2} \\
d=\frac{1}{2} & \vspn \set{\psi_2, G_0 \psi_2, L_{-1} G_0 \psi_2, G_{-1} \psi_2, G_{-1} G_0 \psi_2} & \vspn \set{\psi_2, G_0 \psi_2} \\
d=1 & \vspn \set{\psi_2, L_{-1} \psi_2, G_0 \psi_2, L_{-1} G_0 \psi_2, G_{-1} \psi_2, G_{-1} G_0 \psi_2, L_{-1} G_{-1} G_0 \psi_2, G_{-2} G_0 \psi_2} & \vspn \set{\psi_2, L_{-1} \psi_2, G_0 \psi_2}
\end{tabular}
\caption{Two tables indicating a few of the twisted truncated subspaces when $\Mod{N}$ is a generic Verma module and $\Mod{M}$ is Ramond with $\eps_1 = \pm \frac{1}{2}$.} \label{fig:Truncations}
\end{center}
\end{figure}
}

The above development, using the master equations to first express the states of $\Mod{M} \otimes \Mod{N}$ as linear combinations of elements of $\spsub{\Mod{M}} \otimes \Mod{N}$ and then as linear combinations of elements of $\spsub{\Mod{M}} \otimes \Mod{N}^{(d)}$, results in the following generalisation of \eqref{eq:OldKey}:
\begin{equation} \label{eq:NewKey}
\brac{\Mod{M} \fuse \Mod{N}}^d \subseteq \spsub{\Mod{M}} \otimes_{\CC} \Mod{N}^{(d)}.
\end{equation}
However, the validity of this inclusion hinges on a subtle point --- the second step of this process may introduce terms $\psi_1 \otimes \psi_2$ in which $\psi_1 \notin \spsub{\Mod{M}}$, in which case one has to start again from the first step.  If this repetition can be shown to always terminate, then \eqref{eq:NewKey} follows.  In the case of untwisted fusion products, one can often apply elementary arguments to conclude that termination is guaranteed \cite{GabFus94} (see \cite{RidFus10} for cases where the process encounters infinite regression).

Here, our approach to this question of termination is unashamedly practical:  we have implemented the twisted \NGK{} fusion algorithm on a computer and have observed that our implementation terminates, for all examples we have tested, if we set the twist parameters to $0$ or $-\frac{1}{2}$ for \ns{} or Ramond modules, respectively.  We will therefore defer a proper consideration of the termination question to future work.

\section{Explicit fusion products} \label{sec:Examples}

In this section, we present two explicit computations using the twisted \NGK{} fusion algorithm.  For an example illustrating the fusion of two \ns{} modules using the untwisted algorithm, see \cite[Sec.~4]{CanFusI15}.  We first describe the fusion of two Ramond modules because the product may then be identified straightforwardly using the \ns{} theory developed in \cite{CanFusI15}.  The second example fuses a \ns{} module with a Ramond module, hence the result is Ramond.  The identification in this case relies upon generalising the basic theory of staggered modules \cite{RohRed96,RidSta09,CreLog13} to the Ramond algebra.  We outline the required features of this generalisation in \cref{app:RStag}.

\subsection{Example:  fusing Ramond with Ramond} \label{sec:FusRR}

We first consider the fusion of the Kac Module $\Kac{2,1}$ with itself at central charge $c=0$ ($p=2$ and $p'=4$).  Note that $\Kac{2,1}$ is generated by a Ramond \hwv{} $u$ of conformal weight $h_{2,1} = \tfrac{9}{16}$.  We may therefore identify $\Kac{2,1}$ with the quotient of the Verma module $\Ver{2,1}$ by the submodule generated by its depth $1$ \svs{} (one bosonic and one fermionic).  Thus, we have
\begin{equation} \label{SV:K21}
\brac{L_{-1} - \frac{4}{3}G_{-1}G_{0}} u = 0, \qquad \brac{L_{-1}G_0 - \frac{3}{4}G_{-1}} u = 0
\end{equation}
in $\Kac{2,1}$.  We will assume, for definiteness, that $u$ is bosonic.

Our first task is to determine the composition factors of the fusion product $\Kac{2,1} \fuse \Kac{2,1}$ using the Verlinde formula.  Specifically, \eqref{GrFR:K21xKrsR} gives the Grothendieck fusion rule
\begin{equation}
\Gr{\Kac{2,1} \fuse \Kac{2,1}} = \Gr{\Kac{2,1}} \Grfuse \Gr{\Kac{2,1}} = \Gr{\Kac{1,1}^+} + \Gr{\Kac{1,1}^-} + \Gr{\Kac{3,1}^+} + \Gr{\Kac{3,1}^-}.
\end{equation}
However, each of the Kac modules appearing on the \rhs{} is reducible, with two (simple) composition factors each, so we learn that the fusion product has eight composition factors in all:
\begin{equation} \label{CompFact:K21xK21}
\Gr{\Kac{2,1} \fuse \Kac{2,1}} = \Gr{\NSIrr{0}^+} + 2 \, \Gr{\NSIrr{3/2}^-} + \Gr{\NSIrr{5}^+} + \Gr{\NSIrr{0}^-} + 2 \, \Gr{\NSIrr{3/2}^+} + \Gr{\NSIrr{5}^-}.
\end{equation}
Our goal is now to determine how these factors are arranged structurally in the fusion product.  Note that parity considerations force this product to decompose into the direct sum of two modules, one of which has composition factors $\NSIrr{0}^+$, $\NSIrr{3/2}^-$, $\NSIrr{3/2}^-$ and $\NSIrr{5}^+$ (the factors of the other are obtained from these by applying the parity-reversal functor $\Pi$).

It is convenient, at this point, to note that the composition factors of the Verma modules $\NSVer{1,1}^{\pm}$ (and the Fock spaces $\NSFock{1,1}^{\pm}$) include all of those that appear in \eqref{CompFact:K21xK21}.  Specifically, we have
\begin{equation}
\Gr{\NSVer{1,1}^{\pm}} = \Gr{\NSFock{1,1}^{\pm}} = \Gr{\NSIrr{0}^{\pm}} + \Gr{\NSIrr{1/2}^{\mp}} + \Gr{\NSIrr{3/2}^{\mp}} + \Gr{\NSIrr{3}^{\pm}} + \Gr{\NSIrr{5}^{\pm}} + \cdots
\end{equation}
in the Grothendieck ring.  This information will be useful for certain arguments involving descendant state counting in the calculations that follow.

To identify the structure of $\Kac{2,1} \fuse \Kac{2,1}$, given \eqref{CompFact:K21xK21}, we use the twisted \NGK{} algorithm to compute truncated subspaces of this fusion product.  We will choose the twist parameters for the $G_k$ to be $\eps_1 = \eps_2 = -\frac{1}{2}$ (we always take those for the $L_n$ to be $\eps_1 = \eps_2 = 0$), so that the twisted special subspace is given by the following quotient of $\Kac{2,1}$:
\begin{equation} \label{eq:K21SS}
\spsub{\Kac{2,1}} = \frac{\Kac{2,1}}{\left\langle L_m, \ G_n \st m \le -2, \ n \le -1 \right\rangle \: \Kac{2,1}} = \vspn \set{u, G_0 u}.
\end{equation}
The relations \eqref{SV:K21} are responsible for the finite-dimensionality of this subspace --- the twisted special subspace \eqref{eq:VermaSS} of $\spsub{\Ver{2,1}}$ is infinite-dimensional.  We moreover note that $\spsub{\Kac{2,1}}$ would have remained infinite-dimensional if we had chosen $\eps_1 = +\frac{1}{2}$ instead.  The depth $0$ twisted truncated subspace of $\Kac{2,1}$ is also given by
\begin{equation} \label{eq:K21d=0}
\Kac{2,1}^{(0)} = \vspn \set{u, G_0 u},
\end{equation}
as may be seen by combining the result for $\Ver{2,1}$, given in \cref{fig:Truncations}, with the relations \eqref{SV:K21}.

The depth $0$ truncation of the fusion product is therefore at most four-dimensional:
\begin{equation} \label{eq:BasisK21xK21}
\sqbrac{\Kac{2,1} \fuse \Kac{2,1}}^0 \subseteq \spsub{\Kac{2,1}} \otimes_{\CC} \Kac{2,1}^{(0)} = \vspn \bfset{u \otimes u, G_0 u \otimes G_0 u}{G_0 u \otimes u, u\otimes G_0 u}.
\end{equation}
Here, we separate the bosonic and fermionic vectors with a vertical bar.  The composition factors $\NSIrr{0}^+$ and $\NSIrr{0}^-$ contribute their \hwvs{} to this truncated subspace.  However, these will generate at most two of the four factors of the forms $\NSIrr{3/2}^-$ and $\NSIrr{3/2}^+$ as descendants, hence at least two of these factors must contribute their \hwvs{} to the depth $0$ truncation.  This gives at least four independent states in the depth $0$ truncation, so we conclude that this truncation is precisely four-dimensional.  The inclusion \eqref{eq:BasisK21xK21} is therefore an equality --- there are no spurious states to find.  Moreover, this tells us that the composition factors $\NSIrr{3/2}^-$, $\NSIrr{3/2}^+$, $\NSIrr{5}^+$ and $\NSIrr{5}^-$ must appear as descendants of the factors $\NSIrr{0}^+$, $\NSIrr{0}^-$, $\NSIrr{3/2}^-$ and $\NSIrr{3/2}^+$, respectively.  We summarise this conclusion in the following structure diagram:\footnote{To see that this diagram is the only one consistent with our reasoning, note that the \ns{} \hw{} $0$ corresponds to interior type modules.  A (sub)structure such as $0 \lra \frac{3}{2} \lra 5$ is therefore impossible --- it would indicate a \hwm{}, hence a quotient of the Verma module $\NSVer{0}^+$, but the braided \sv{} structure of the latter rules this out.}
\begin{equation} \label{pic:K21xK21d=0}
\parbox[c]{0.45\textwidth}{
\scalebox{0.85}{
\begin{tikzpicture}[->,node distance=1cm,>=stealth',semithick]
  \node[] (1) {$0$:};
  \node[] (1a) [below of=1] {$\frac{3}{2}$:};
  \node[] (1b) [below of=1a] {$5$:};
  \node[sv,label=left:{$+$}] (2) [right = 1cm of 1] {};
  \node[sv,label=left:{$-$}] (2a) [right = 1cm of 1a] {};
  \path[] (2) edge (2a);
  \node[sv,label=right:{$-$}] (3a) [right = 2.5cm of 1a] {};
  \node[sv,label=right:{$+$}] (3b) [right = 2.5cm of 1b] {};
  \path[] (3a) edge (3b);
  \node[right = 3.75cm of 1a] {$\bigoplus$};
  \node[sv,label=left:{$-$}] (4) [right = 5.5cm of 1] {};
  \node[sv,label=left:{$+$}] (4a) [right = 5.5cm of 1a] {};
  \path[] (4) edge (4a);
  \node[sv,label=right:{$+$}] (5a) [right = 7cm of 1a] {};
  \node[sv,label=right:{$-$}] (5b) [right = 7cm of 1b] {};
  \path[] (5a) edge (5b);
\end{tikzpicture}
\ .}
}
\end{equation}%$
As in \cref{sec:Back}, composition factors (subsingular vectors) are denoted by black circles and we indicate their parities and conformal weights in a (hopefully) obvious fashion.  However, this diagram is not yet complete --- we cannot say at this point whether the two summands of $\Kac{2,1}\fuse \Kac{2,1}$ may be further decomposed or not.  To determine this, we need to compute truncated subspaces of depth $d>0$.

Before proceeding, however, let us quickly check the above conclusion by showing explicitly that the twisted \NGK{} fusion algorithm gives the correct eigenvalues and eigenvector parities for the action of $L_0$ on the depth $0$ truncation (we note that the other \ns{} modes do not act on this truncated subspace).  For this, we will use the following two formulae along with the \sv{} equations \eqref{SV:K21}:
\begin{subequations}
\begin{align}
\coproduct{L_0} &= w L_{-1} \otimes \wun + L_0 \otimes \wun + \wun \otimes L_0, \label{M1:L0} \\
G_{-1} \otimes \wun &= -\frac{1}{2} w^{-1} G_0 \otimes \wun -\mu_1 w^{-1/2}(-w)^{-1/2} \wun \otimes G_0 + \cdots. \label{M3:G-1}
\end{align}
\end{subequations}
The first is \eqref{eq:Master1} with $\tS_n = L_0$, $\eps_1 = 0$ and $\mu_1 = 1$; the second is \eqref{eq:Master3} with $S_n = G_{1/2}$ and $\eps_1 = \eps_2 = -\frac{1}{2}$.  We have noted that $\coproduct{\tG_{-1/2-j-k}} = 0$, because we are computing to depth $0$, and the omitted terms in \eqref{M3:G-1} correspond to terms that annihilate each of the states encountered in the computations that follow.

The action of $L_0$ on the depth $0$ truncated subspace is now computed to be
\begin{equation}
\begin{aligned}
\coproduct{L_0} u \otimes u &= w L_{-1} u \otimes u + \frac{9}{8} u\otimes u = \frac{4}{3} w G_{-1}G_0 u\otimes u + \frac{9}{8} u\otimes u \\
&= \frac{3}{4} u\otimes u + \frac{4}{3} w^{1/2}(-w)^{-1/2} G_0 u\otimes G_0 u, \\
\coproduct{L_0} G_0 u \otimes G_0 u &= -\frac{27}{64} w^{1/2}(-w)^{-1/2} u \otimes u + \frac{3}{4} G_0 u \otimes G_0 u, \\
\coproduct{L_0} G_0 u \otimes u &= \frac{3}{4} G_0 u \otimes  u -\frac{3}{4} w^{1/2}(-w)^{-1/2} u \otimes G_0 u, \\
\coproduct{L_0} u \otimes G_0 u &= \frac{3}{4} w^{1/2}(-w)^{-1/2} G_{0} u \otimes u + \frac{3}{4} u \otimes G_0 u.
\end{aligned}
\end{equation}
With respect to the ordered basis \eqref{eq:BasisK21xK21}, $L_0$ is represented by the matrix
\begin{equation}
\coproduct{L_0} = 
\begin{amatrix}{cc|cc}
\frac{3}{4} & \frac{4}{3} w^{1/2}(-w)^{-1/2} & 0 & 0 \\
-\frac{27}{64} w^{1/2}(-w)^{-1/2} & \frac{3}{4} & 0 & 0 \\
\hline
0 & 0 & \frac{3}{4} & -\frac{3}{4}w^{1/2}(-w)^{-1/2} \\
0 & 0 & \frac{3}{4} w^{1/2}(-w)^{-1/2} & \frac{3}{4}
\end{amatrix}
,
\end{equation}
where the block structure confirms that $L_0$ is bosonic with two bosonic and two fermionic eigenvectors.  We note that each diagonal block has trace $\frac{3}{2}$ and determinant $0$, hence that the eigenvalues of $\coproduct{L_0}$ are $0$ and $\frac{3}{2}$, each with multiplicity two corresponding to one bosonic and one fermionic eigenvector, as concluded above.

Having checked our reasoning, we turn now to truncated subspaces of greater depth.  Specifically, we shall examine the depth $\frac{3}{2}$ truncation of $\Kac{2,1} \fuse \Kac{2,1}$.  The twisted special subspace $\spsub{\Kac{2,1}}$ is still given by \eqref{eq:K21SS}, but the twisted truncated subspace $\Kac{2,1}^{(3/2)}$ needs calculating.  That of a Ramond Verma module $\RVer{}$, generated by a \hwv{} $\psi_2$, turns out to be $13$-dimensional (compare with \cref{fig:Truncations}):
\begin{align}
\RVer{}^{(3/2)} = \vspn &\bigl\{ \psi_2, L_{-1} \psi_2, G_0 \psi_2, L_{-1} G_0 \psi_2, L_{-1}^2 G_0 \psi_2, L_{-2} G_0 \psi_2, G_{-1} \psi_2, \bigr. \notag \\
&\phantom{\big\{} \bigl. L_{-1} G_{-1} \psi_2, G_{-2} \psi_2, G_{-1} G_0 \psi_2, L_{-1} G_{-1} G_0 \psi_2, G_{-2} G_0 \psi_2, G_{-2} G_{-1} G_0 \psi_2 \bigr\}.
\end{align}
The \sv{} relations \eqref{SV:K21} and their $L_{-1}$, $G_{-1}$ and $G_{-2}$ descendants cut this down, using the definition \eqref{eq:DefTwTruncSubSp} of truncated subspaces, so that $\Kac{2,1}^{(3/2)}$ is only six-dimensional.  It follows that the depth $\frac{3}{2}$ truncated fusion product is at most $12$-dimensional.

We compare this with the dimension obtained from the composition factors \eqref{CompFact:K21xK21} and the partial structure \eqref{pic:K21xK21d=0}.  Each of the two conformal weight $0$ \hwvs{} has a single descendant, to depth $\frac{3}{2}$:  the \sv{} of conformal weight $\frac{3}{2}$ (the other possible descendants of weight $\frac{1}{2}$, $1$ and $\frac{3}{2}$ must not appear as there is no composition factor isomorphic to $\NSIrr{1/2}^{\pm}$).  Similarly, each of the weight $\frac{3}{2}$ vectors that appeared in the depth $0$ truncation must have three descendants, to depth $\frac{3}{2}$ (the \sv{} of weight $3$ cannot be present as there is no composition factor isomorphic to $\NSIrr{3}^{\pm}$).  As this already amounts to $12$ independent states, we conclude that the depth $\frac{3}{2}$ truncated fusion product is exactly $12$-dimensional, hence that there are again no spurious states to find.

Computing the action of $L_0$ on the depth $\frac{3}{2}$ truncated fusion product, we find two Jordan blocks (one bosonic and one fermionic) for the eigenvalue $\frac{3}{2}$.  The calculations in this $12$-dimensional space become somewhat tedious, so we omit the details and just report the results.  In particular, these Jordan blocks allow us to conclude immediately that the fusion product $\Kac{2,1} \fuse \Kac{2,1}$ decomposes as the direct sum of two \ns{} \emph{staggered modules} (see \cite[App.~B]{CanFusI15}).  The full structure diagram is therefore
\begin{equation} \label{pic:K21xK21d=3/2}
\parbox[c]{0.45\textwidth}{
\scalebox{0.85}{
\begin{tikzpicture}[->,node distance=1cm,>=stealth',semithick]
  \node[] (1) {$0$:};
  \node[] (1a) [below of=1] {$\frac{3}{2}$:};
  \node[] (1b) [below of=1a] {$5$:};
  \node[sv,label=left:{$+$}] (2) [right = 1cm of 1] {};
  \node[sv,label=left:{$-$}] (2a) [right = 1cm of 1a] {};
  \path[] (2) edge (2a);
  \node[sv,label=right:{$-$}] (3a) [right = 2.5cm of 1a] {};
  \node[sv,label=right:{$+$}] (3b) [right = 2.5cm of 1b] {};
  \path[] (3a) edge (3b);
  \path[] (3a) edge (2);
  \path[] (3a) edge (2a);
  \path[] (3b) edge (2a);
  \node[right = 3.75cm of 1a] {$\bigoplus$};
  \node[sv,label=left:{$-$}] (4) [right = 5.5cm of 1] {};
  \node[sv,label=left:{$+$}] (4a) [right = 5.5cm of 1a] {};
  \path[] (4) edge (4a);
  \node[sv,label=right:{$+$}] (5a) [right = 7cm of 1a] {};
  \node[sv,label=right:{$-$}] (5b) [right = 7cm of 1b] {};
  \path[] (5a) edge (5b);
  \path[] (5a) edge (4);
  \path[] (5a) edge (4a);
  \path[] (5b) edge (4a);
\end{tikzpicture}
\ ,}
}
\end{equation}%$
where the horizontal arrows indicate the Jordan blocks in $\coproduct{L_0}$.

To be more precise, each staggered module appearing in this fusion product may be characterised through the non-split short exact sequence
\begin{equation} \label{es:K11K31}
\dses{\Kac{1,1}^{\pm}}{}{\Stag{2,1}{1,0}(\beta)^{\pm}}{}{\Kac{3,1}^{\mp}},
\end{equation}
where we follow the notation for \ns{} staggered modules outlined in \cite[App.~B]{CanFusI15}.  In particular, the parity label matches that of the states of minimal conformal weight and $\beta \in \CC$ is the \emph{logarithmic coupling} \cite{RidPer07}, computed as follows:  First, let $x^{\pm}$ denote the \hwv{} of conformal weight $0$, so that the \sv{} $U x^{\pm}$, where $U = L_{-1} G_{-1/2} - \frac{1}{2} G_{-3/2}$, is the $L_0$-eigenvector in the Jordan block of eigenvalue $\frac{3}{2}$.  Let $y^{\mp}$ be a Jordan partner to $U x^{\pm}$, so $(L_0 - \frac{3}{2}) y^{\mp} = U x^{\pm}$ and determine $\beta$ from $U^{\dag} y^{\mp} = \beta x^{\pm}$.  Performing this calculation explicitly in the depth $\frac{3}{2}$ truncation of $\Kac{2,1} \fuse \Kac{2,1}$, we obtain
\begin{equation}
\coproduct{U^{\dag}} y^{\mp} = \brac{\coproduct{G_{1/2}} \coproduct{L_1} - \frac{1}{2} \coproduct{G_{3/2}}} y^{\mp} = \frac{3}{8} x^{\pm}.
\end{equation}
The fusion product is therefore identified as\footnote{Actually, the methods of \cite{RidLog07} may be used to show that there is a unique staggered module, up to isomorphism, satisfying \eqref{es:K11K31}.  Strictly speaking, the value of the logarithmic coupling is therefore not needed to completely identify the fusion product.}
\begin{equation} \label{FR:K21xK21}
\Kac{2,1} \fuse \Kac{2,1} = \Stag{2,1}{1,0}(\tfrac{3}{8})^+ \oplus \Stag{2,1}{1,0}(\tfrac{3}{8})^-.
\end{equation}
This logarithmic coupling is confirmed by the heuristic limit formula \cite[Eq.~(B.5)]{CanFusI15}, originally obtained for Virasoro logarithmic minimal models in \cite{VasInd11,GaiLat13}.

Whilst this is not needed for the identification \eqref{FR:K21xK21}, let us remark that our calculations have justified every arrow in the structure diagram \eqref{pic:K21xK21d=3/2} except those pointing from the \ssvs{} of conformal weight $5$ to the \svs{} of weight $\frac{3}{2}$.  To verify these arrows explicitly with the \NGK{} fusion algorithm would require computing to depth $5$ which is infeasible with our current implementation.  However, if such an arrow exists, meaning that the staggered module has no \ssv{} of conformal weight $5$ that is actually singular, then it clearly points to either the \hwv{} of weight $0$ or the \sv{} of weight $\frac{3}{2}$.  The former is ruled out by the \ns{} generalisation of the Projection Lemma \cite[Lem.~5.1]{RidSta09}, so we only need check that $\Stag{2,1}{1,0}(\frac{3}{8})^{\pm}$ possesses no \sv{} of weight $5$.  This was explicitly verified by coding the general form of the weight $5$ \ssvs{} in $\Stag{2,1}{1,0}(\frac{3}{8})^{\pm}$ using a computer implementation of \ns{} staggered modules.

\subsection{Example:  fusing Ramond with \ns{}} \label{sec:FusNSR}

Our second example addresses the fusion of the Ramond module $\Kac{2,1}$ with the \ns{} module $\Kac{2,2}^+$, again at central charge $c=0$ ($p=2$ and $p'=4$). The Grothendieck fusion rule \eqref{GrFR:K21xKrsNS} gives
\begin{equation}
\Gr{\Kac{2,1} \fuse \Kac{2,2}^+} = \Gr{\Kac{2,1}} \Grfuse \Gr{\Kac{2,2}^+} = \Gr{\Kac{1,2}} + \Gr{\Kac{3,2}},
\end{equation}
hence the fusion product has five composition factors in all:
\begin{equation}
\Gr{\Kac{2,1} \fuse \Kac{2,2}^+} = \Gr{\RIrr{0}^+} + \Gr{\RIrr{0}^-} + 2 \, \Gr{\RIrr{1}} + \Gr{\RIrr{4}}.
\end{equation}
Here, we recall that $h_{1,2} = 0 = c/24$ is the unique conformal weight for which a simple Ramond \hwm{} is not invariant under parity-reversal ($G_0$ acts as $0$ on the \hwv{}).  For convenience, we compare this with the composition factors of $\RVer{1,2}^{\pm}$ and $\RFock{1,2}$:
\begin{equation}
\begin{aligned}
\Gr{\RVer{1,2}^{\pm}} &= \Gr{\RIrr{0}^{\pm}} + \Gr{\RIrr{1}} + \Gr{\RIrr{4}} + \cdots, \\
\Gr{\RFock{1,2}} &= \Gr{\RIrr{0}^+} + \Gr{\RIrr{0}^-} + 2 \, \Gr{\RIrr{1}} + 2 \, \Gr{\RIrr{4}} + \cdots.
\end{aligned}
\end{equation}

To determine the structure of the fusion product $\Kac{2,1} \fuse \Kac{2,2}^+$, we again turn to the twisted \NGK{} algorithm, initially for depth $0$, choosing $\eps_1 = -\frac{1}{2}$ and $\eps_2 = 0$ for the $G_k$.  Letting $u$ denote the bosonic \hwv{} of $\Kac{2,1}$ of conformal weight $h_{2,1} = \frac{9}{16}$, as in \cref{sec:FusRR}, we have the same \sv{} relations \eqref{SV:K21} as before.  Let $v$ denote the (bosonic) \hwv{} of $\Kac{2,2}^+$ of conformal weight $h_{2,2} = \frac{3}{16}$.  Then, it is easy to check the following \sv{} relation:
\begin{equation} \label{SV:K22}
\brac{L_{-1}^2 - \frac{1}{4} L_{-2} - G_{-3/2} G_{-1/2}} v = 0.
\end{equation}
The twisted special subspace of $\Kac{2,1}$ was given in \eqref{eq:K21SS} and the depth $0$ twisted truncated subspace of $\Kac{2,2}^+$ is given in \cref{fig:Truncations}:
\begin{equation}
{\Kac{2,2}^+}^{(0)} = \vspn \set{v, G_{-1/2} v}.
\end{equation}
This is identical to the truncated subspace of the Verma module in view of the \sv{} relation \eqref{SV:K22}.

It follows that the depth $0$ truncation of $\Kac{2,1} \fuse \Kac{2,2}^+$ is at most four-dimensional:
\begin{equation} \label{eq:BasisK12xK22}
\sqbrac{\Kac{2,1} \fuse \Kac{2,2}}^0 \subseteq \vspn \bfset{u \otimes v, G_0 u \otimes G_{-1/2} v}{u\otimes  G_{-1/2}v, G_0 u \otimes v}.
\end{equation}
The composition factors $\RIrr{0}^{\pm}$ will contribute two \hwvs{} to this truncation and each of these will generate a centre-type \hwm{}.  It is thus possible that both the composition factors of type $\RIrr{1}$, and also that of type $\RIrr{4}$, may be descended from these \hwvs{} and hence be set to zero in the depth zero truncation.  In other words, there may exist up to two spurious states in \eqref{eq:BasisK12xK22}.

Spurious states may be determined from non-trivial relations, in particular from the \sv{} relations \eqref{SV:K21} and \eqref{SV:K22}.  The former were used to determine the twisted special subspace of $\Kac{2,1}$, so we must use the latter in our search.  Taking $\tS_n = L_{-1}$, $\eps_1 = 0$ and $\mu_1 = 1$ in \eqref{eq:Master1} gives
\begin{equation} \label{eq:CoprodL-1}
0 = \coproduct{L_{-1}} = L_{-1} \otimes \wun + \wun \otimes L_{-1} \qquad \Ra \qquad 
0 = \coproduct{L_{-1}^2} = L_{-1}^2 \otimes \wun + 2 L_{-1} \otimes L_{-1} + \wun \otimes L_{-1}^2,
\end{equation}
as we are computing to depth $0$.  Thus,
\begin{align} \label{eq:ToBeSimplified}
0 &= \coproduct{L_{-1}^2} u \otimes v - \coproduct{L_{-1}} L_{-1} u \otimes v = L_{-1} u \otimes L_{-1} v + u \otimes L_{-1}^2 v \notag \\
&= \frac{4}{3} G_{-1} G_0 u \otimes L_{-1} v + \frac{1}{4} u \otimes L_{-2} v + u \otimes G_{-3/2} G_{-1/2} v,
\end{align}
where we have used \eqref{SV:K21} and \eqref{SV:K22}.  To simplify the first term of \eqref{eq:ToBeSimplified}, take \eqref{eq:Master3} with $S_n = G_{-1}$, $\eps_1 = -\frac{1}{2}$ and $\eps_2 = 0$:
\begin{equation} \label{eq:G-1}
G_{-1} \otimes \wun = \mu_1 \sqbrac{-(-w)^{-1/2} \wun \otimes G_{-1/2} + \frac{1}{2} (-w)^{-3/2} \wun \otimes G_{1/2} + \cdots}.
\end{equation}
Applying \eqref{eq:G-1}, \eqref{eq:CoprodL-1} and \eqref{SV:K21} in succession, twice, then \eqref{eq:G-1} once again, as well as the commutation relations \eqref{eq:CommN=1}, we deduce that
\begin{equation}
\frac{4}{3} G_{-1} G_0 u \otimes L_{-1} v = -\frac{3}{16} w^{-2} u \otimes v - 2 (-w)^{-3/2} G_0 u \otimes G_{-1/2} v.
\end{equation}
For the second term of \eqref{eq:ToBeSimplified}, note that applying \eqref{eq:Master2} with $\tS_n = L_{-2}$, $\eps_1 = 0$ and $\mu_1 = 1$ to $u \otimes v$ gives
\begin{equation}
\frac{1}{4} u \otimes L_{-2} v = -\frac{1}{4} w^{-1} L_{-1} u \otimes v + \frac{9}{64} w^{-2} u \otimes v = \frac{9}{64} w^{-2} u \otimes v + \frac{1}{3} (-w)^{-3/2} G_0 u \otimes G_{-1/2} v,
\end{equation}
using \eqref{SV:K21} and \eqref{eq:G-1}.  Finally, setting $\tS_n = \tG_{-1}$, $\eps_1 = -\frac{1}{2}$ and $\eps_2 = 0$ in \eqref{eq:Master2} yields
\begin{equation}
0 = \coproduct{\tG_{-1}} = w^{-1} G_0 \otimes \wun + \mu_1 \sqbrac{(-w)^{1/2} \wun \otimes G_{-3/2} + \frac{1}{2} (-w)^{-1/2} \wun \otimes G_{-1/2} - \frac{1}{8} (-w)^{-3/2} \wun \otimes G_{1/2} + \cdots},
\end{equation}
which simplifies the third term of \eqref{eq:ToBeSimplified} (again using \eqref{SV:K21} and \eqref{eq:G-1}):
\begin{equation}
u \otimes G_{-3/2} G_{-1/2} v = \frac{3}{64} w^{-2} u \otimes v + \frac{5}{3} (-w)^{-3/2} G_0 u \otimes G_{-1/2} v.
\end{equation}
With these simplifications, the \rhs{} of \eqref{eq:ToBeSimplified} is easily checked to vanish identically.  This means that we have not obtained a spurious state.  Similar calculations, starting from applying $\coproduct{L_{-1}^2} = 0$ to the other vectors in \eqref{eq:BasisK12xK22} and then using \eqref{SV:K22} and its descendants, also fail to find spurious states.  This strongly suggests that there are no spurious states to find and that the inclusion \eqref{eq:BasisK12xK22} is actually an equality.

Granted this, we can now determine the action of $L_0$ and $G_0$ on the depth $0$ truncation of $\Kac{2,1} \fuse \Kac{2,2}^+$ (the other $N=1$ modes do not act).  These calculations require \eqref{SV:K21}, \eqref{M1:L0}, \eqref{eq:G-1} and
\begin{equation}
\coproduct{G_0} = \coproduct{\tG_0} = G_0 \otimes \wun + \cdots + \mu_1 \sqbrac{(-w)^{1/2} \wun \otimes G_{-1/2} + \frac{1}{2} (-w)^{-1/2} \wun \otimes G_{1/2} + \cdots},
\end{equation}
the first equality being \eqref{eq:STilde} for depth $0$ and the second being \eqref{eq:Master1} with $\tS_n = \tG_0$, $\eps_1 = -\frac{1}{2}$ and $\eps_2 = 0$.  The results, with respect to the ordered basis \eqref{eq:BasisK12xK22}, are
\begin{subequations}
\begin{align}
\coproduct{L_0} &= 
\begin{amatrix}{cc|cc}
\frac{3}{4} & -\frac{9}{64}(-w)^{-1/2} & 0 & 0 \\
-\frac{4}{3}(-w)^{1/2} & \frac{1}{4} & 0 & 0 \\
\hline
0 & 0 & \frac{1}{4}	& \frac{3}{4} (-w)^{1/2} \\
0 & 0 & \frac{1}{4}(-w)^{-1/2} & \frac{3}{4}
\end{amatrix}
, \\
\coproduct{G_0} &= 
\begin{amatrix}{cc|cc}
0 & 0 & \frac{3}{16} (-w)^{-1/2} & \frac{9}{16}\\
0 & 0 & -\frac{1}{3} & -(-w)^{1/2} \\
\hline
(-w)^{1/2} & -\frac{3}{16} & 0 & 0 \\
1 & -\frac{3}{16}(-w)^{-1/2} & 0 & 0
\end{amatrix}
.
\end{align}
\end{subequations}
The eigenvalues of $\coproduct{L_0}$ are easily found to be $0$ and $1$, each occurring with multiplicity $2$ and an eigenvector of each parity.  Changing to an ordered basis of (appropriately normalised) definite parity $L_0$-eigenvectors, these matrices become
\begin{equation}
\coproduct{L_0} = 
\begin{amatrix}{cc|cc}
0 & 0 & 0 & 0 \\
0 & 1 & 0 & 0 \\
\hline
0 & 0 & 0 & 0 \\
0 & 0 & 0 & 1
\end{amatrix}
, \qquad \coproduct{G_0} = 
\begin{amatrix}{cc|cc}
0 & 0 & 0 & 0 \\
0 & 0 & 0 & 1 \\
\hline
0 & 0 & 0 & 0 \\
0 & 1 & 0 & 0
\end{amatrix}
.
\end{equation}
$\coproduct{G_0}$ therefore annihilates both the conformal weight $0$ vectors while swapping those of conformal weight $1$.

This analysis shows that one of the $\RIrr{1}$ factors is not composed of descendant states; the other copy of $\RIrr{1}$ is descended from the $\RIrr{0}^+$ or $\RIrr{0}^-$ factor, or from both.  Similarly, the $\RIrr{4}$ factor is descended from one of the $\RIrr{1}$ factors, but we cannot as yet say which one.  To determine the full structure of the fusion product, we will again have to delve deeper with the \NGK{} fusion algorithm.

Continuing the analysis to depth $1$, the twisted special subspace $\spsub{\Kac{2,1}}$ does not change, but the depth $1$ twisted truncated subspace ${\Kac{2,2}^+}^{(1)}$ differs from the depth $1$ Verma subspace given in \cref{fig:Truncations} because of the \sv{} relation \eqref{SV:K22}:
\begin{equation}
{\Kac{2,2}^+}^{(1)} = \vspn \set{v, L_{-1} v, G_{-1/2} v, L_{-1} G_{-1/2} v, G_{-3/2} v}.
\end{equation}
The depth $1$ truncation of $\Kac{2,1} \fuse \Kac{2,2}^+$ is therefore at most $10$-dimensional.  However, the $\RIrr{1}$ that is not a descendant must contribute six states to the depth $1$ truncation, two of conformal weight $1$ and four of conformal weight $2$.  Similarly, the $\RIrr{0}^{\pm}$ must contribute two states of conformal weight $0$ and two states of conformal weight $1$ belonging to the descendant $\RIrr{1}$ factor.  As this is ten states in all, we see that the dimension of the depth $1$ truncation of $\Kac{2,1} \fuse \Kac{2,2}^+$ is precisely ten --- there are no spurious states to find.\footnote{The fact that our calculations in this section have not required spurious states is a reflection of the relatively simple examples that we have chosen to detail here.  The more general computations of \cref{sec:Results} frequently required identifying many spurious states.}

We first compute $\coproduct{L_0}$, using \eqref{eq:Master}, on the depth $1$ truncation of $\Kac{2,1} \fuse \Kac{2,2}^+$.  Its (generalised) eigenvalues are $0$, $1$ and $2$, appearing with multiplicities $2$, $4$ and $4$, respectively, and the dimensions of the bosonic and fermionic subspaces are equal in each eigenspace.  There are two rank $2$ Jordan blocks, one bosonic and one fermionic, of eigenvalue $1$, hence the fusion product is a Ramond staggered module in the sense of \cref{app:RStag}.  Let $x^+$ ($x^-$) denote the bosonic (fermionic) eigenvector of conformal weight $0$.  Then, explicitly computing $\coproduct{L_{-1}} x^{\pm}$ and $\coproduct{G_{-1}} x^{\pm}$, again using \eqref{eq:Master}, shows that each result is non-zero, hence that the $\RIrr{1}$ factor must be descended from both $\RIrr{0}^+$ and $\RIrr{0}^-$.  It follows that we may normalise $x^+$ and $x^-$ so that
\begin{equation}
\coproduct{L_{-1}} x^+ = \frac{1}{2} \coproduct{G_{-1}} x^-, \qquad \coproduct{L_{-1}} x^- = \frac{1}{2} \coproduct{G_{-1}} x^+,
\end{equation}
recalling that $\coproduct{G_0} x^+ = \coproduct{G_0} x^- = 0$.

Because the fusion product is a staggered module, it follows from the theory outlined in \cref{app:RStag} that the composition factor $\RIrr{4}$ cannot be descended from the factors $\RIrr{0}^{\pm}$.  We may therefore draw the structure diagram of $\Kac{2,1} \fuse \Kac{2,2}^+$ as follows:
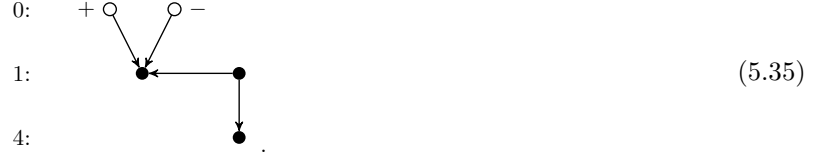
\begin{equation} \label{pic:K21XK22}
\parbox[c]{0.25\textwidth}{
\scalebox{0.85}{
\begin{tikzpicture}[->,node distance=1cm,>=stealth',semithick]
  \node[] (1) {$0$:};
  \node[] (1a) [below of=1] {$1$:};
  \node[] (1b) [below of=1a] {$4$:};
  \node[osv,label=left:{$+$}] (2) [right = 1cm of 1] {};
   \node[osv,label=right:{$-$}] (2a) [right = 2cm of 1] {};
  \node[sv] (2b) [right = 1.5cm of 1a] {};
  \path[] (2) edge (2b);
    \path[] (2a) edge (2b);
  \node[sv] (3) [right = 3cm of 1a] {};
  \node[sv] (3a) [right = 3cm of 1b] {};
  \path[] (3) edge (2b);
  \path[] (3) edge (3a);
\end{tikzpicture}
\ .}}
\end{equation}%$
Here, we indicate the composition factors (\ssvs{}) corresponding to conformal weight $\frac{c}{24} = 0$ with white circles, as in \cref{sec:Back}.  Each black circle corresponds to two composition factors (\ssvs{}), one of each parity.  In particular, the rightmost circle of weight $1$ accounts for the two weight $1$ states found in the depth $0$ analysis.  This identifies the fusion product as a staggered module with exact sequence
\begin{equation}
\dses{\Kac{1,2}}{}{\Kac{2,1} \fuse \Kac{2,2}^+}{}{\Kac{3,2}}.
\end{equation}
It only remains to determine the logarithmic couplings.

Choose $y^+$ and $y^-$ to be Jordan partners of $\coproduct{L_{-1}} x^+$ and $\coproduct{L_{-1}} x^-$, respectively.  As the latter are \svs{}, the logarithmic couplings $\beta^{\pm}$ defined by
\begin{equation}
\coproduct{L_1} y^{\pm} = \beta^{\pm} x^{\pm}
\end{equation}
are independent of the choices of $y^{\pm}$.  As is discussed in \cref{app:RStag}, it appears that these logarithmic couplings could be independent, hence we must measure them both.  However, explicit calculation confirms that they coincide (as one might have expected):
\begin{equation}
\beta^+ = \beta^- = \frac{3}{16}.
\end{equation}
The fusion rule is therefore
\begin{equation}
\Kac{2,1} \fuse \Kac{2,2}^+ = \Stag{2,2}{1,0}(\tfrac{3}{16},\tfrac{3}{16}).
\end{equation}

\section{Further results} \label{sec:Results}

Here we posit two conjectures for the fusion rules involving \ram{} modules and discuss the evidence that we have obtained for each conjecture, using a computer implementation of the twisted \NGK{} fusion algorithm. Similar conjectures were first presented in \cite{CanFusI15} for the \ns{} sector. We use the \NGK{} algorithm in conjunction with the fermionic Verlinde formula for each fusion product, alongside the theory of staggered modules. This significantly reduces the depths required to completely identify the fusion product, though we have found that the complexity of the twisted algorithm limits the Ramond calculations to depths at most $2$. As in \cite{CanFusI15}, we consider fusion rules between Kac modules of central charges $c=\tfrac{3}{2}$, $-\tfrac{5}{2}$, $-\tfrac{81}{10}$, $0$, $-\tfrac{21}{4}$ and $\tfrac{7}{10}$, corresponding to $(p,p')=(1,1)$, $(1,3)$, $(1,5)$, $(2,4)$, $(2,8)$ and $(3,5)$, respectively.

\subsection{Fusing $\Kac{r,1}$ with $\Kac{1,s}$}

The notion of lattice fusion \cite{PeaLog06,PeaLog14,MorKac15}, whereby fusion rules may be predicted from calculations in a statistical lattice model, requires that the fusion product of a ``first row'' module with a ``first column'' module be given by
\begin{equation} \label{FR:Expected}
\Kac{r,1}\fuse \Kac{1,s} = \Kac{r,s}.
\end{equation}
This was verified in \cite{CanFusI15}, in many examples where both modules were \ns{}, with the result (including parity) being
\begin{subequations}
\begin{equation}
\Kac{r,1}^+\fuse \Kac{1,s}^+ = \Kac{r,s}^+ \qquad \text{(\(r\), \(s\) odd),}
\end{equation}
in agreement with \eqref{GrFR:NSxNS}.  We have also verified this, using the twisted fusion algorithm, for \ns{} by Ramond fusions:
\begin{equation}
\Kac{r,1}^+\fuse \Kac{1,s} = \Kac{r,s} \qquad \text{(\(r\) odd, \(s\) even).}
\end{equation}
However, the result for Ramond by Ramond fusion rules does not quite accord with \eqref{FR:Expected}:
\begin{equation} \label{FR:Kr1xK1sRR}
\Kac{r,1}\fuse \Kac{1,s} = \Kac{r,s}^+ \oplus \Kac{r,s}^- \qquad \text{(\(r\), \(s\) even).}
\end{equation}
\end{subequations}
We expect that this slight disagreement with expectations can be accommodated within the formalism of lattice fusion products.  Unfortunately, lattice fusion products involving Ramond Kac modules do not seem to have been considered in \cite{PeaLog14}.

To illustrate the above conjectured fusion rules, we consider the Ramond by Ramond example $\Kac{2,1}\fuse \Kac{1,4}$, for $(p,p')=(1,3)$ ($c=-5/2$). The Grothendieck fusion rule \eqref{GrFR:RxR} says that the result has the same composition factors as $\Kac{2,4}^+$ and $\Kac{2,4}^-$, namely $\Irr{0}^{\pm}$, $\Irr{1/2}^{\pm}$ and $\Irr{5/2}^{\pm}$ with each parity appearing once.  A depth $0$ calculation reveals both copies of $\Irr{0}$ and both copies of $\Irr{1/2}$, hence that the structure is one of the following possibilities:
\begin{equation}
\parbox[c]{0.6\textwidth}{
\begin{tikzpicture}[->,scale = 0.8, transform shape,>=stealth',semithick]
  \node (1) [] {$0$:};
  \node    (1a) [below = 0.5cm of 1] {$\tfrac{1}{2}$:};
  \node    (1b) [below = 0.5cm of 1a] {$\tfrac{5}{2}$:};
  \node[sv,label=left:{$+$}] (2) [right = 1.5cm of 1] {};
  \node[sv,label=left:{$-$}] (2a) [right = 1.5cm of 1a] {};
  \node[sv,label=left:{$-$}] (2b) [right = 1.5cm of 1b] {};
  \path[] (2a) edge (2);
  \path[] (2a) edge (2b);
  \node (p) [right = 0.75cm of 2a.center] {$\oplus$};
  \node[sv,label=right:{$-$}] (3) [right = 2cm of 2.center] {};
  \node[sv,label=right:{$+$}] (3a) [right = 2cm of 2a.center] {};
  \node[sv,label=right:{$+$}] (3b) [right = 2cm of 2b.center] {};
  \path[] (3a) edge (3);
  \path[] (3a) edge (3b);
  \node (or) [right = 1.25cm of 3a.center] {\scalebox{1.25}{or}};
  \node[sv,label=left:{$+$}] (4) [right = 3cm of 3.center] {};
  \node (o) [below = 0.25cm of 4.center] {$\oplus$};
  \node[sv,label=left:{$-$}] (4a) [right = 3cm of 3a.center] {};
  \node[sv,label=left:{$-$}] (4b) [right = 3cm of 3b.center] {};
  \path[] (4a) edge (4b);
  \node (p') [right = 0.75cm of 4a.center] {$\oplus$};
  \node[sv,label=right:{$-$}] (5) [right = 2cm of 4.center] {};
  \node (o') [below = 0.25cm of 5.center] {$\oplus$};
  \node[sv,label=right:{$+$}] (5a) [right = 2cm of 4a.center] {};
  \node[sv,label=right:{$+$}] (5b) [right = 2cm of 4b.center] {};
  \path[] (5a) edge (5b);
\end{tikzpicture}
\ .}
\end{equation}%$
Here, we have omitted two additional possibilities by supposing that fusion is preserved by the parity reversal functor $\Pi$ (as $\Kac{2,1}$ and $\Kac{1,4}$ are $\Pi$-invariant, so is their fusion product).  In any case, a depth $1/2$ calculation now confirms that the left structure is correct, hence that
\begin{equation}
\Kac{2,1} \fuse \Kac{1,4} = \Kac{2,4}^+ \oplus \Kac{2,4}^-,
\end{equation}
as predicted by \eqref{FR:Kr1xK1sRR}.

\subsection{Fusing near the edge} \label{sec:Edge}

The fusion rules for general Kac modules are not accessible with the technology developed here.  Instead, we content ourselves with conjecturing fusion rules that involve Kac modules whose labels $r$ and $s$ are not both large.  To make this more precise, consider the Grothendieck fusion rules
\begin{equation} \label{GrFR:Kr1orK1s}
\Gr{\Kac{r,1}} \Grfuse \Gr{\Kac{r',s'}} = \sideset{}{'} \sum_{r'' = \abs{r-r'}+1}^{r+r'-1} \Gr{\Kac{r'',s'}}, \qquad
\Gr{\Kac{1,s}} \Grfuse \Gr{\Kac{r',s'}} = \sideset{}{'} \sum_{s'' = \abs{s-s'}+1}^{s+s'-1} \Gr{\Kac{r',s''}},
\end{equation}
which follow from \eqref{GrFR:KxK}, except that we omit parities and the additional parity-reversed terms in the Ramond by Ramond case, for brevity.  The primed sums indicate, as usual, that the index increases in steps of two.  If the labels $r''$ and $s'$ ($r'$ and $s''$) appearing on the \rhs{}s of these rules are not both large, meaning that they satisfy either $r'' \le p$ or $s' \le p'$ ($r' \le p$ or $s'' \le p'$), then we conjecture that the following prescription identifies the fusion product (up to any logarithmic couplings):
\begin{enumerate}[leftmargin=*,label=\arabic*)]
\item Write down a list of all Kac modules $\Kac{r'',s'}$ ($\Kac{r',s''}$) appearing in the decomposition \eqref{GrFR:Kr1orK1s} of the corresponding Grothendieck fusion product, including parities and any omitted terms, in order of increasing $r''$ ($s''$). \label{it:KacList}
\item Starting from the \emph{smallest} value of $r''$ ($s''$), check whether there exists a $\Kac{\rho'',s'}$ ($\Kac{r',\sigma''}$) in the list which is the reflection of $\Kac{r'',s'}$ ($\Kac{r',s''}$) about the next boundary.  This means that $\rho''$ ($\sigma''$) must satisfy $0 < \rho'' - r'' < 2p$ ($0 < \sigma'' -s'' < 2p'$) and $\Kac{\frac{1}{2} (r'' + \rho''),s'}$ ($\Kac{r',\frac{1}{2} (s'' + \sigma'')}$) must be of boundary or corner type.
\item If there does, and if the relative parities of $\Kac{r'',s'}$ and $\Kac{\rho'',s'}$ ($\Kac{r',s''}$ and $\Kac{r',\sigma''}$) allow one to combine them into a single indecomposable, then replace $\Kac{r'',s'}$ and $\Kac{\rho'',s'}$ ($\Kac{r',s''}$ and $\Kac{r',\sigma''}$) in the list by the staggered module $\Stag{\frac{1}{2} (\rho'' + r''), s'}{\frac{1}{2} (\rho'' - r''),0}$ ($\Stag{r',\frac{1}{2} (\sigma'' + s'')}{0,\frac{1}{2} (\sigma'' - s'')}$).  The parity of the staggered module is defined to be that of its ground states (states of minimal conformal weight).
\item Repeat with $\Kac{r'',s'}$ ($\Kac{r',s''}$), where $r''$ ($s''$) is the next-highest value.  Once all values are exhausted, the list consists of the direct summands of the fusion product.
\end{enumerate}
This prescription generalises those proposed in \cite{EbeVir06,RasFus07b,RidPer07} for the Virasoro logarithmic minimal models and is a straightforward extension of the \ns{} prescription of \cite{CanFusI15}.  Note that any logarithmic couplings are not determined and must therefore be computed independently.

We illustrate this prescription with two examples.  First, we consider the \ns{} by Ramond fusion product $\Kac{3,1}^+ \fuse \Kac{3,4}$ for $(p,p') = (4,6)$ ($c=1$).  The Grothendieck fusion rule is
\begin{equation}
\Gr{\Kac{3,1}^+} \Grfuse \Gr{\Kac{3,4}} = \Gr{\Kac{1,4}} + \Gr{\Kac{3,4}} + \Gr{\Kac{5,4}},
\end{equation}
so the above prescription requires us to start with $\Kac{1,4}$.  Its reflection about the boundary at $r=4$ is $\Kac{7,4}$, see \cref{fig:KacTables}, which is not in our list.  We therefore move on to $\Kac{3,4}$ whose reflection is $\Kac{5,4}$.  Because this reflection is in the list, we replace $\Kac{3,4}$ and $\Kac{5,4}$ by a Ramond staggered module $\Stag{4,4}{1,0}$.  Since $h_{5,4} = \frac{17}{16} > \frac{1}{16} = h_{3,4} \neq \frac{c}{24}$, there is a single logarithmic coupling $\beta$ to determine (\cref{app:RStag}).  We could try to compute this by performing a depth $1$ twisted \NGK{} calculation, but it is more efficient to use the (heuristic) formula \cite[Eq.~(B.5)]{CanFusI15}, originally derived for Virasoro logarithmic minimal models in \cite{VasInd11,GaiLat13}.  In this way, we arrive at the predicted fusion rule
\begin{equation}
\Kac{3,1}^+ \fuse \Kac{3,4} = \Kac{1,4} \oplus \Stag{4,4}{1,0}(\tfrac{5}{9}).
\end{equation}
Unfortunately, this \NGK{} calculation turned out to be infeasible with our current implementation.

Our second example is the Ramond by Ramond fusion product $\Kac{1,4} \fuse \Kac{1,4}$ for $(p,p') = (1,3)$ ($c=-\frac{5}{2}$).  This time the Grothendieck fusion rule gives modules of both parities:
\begin{equation}
\Gr{\Kac{1,4}} \Grfuse \Gr{\Kac{1,4}} = \Gr{\Kac{1,1}^+} + \Gr{\Kac{1,1}^-} + \Gr{\Kac{1,3}^+} + \Gr{\Kac{1,3}^-} + \Gr{\Kac{1,5}^+} + \Gr{\Kac{1,5}^-} + \Gr{\Kac{1,7}^+} + \Gr{\Kac{1,7}^-}.
\end{equation}
Since $\Kac{1,1}$ reflects onto $\Kac{1,5}$, we predict that the fusion rule is actually
\begin{equation}
\Kac{1,4} \fuse \Kac{1,4} = {\Stag{1,3}{0,2}}^+ \oplus {\Stag{1,3}{0,2}}^- \oplus \Kac{1,3}^+ \oplus \Kac{1,3}^- \oplus \Kac{1,7}^+ \oplus \Kac{1,7}^-,
\end{equation}
noting that there is no logarithmic coupling to compute because $h_{1,1} = h_{1,5}$.  We remark that in this case, $\Kac{1,5}$ reflects onto $\Kac{1,7}$, so one might have expected staggered modules of the form $\Stag{1,6}{0,1}(\beta)^{\pm}$ (or more complicated indecomposables involving three Kac modules).  However, the above prescription requires us to test for reflections in order of smallest to largest label.  The staggered modules ${\Stag{1,3}{0,2}}^{\pm}$ are confirmed by a depth $\frac{1}{2}$ \NGK{} calculation.

We mention the following consequence of this conjectured prescription for fusion rules when $(p,p')=(1,1)$ ($c=\frac{3}{2}$).  Then, the Fock spaces $\Fock{r,s}$ are all of corner type, hence they are all semisimple.  The same is therefore true for the Kac modules $\Kac{r,s}$ as well.  Indeed, every Kac module is a direct sum of the simple modules $\Kac{r,1}$ (or, equivalently, $\Kac{1,s}$).  The above prescription applies to these modules and, because all modules are corner type, there are no reflections to consider, hence no staggered modules appear in the Kac fusion products.  The fusion product of two $c=\frac{3}{2}$ Kac modules is therefore semisimple, hence the result may be obtained by lifting the Grothendieck fusion product \eqref{GrFR:KxK}:
\begin{equation}
\Kac{r,s} \fuse \Kac{r',s'} = \sideset{}{'} \bigoplus_{r''=\abs{r-r'}+1}^{r+r'-1} \sideset{}{'} \bigoplus_{s''=\abs{s-s'}+1}^{s+s'-1} \Kac{r'',s''} \qquad \text{(\(c = \tfrac{3}{2}\)).}
\end{equation}
Nevertheless, staggered modules do exist at $c=\frac{3}{2}$.  This follows from the $N=1$ analogue of \cite[Prop.~7.5]{RidSta09}.

We conclude by listing a selection of the \ns{} by Ramond and Ramond by Ramond fusion rules that we have been able to obtain using our implementation of the twisted \NGK{} fusion algorithm (see \cite[App.~C]{CanFusI15} for our \ns{} by \ns{} results).  Each fusion rule is consistent with the prescription conjectured above and each is selected because it yields at least one staggered module.  This therefore serves to record the logarithmic couplings of these staggered modules.  In each case, the value of the logarithmic coupling has been independently confirmed using \cite[Eq.~(B.5)]{CanFusI15} which also appears to work in the Ramond sector, even when centre type modules are involved.  We mention that we have also computed many examples in which the fusion product decomposes into a direct sum of Kac modules, again in accordance with the above prescription.

\subsection*{$\bm{(p,p') = (1,3)}$ ($\bm{c=-\frac{5}{2}}$)}

\begin{equation}
\begin{aligned}
  \Kac{1,2}\fuse \Kac{1,3}^+ &= \Stag{1,3}{0,1}, \\
  \Kac{1,2}\fuse \Kac{1,9}^+ &= \Stag{1,9}{0,1}(-8),
\end{aligned}
\qquad
\begin{aligned}
  \Kac{1,2}\fuse \Kac{1,6} &= \Stag{1,6}{0,1}(-2)^+ \oplus \Stag{1,6}{0,1}(-2)^-.
\end{aligned}
\end{equation}

\subsection*{$\bm{(p,p') = (1,5)}$ ($\bm{c=-\frac{81}{10}}$)}

\begin{equation}
\begin{aligned}
  \Kac{1,2}\fuse \Kac{1,5}^+ &=  \Stag{1,5}{0,1},
\end{aligned}
\qquad
\begin{aligned}
  \Kac{1,2}\fuse \Kac{1,10} &= \Stag{1,10}{0,1}(-4)^+ \oplus \Stag{1,10}{0,1}(-4)^-.
\end{aligned}
\end{equation}

\subsection*{$\bm{(p,p') = (2,4)}$ ($\bm{c=0}$)}

\begin{equation}
\begin{aligned}
  \Kac{2,1}\fuse \Kac{2,2}^+ &= \Stag{2,2}{1,0}(\tfrac{3}{16},\tfrac{3}{16}), \\
  \Kac{2,1}\fuse \Kac{2,4}^+ &= \Stag{2,4}{1,0}, \\
  \Kac{1,3}^+\fuse \Kac{1,4} &= \Stag{1,4}{0,2}(-\tfrac{3}{16},-\tfrac{3}{16}) \oplus \Kac{1,4}\\
  \Kac{1,3}^+\fuse \Kac{2,3} &= \Kac{2,1} \oplus \Stag{2,4}{0,1},
\end{aligned}
\qquad
\begin{aligned}
  \Kac{2,1}\fuse \Kac{2,1} &= \Stag{2,1}{1,0}(\tfrac{3}{8})^+ \oplus \Stag{2,1}{1,0}(\tfrac{3}{8})^-, \\
  \Kac{2,1}\fuse \Kac{2,3} &= \Stag{2,3}{1,0}(\tfrac{1}{2})^+ \oplus \Stag{2,3}{1,0}(\tfrac{1}{2})^-.
\end{aligned}
\end{equation}

\subsection*{$\bm{(p,p') = (2,8)}$ ($\bm{c=-\frac{21}{4}}$)}

\begin{equation}
\begin{aligned}
  \Kac{2,1}\fuse \Kac{2,6}^+ &= \Stag{2,6}{1,0}(\tfrac{15}{16}), \\
  \Kac{1,2}\fuse \Kac{2,8}^+ &= \Stag{2,8}{0,1},
\end{aligned}
\qquad
\begin{aligned}
  \Kac{2,1}\fuse \Kac{2,7} &= \Stag{2,7}{1,0}(\tfrac{3}{4})^+ \oplus \Stag{2,7}{1,0}(\tfrac{3}{4})^-, \\
  \Kac{1,2}\fuse \Kac{1,8} &= \Stag{1,8}{0,1}(-3)^+ \oplus \Stag{1,8}{0,1}(-3)^-.
\end{aligned}
\end{equation}

\subsection*{$\bm{(p,p') = (3,5)}$ ($\bm{c=\frac{7}{10}}$)}

\begin{equation}
\begin{aligned}
  \Kac{2,1}\fuse \Kac{3,5}^+ &= \Stag{3,5}{1,0}, \\
  \Kac{1,2}\fuse \Kac{1,5}^+ &= \Stag{1,5}{0,1}(-\tfrac{16}{9}), \\
  \Kac{1,2}\fuse \Kac{3,5}^+ &= \Stag{3,5}{0,1},
\end{aligned}
\qquad
\begin{aligned}
  \Kac{2,1}\fuse \Kac{3,4} &= \Stag{3,4}{1,0}(\tfrac{2}{5})^+ \oplus \Stag{3,4}{1,0}(\tfrac{2}{5})^-, \\
  \Kac{1,2}\fuse \Kac{2,5} &= \Stag{2,5}{0,1}(-\tfrac{2}{3})^+ \oplus \Stag{2,5}{0,1}(-\tfrac{2}{3})^-.
\end{aligned}
\end{equation}

\noindent We did not attempt to compute fusion rules involving the centre modules $\Kac{1,2}$, for $(p,p') = (2,4)$, or $\Kac{1,4}$, for $(p,p') = (2,8)$, as their exceptional structures would have required rewriting much of the computer implementation.  The results are, nevertheless, also expected to conform with the above conjectured fusion prescription.

\section{Discussion and conclusions} \label{sec:Conc}

We have seen that fusion rules for $N=1$ superconformal logarithmic minimal models may be profitably explored using generalisations of the methods that have been applied so successfully to the Virasoro logarithmic minimal models in the past.  However, these generalised methods become significantly more complicated when the Ramond sector is considered.  In particular, we have developed and implemented a twisted \NGK{} fusion algorithm, observing a sizeable increase in computational complexity as compared with its untwisted version.  This was expected of course:  the multivalued nature of the fields in the Ramond sector always requires expending more effort, though the results that one derives with this extra effort often end up being simple generalisations of the corresponding \ns{} results.

The fusion results that we have successfully obtained here are no exception.  The Grothendieck fusion rules for Kac modules, which we obtained from a fermionic analogue of the standard Verlinde formula, show considerably regularity across the sectors.  This regularity was also confirmed in the many examples of genuine Kac module fusion rules that we explicitly calculated using the twisted \NGK{} algorithm.  There are two observations relating to these results that we would like to discuss here:  the necessity of noting the global parity of an $N=1$ module and the fact that Ramond by Ramond fusion always returns two copies of each direct summand, one of each parity.

It is, unfortunately, common practice to ignore global parity labels for ($\ZZ_2$-graded) modules over \vosas{}.\footnote{Global parities are considered explicitly in, for example, the works of Iohara and Koga \cite{IohRepI03,IohRepII03,IohStr06}.  However, they too ignore these parities when stating the fusion rules of the $N=1$ superconformal (non-logarithmic) minimal models \cite{IohFus01,IohFus09}.}  Whilst this is largely defensible when considering individual modules, it is less so when computing fusion products.  In particular, we have seen that global parity played an important role in the derivation of the Grothendieck fusion rules and its explicit consideration would be essential in a correlation function based approach to fusion.  Moreover, the \ns{} results reported here, and in \cite{CanFusI15}, illustrate that the global parities of the direct summands of a fusion product are not constant, in general.

There are also conceptual reasons to consider global parities.  We have seen that fusing two Ramond Kac modules always results in \ns{} modules that appear in pairs, where the modules in each pair come with the same multiplicity and only differ by their global parities.  This pairing was also observed for the Ramond fusion of the free fermion (see \cref{app:VerRat}) and we expect it to hold quite generally as a consequence of assuming that fusion respects the parity-reversal functor $\Pi$:
\begin{equation} \label{eq:Assumption}
(\Pi \Mod{M}) \fuse \Mod{N} = \Pi (\Mod{M} \fuse \Mod{N}) = \Mod{M} \fuse (\Pi \Mod{N}).
\end{equation}
If we had ignored global parities, then we might have erroneously concluded that fusing a Ramond module with itself gives the vacuum module as a (submodule of a) direct summand whose multiplicity is $2$, contrary to the usual expectations of self-conjugate modules.

A second conceptual reason not to ignore global parities is the modular properties of the quantum state space.  Consider the free fermion whose quantum state space has the form
\begin{equation}
(\mathsf{NS}^+ \otimes \mathsf{NS}^+) \oplus (\mathsf{NS}^- \otimes \mathsf{NS}^-) \oplus (\mathsf{R} \otimes \mathsf{R}).
\end{equation}
Ignoring global parities might lead one to miss the multiplicity of $2$ for $\abs{\ch{\mathsf{NS}}}^2$ and $\abs{\sch{\mathsf{NS}}}^2$ in the partition function and superpartition function, respectively.  This factor is, of course, necessary to compensate for the factor of $\sqrt{2}$ that arises in the Ramond character, see \eqref{eq:FFChars}, when showing that the sum of the partition and superpartition function is modular invariant, as it should be (this sum is four times the partition function of the bosonic orbifold, the Virasoro minimal model $\MinMod{3}{4}$).

Parity issues aside, the fusion results presented here also confirm the expectation that the $N=1$ superconformal logarithmic minimal models will exhibit logarithmic behaviour in the Ramond sector (this was confirmed for the \ns{} sector in \cite{PeaLog14,CanFusI15}).  More precisely, fusing appropriate \ns{} Kac modules and Ramond Kac modules generates reducible but indecomposable $N=1$ Ramond modules on which $L_0$ acts non-semisimply.  These Ramond staggered modules have similar structures to their \ns{} counterparts, except when the ground states have conformal weight $h = \frac{c}{24}$.  In the latter case, it appears that \emph{two} logarithmic couplings are required to completely fix the isomorphism class of the staggered module.  However, these couplings are expected to be equal when the staggered module is generated by fusing Kac modules (and this indeed follows if we assume that \eqref{eq:Assumption} is valid).  Whilst this equality is a very natural expectation, we mention that it appears to have non-trivial consequences for the corresponding scalar products and correlation functions (see \cref{app:RStag}).

We remark that Ramond staggered modules always possess a non-semisimple action of $G_0$, for a semisimple action of $G_0$ would imply that the action of $L_0 = G_0^2 + \frac{c}{24}$ is also semisimple.  The converse is not true, of course, but our results suggest that fusing Kac modules never generates Ramond modules with a semisimple $L_0$-action but a non-semisimple $G_0$-action, such as the pre-Verma modules $\PreVer{c/24}^{\pm}$ of \cref{sec:Verma}.  Nevertheless, such modules are $N=1$ modules so they may yet play some role in physical applications.

We conclude with a brief outlook for future research directions.  One of our motivations, besides the lattice calculations of \cite{PeaLog14}, for exploring $N=1$ superconformal field theories is to develop technology and gain intuition that may profitably be exploited in similar studies of $N>1$ theories.  In particular, the nilpotent action of $G_0$ on states of conformal weight $h = \frac{c}{24}$ serves as an accessible starting point for exploring the difficulties caused by nilpotent fermions (which are legion in the $N>1$ superconformal algebras).  Developing logarithmic technology here seems prudent given that the non-unitary $N=2$ superconformal minimal models are surely logarithmic.  We hope to report on such $N>1$ studies in the future.

One can also try to approach the $N>1$ superconformal minimal models, logarithmic or otherwise, through relations with other \cfts{}.  Here, we have in mind those with affine Kac-Moody superalgebra symmetries, certain of which give superconformal models upon quantum hamiltonian reduction \cite{KacQua03}.  Unfortunately, these affine superalgebra theories are not well understood at present, see \cite{RozQua92,SalGL106,SalSU207,GotWZN07,CreRel11,CreWAl11} for some limited progress, though one might expect that the affine symmetry might yet lead to beautiful general features.  Again, affine superalgebra theories are almost always logarithmic and their elucidation is expected to yield important insights into superconformal minimal models, general \lcfts{} and their myriad applications.  We also hope to report on these affine theories in the future.

\appendix

\section{A fermionic Verlinde formula} \label{app:VerlindeFormula}

In this section, we provide a derivation of the fermionic Verlinde formula \eqref{eq:Verlinde}.  This extends the results of \cite{EhoFus94} and rests on one main assumption, that the standard module formalism of \cite{CreLog13,RidVer14} applies to the bosonic orbifold of the $N=1$ algebra.  As the application at hand only involves \ns{} modules of non-vanishing supercharacter and Ramond modules of vanishing supercharacter, we will incorporate these facts into the derivation and thereby simplify it.

\subsection{The derivation} \label{sec:FermVer}

We start by noting that the $N=1$ \vosa{} admits an order $2$ automorphism that fixes the even (bosonic) elements and negates the odd (fermionic) ones.  The corresponding orbifold is the \voa{} spanned by the bosonic elements.  The assumption that this bosonic orbifold admits a collection of standard modules implies that the Grothendieck fusion rules of the orbifold may be computed using the standard Verlinde formula.  We compute these rules in terms of $N=1$ data, thereby reconstructing a Verlinde formula for the $N=1$ algebras.

The derivation is conveniently cast in the language of induction and restriction.  To start, we note that the vacuum module $\Kac{1,1}^+$ of the $N=1$ algebra restricts to the direct sum of its bosonic states, which form the vacuum module $\Mod{I}^+$ of the bosonic orbifold, and its fermionic states, which form a module $\Mod{J}^-$ over the orbifold algebra.  Conversely, inducing either orbifold module to an $N=1$ module recovers the $N=1$ vacuum module.  Thus,
\begin{equation} \label{eq:ResVacMod}
\Res{\Kac{1,1}^+} = \Mod{I}^+ \oplus \Mod{J}^-, \qquad \Ind{\Mod{I}^+} = \Ind{\Mod{J}^-} = \Kac{1,1}^+.
\end{equation}
It is sometimes convenient to remember the parities inherited in this fashion by orbifold modules, even though the natural $\ZZ_2$-grading on the bosonic orbifold algebra is trivial (parity is meaningless for bosonic algebras).  For example, such considerations show that $\Mod{J}^-$ is a simple current of order two:
\begin{equation}
\Ind{\brac{\Mod{J}^- \fuse \Mod{J}^-}} = \Ind{\Mod{J}^-} \fuse \Ind{\Mod{J}^-} = \Kac{1,1}^+ \fuse \Kac{1,1}^+ = \Kac{1,1}^+ \qquad \Ra \qquad \Mod{J}^- \fuse \Mod{J}^- = \Mod{I}^+.
\end{equation}
Here, we have used \cite[Eq.~(3.3)]{RidVer14} to compute the induction of the fusion product on the \lhs{} and noted that the only alternative conclusion would be that $\Mod{J}^- \fuse \Mod{J}^- = \Mod{J}^-$ (which violates parity).  It follows that the $N=1$ algebra is the simple current extension, by $\Mod{J}^-$, of the orbifold algebra.

This restriction and induction generalises to $N=1$ modules (we consider only Fock spaces and Kac modules for simplicity) and the corresponding orbifold modules as follows.  In the \ns{} sector, we may take restriction and induction to act as
\begin{equation} \label{eq:IndResNS}
\Res{\Mod{M}^{\pm}} = \Orb{}^+ \oplus (\Mod{J}^- \times \Orb{}^+), \qquad \Ind{\Orb{}^+} = \Mod{M}^+, \quad \Ind{(\Mod{J}^- \times \Orb{}^+)} = \Mod{M}^-,
\end{equation}
where we take $\Orb{}^+$ to consist of the bosonic states of $\Mod{M}^{\pm}$ (regardless of the latter's parity).  This means that the ground states of $\Mod{M}^+$ may be identified with those of $\Orb{}^+$, while the ground states of $\Mod{M}^-$ should be identified with those of $\Mod{J}^- \times \Orb{}^+$.  In the Ramond sector, things are more straightforward because there is no need to indicate the parity.  The restriction and induction functors act as
\begin{equation} \label{eq:IndResR}
\Res{\Mod{M}} = \Orb{} \oplus \Orb{}, \qquad \Ind{\Orb{}} = \Mod{M},
\end{equation}
because $\Orb{} \fuse \Mod{J}^- = \Orb{}$.  From here on, we will omit parity labels on all orbifold modules.

Define the orbifold modules $\NSOrb{\lambda}$, $\Mod{J} \fuse \NSOrb{\lambda}$ and $\ROrb{\lambda}$, for $\lambda \in \RR$, to consist of the bosonic states of $\NSFock{\lambda}^+$, $\NSFock{\lambda}^-$ and $\RFock{\lambda}$, respectively, as in \eqref{eq:IndResNS} and \eqref{eq:IndResR}.  We will assume that these define a set of standard modules for the orbifold \voa{}.  In particular, their characters are linearly independent.  The correspondences $\NSOrb{\lambda} \leftrightarrow \NSFock{\lambda}^+$, $\Mod{J} \fuse \NSOrb{\lambda} \leftrightarrow \NSFock{\lambda}^-$ and $\ROrb{\lambda} \leftrightarrow \RFock{\lambda}$ are each one-to-one.  However, the analogous correspondence between characters is only two-to-one in the \ns{} sector because characters do not distinguish $\NSFock{\lambda}^+$ from $\NSFock{\lambda}^-$.  The correspondence for \ns{} supercharacters is likewise two-to-one as $\sch{\NSFock{\lambda}^+} = -\sch{\NSFock{\lambda}^-}$, though that of the Ramond characters remains one-to-one.  It follows that in the \ns{} sector, integrating over all the standard orbifold characters, $\ch{\NSOrb{\lambda}}$ and $\ch{\Mod{J} \fuse \NSOrb{\lambda}}$, is equivalent to integrating over all the standard $N=1$ characters, $\ch{\NSFock{\lambda}}$, \emph{twice}, and the same is true for supercharacters.  However, integrating over the $\ch{\ROrb{\lambda}}$ is the same as integrating over all the $\ch{\RFock{\lambda}}$.

With this preparation, we can now relate the S-matrix entries, in the basis of standard characters, of the $N=1$ theory to those of its bosonic orbifold.  For this, it is convenient to introduce the \emph{monodromy charge} $Q(\Orb{})$ of an orbifold module $\Orb{}$ (with respect to the simple current $\Mod{J}$) \cite{SchExt89}:
\begin{equation}
Q(\Orb{}) = h(\Orb{}) + h(\Mod{J}) - h(\Mod{J} \fuse \Orb{}) =
\begin{cases}
0 & \text{if \(\Ind{\Orb{}}\) is \ns{},} \\
\frac{1}{2} & \text{if \(\Ind{\Orb{}}\) is Ramond.}
\end{cases}
\end{equation}
Here, $h(\Orb{})$ denotes the conformal weight of the ground states of $\Orb{}$ so, in particular, $h(\Mod{J}) = \frac{3}{2}$.  The point is that the monodromy charge governs how the simple current acts on the S-matrix entries \cite{SchSimp90}:
\begin{equation} \label{eq:SConSMat}
\OrbSmat{\Mod{J} \fuse \Orb{}}{\Orb{}'} = \ee^{2 \pi \ii Q(\Orb{}')} \OrbSmat{\Orb{}}{\Orb{}'} =
\begin{cases}
+\OrbSmat{\Orb{}}{\Orb{}'} & \text{if \(\Ind{\Orb{}'}\) is \ns{},} \\
-\OrbSmat{\Orb{}}{\Orb{}'} & \text{if \(\Ind{\Orb{}'}\) is Ramond.}
\end{cases}
\end{equation}
Here, $\Orb{}$ and $\Orb{}'$ are standard orbifold modules ($\NSOrb{\lambda}$, $\Mod{J} \fuse \NSOrb{\lambda}$, $\ROrb{\lambda}$) and we denote the S-matrix of the bosonic orbifold by $\OrbmodS$ to distinguish it from $\modS$, the S-matrix of the $N=1$ algebra.  The analogous relation for $\OrbSmat{\Orb{}}{\Mod{J} \times \Orb{}'}$ now follows from the symmetry of the orbifold S-matrix in the standard basis \cite{RidVer14}.

From $\Res{\NSFock{\lambda}^+} = \NSOrb{\lambda} \oplus (\Mod{J} \times \NSOrb{\lambda})$ (we choose positive parities as in \cref{sec:Mod}), we deduce that
\begin{equation}
\Sch{\NSFock{\lambda}^+}
= \int \brac{\OrbSmat{\NSOrb{\lambda}}{\Orb{}} + \OrbSmat{\Mod{J} \fuse \NSOrb{\lambda}}{\Orb{}}} \ch{\Orb{}} \: \dd \Orb{}
= \int_{\Ind{\Orb{}} \in \text{NS}} 2 \, \OrbSmat{\NSOrb{\lambda}}{\Orb{}} \ch{\Orb{}} \: \dd \Orb{},
\end{equation}
where the first integral is over all the standard orbifold modules and the second is over the $\NSOrb{\mu}$ and $\Mod{J} \fuse \NSOrb{\mu}$ that induce to \ns{} $N=1$ modules.  It follows that
\begin{align}
\Sch{\NSFock{\lambda}^+}
&= \int_{-\infty}^{\infty} 2 \, \Bigl( \OrbSmat{\NSOrb{\lambda}}{\NSOrb{\mu}} \ch{\NSOrb{\mu}} + \OrbSmat{\NSOrb{\lambda}}{\Mod{J} \fuse \NSOrb{\mu}} \ch{\Mod{J} \fuse \NSOrb{\mu}} \Bigr) \: \dd \mu \notag \\
&= \int_{-\infty}^{\infty} 2 \, \OrbSmat{\NSOrb{\lambda}}{\NSOrb{\mu}} \Bigl( \ch{\NSOrb{\mu}} + \ch{\Mod{J} \fuse \NSOrb{\mu}} \Bigr) \: \dd \mu \notag \\
&= \int_{-\infty}^{\infty} 2 \, \OrbSmat{\NSOrb{\lambda}}{\NSOrb{\mu}} \ch{\NSFock{\mu}^+} \: \dd \mu,
\end{align}
where we have used \eqref{eq:SConSMat} to simplify the S-transform.  We conclude that
\begin{subequations} \label{eq:SFerm=SBos}
\begin{align}
\Smat{\NSFock{\lambda}^+}{\NSFock{\mu}^+} &= 2 \, \OrbSmat{\NSOrb{\lambda}}{\NSOrb{\mu}}.
\intertext{Similar calculations result in}
\Smat{\RFock{\lambda}}{\overline{\NSFock{\mu}^+}} &= 2 \, \OrbSmat{\ROrb{\lambda}}{\NSOrb{\mu}}, \\
\Smat{\overline{\NSFock{\lambda}^+}}{\RFock{\mu}} &= \OrbSmat{\NSOrb{\lambda}}{\ROrb{\mu}},
\end{align}
\end{subequations}
recalling that the bar indicates the supercharacter.  The other S-matrix entries vanish.

There are almost identical relations holding for the S-matrix entries involving the Kac modules.  As restricting an $N=1$ module to its bosonic or fermionic orbifold submodule defines an exact functor, the analogues of the Kac modules in the orbifold theory have characters satisfying similar identities to \eqref{eq:KacChar}.  The analogues of \eqref{eq:SFerm=SBos} then follow readily.  In particular, the vacuum S-matrix entries satisfy
\begin{equation} \label{eq:SFerm=SBosVac}
\Smat{\Kac{1,1}^+}{\NSFock{\mu}^+} = 2 \, \OrbSmat{\Mod{I}}{\NSOrb{\mu}}, \qquad
\Smat{\overline{\Kac{1,1}^+}}{\RFock{\mu}} = \OrbSmat{\Mod{I}}{\ROrb{\mu}},
\end{equation}
where we recall that $\Mod{I}$ denotes the vacuum module of the orbifold algebra.

We turn now to the Verlinde product of two orbifold modules, $\Mod{M}$ and $\Mod{N}$, and their $N=1$ inductions:
\begin{align}
\ch{\Ind{\Mod{M}}} \Grfuse \ch{\Ind{\Mod{N}}} &= \ch{\Ind{\Mod{M}} \fuse \Ind{\Mod{N}}} = \ch{\Ind{(\Mod{M} \fuse \Mod{N})}} = \brac{\ch{\Mod{I}} + \ch{\Mod{J}}} \Grfuse \ch{\Mod{M} \fuse \Mod{N}} \notag \\
&= \brac{\ch{\Mod{I}} + \ch{\Mod{J}}} \Grfuse \int \orbfuscoeff{\Mod{M}}{\Mod{N}}{\Orb{}} \ch{\Orb{}} \: \dd \Orb{} = \int \orbfuscoeff{\Mod{M}}{\Mod{N}}{\Orb{}} \ch{\Ind{\Orb{}}} \: \dd \Orb{},
\end{align}
where the integration is over all the standard orbifold modules and the $\orbfuscoeff{\Mod{M}}{\Mod{N}}{\Orb{}}$ denote the Verlinde coefficients of the orbifold algebra.  As $\ch{\Ind{\NSOrb{\nu}}} = \ch{\NSFock{\nu}^+} = \ch{\NSFock{\nu}^-} = \ch{\Ind{(\Mod{J} \fuse \NSOrb{\nu})}}$, we arrive at
\begin{equation}
\ch{\Ind{\Mod{M}}} \Grfuse \ch{\Ind{\Mod{N}}} = \int_{-\infty}^{\infty} \brac{\orbfuscoeff{\Mod{M}}{\Mod{N}}{\NSOrb{\nu}} + \orbfuscoeff{\Mod{M}}{\Mod{N}}{\Mod{J} \fuse \NSOrb{\nu}}} \ch{\NSFock{\nu}^+} \: \dd \nu + \int_{-\infty}^{\infty} \orbfuscoeff{\Mod{M}}{\Mod{N}}{\ROrb{\nu}} \ch{\RFock{\nu}} \: \dd \nu,
\end{equation}
hence the $N=1$ Verlinde coefficients satisfy
\begin{subequations} \label{eq:VerCoeff}
\begin{equation} \label{eq:VerCoeffChar}
\fuscoeff{\Ind{\Mod{M}}}{\Ind{\Mod{N}}}{\NSFock{\nu}^+} = \orbfuscoeff{\Mod{M}}{\Mod{N}}{\NSOrb{\nu}} + \orbfuscoeff{\Mod{M}}{\Mod{N}}{\Mod{J} \fuse \NSOrb{\nu}}, \qquad 
\fuscoeff{\Ind{\Mod{M}}}{\Ind{\Mod{N}}}{\RFock{\nu}} = \orbfuscoeff{\Mod{M}}{\Mod{N}}{\ROrb{\nu}}.
\end{equation}
Repeating this analysis for supercharacters gives instead
\begin{equation} \label{eq:VerCoeffSChar}
\fuscoeff{\overline{\Ind{\Mod{M}}}}{\overline{\Ind{\Mod{N}}}}{\overline{\NSFock{\nu}^+}} = \orbfuscoeff{\Mod{M}}{\Mod{N}}{\NSOrb{\nu}} - \orbfuscoeff{\Mod{M}}{\Mod{N}}{\Mod{J} \fuse \NSOrb{\nu}}.
\end{equation}
\end{subequations}
Of course, any $N=1$ Verlinde coefficient involving a Ramond supercharacter vanishes.

Substituting in the standard Verlinde formula for the orbifold Verlinde coefficients now gives the $N=1$ Verlinde formulae.  We first suppose that $\Mod{M}$ and $\Mod{N}$ are orbifold modules whose inductions belong to the \ns{} sector.  Using \eqref{eq:VerCoeffChar} twice, then \eqref{eq:SFerm=SBos} and \eqref{eq:SFerm=SBosVac}, results in the following $N=1$ Verlinde formula for $\text{NS} \Grfuse \text{NS}$ Verlinde coefficients:
\begin{subequations} \label{eq:FermVerFormulae}
\begin{align}
\fuscoeff{\Ind{\Mod{M}}}{\Ind{\Mod{N}}}{\NSFock{\nu}^+}
&= \int \frac{\OrbSmat{\Mod{M}}{\Orb{}} \OrbSmat{\Mod{N}}{\Orb{}} \brac{\OrbSmat{\NSOrb{\nu}}{\Orb{}} + \OrbSmat{\Mod{J} \fuse \NSFock{\nu}^+}{\Orb{}}}^*}{\OrbSmat{\Mod{I}}{\Orb{}}} \: \dd \Orb{} \notag \\
&= \int_{-\infty}^{\infty} \frac{\OrbSmat{\Mod{M}}{\NSOrb{\rho}} \OrbSmat{\Mod{N}}{\NSOrb{\rho}} \, 2 \, \OrbSmat{\NSOrb{\nu}}{\NSOrb{\rho}}^*}{\OrbSmat{\Mod{I}}{\NSOrb{\rho}}} \: \dd \rho \notag \\
&\mspace{30mu} + \int_{-\infty}^{\infty} \frac{\OrbSmat{\Mod{M}}{\Mod{J} \fuse \NSOrb{\rho}} \OrbSmat{\Mod{N}}{\Mod{J} \fuse \NSOrb{\rho}} \, 2 \, \OrbSmat{\NSOrb{\nu}}{\Mod{J} \fuse \NSOrb{\rho}}^*}{\OrbSmat{\Mod{I}}{\Mod{J} \fuse \NSOrb{\rho}}} \: \dd \rho \notag \\
&= 4 \int_{-\infty}^{\infty} \frac{\OrbSmat{\Mod{M}}{\NSOrb{\rho}} \OrbSmat{\Mod{N}}{\NSOrb{\rho}} \OrbSmat{\NSOrb{\nu}}{\NSOrb{\rho}}^*}{\OrbSmat{\Mod{I}}{\NSOrb{\rho}}} \: \dd \rho \notag \\
&= \int_{-\infty}^{\infty} \frac{\Smat{\Ind{\Mod{M}}}{\NSFock{\rho}^+} \Smat{\Ind{\Mod{N}}}{\NSFock{\rho}^+} \Smat{\NSFock{\nu}^+}{\NSFock{\rho}^+}^*}{\Smat{\Kac{1,1}^+}{\NSFock{\rho}^+}} \: \dd \rho \qquad \text{(\(\Ind{\Mod{M}}, \ \Ind{\Mod{N}} \in \text{NS}\)).}
\intertext{Analogous calculations give the remaining non-vanishing Verlinde coefficients:}
\fuscoeff{\Ind{\Mod{M}}}{\Ind{\Mod{N}}}{\NSFock{\nu}^+}
&= \int_{-\infty}^{\infty} \frac{\Smat{\Ind{\Mod{M}}}{\overline{\NSFock{\rho}^+}} \Smat{\Ind{\Mod{N}}}{\overline{\NSFock{\rho}^+}} \Smat{\NSFock{\nu}^+}{\NSFock{\rho}^+}^*}{\Smat{\Kac{1,1}^+}{\NSFock{\rho}^+}} \: \dd \rho \qquad \text{(\(\Ind{\Mod{M}}, \ \Ind{\Mod{N}} \in \text{R}\)),} \\
\fuscoeff{\Ind{\Mod{M}}}{\Ind{\Mod{N}}}{\RFock{\nu}}
&= \int_{-\infty}^{\infty} \frac{\Smat{\Ind{\Mod{M}}}{\NSFock{\rho}^+} \Smat{\Ind{\Mod{N}}}{\overline{\NSFock{\rho}^+}} \Smat{\RFock{\nu}}{\overline{\NSFock{\rho}^+}}^*}{2 \, \Smat{\Kac{1,1}^+}{\NSFock{\rho}^+}} \: \dd \rho \qquad \text{(\(\Ind{\Mod{M}} \in \text{NS}\), \(\Ind{\Mod{N}} \in \text{R}\)),} \\
\fuscoeff{\overline{\Ind{\Mod{M}}}}{\overline{\Ind{\Mod{N}}}}{\overline{\NSFock{\nu}^+}}
&= \int_{-\infty}^{\infty} \frac{2 \, \Smat{\overline{\Ind{\Mod{M}}}}{\RFock{\rho}} \Smat{\overline{\Ind{\Mod{N}}}}{\RFock{\rho}} \Smat{\overline{\NSFock{\nu}^+}}{\RFock{\rho}}^*}{\Smat{\overline{\Kac{1,1}^+}}{\RFock{\rho}}} \: \dd \rho \qquad \text{(\(\Ind{\Mod{M}}, \ \Ind{\Mod{N}} \in \text{NS}\)).}
\end{align}
\end{subequations}
This completes the derivation of the $N=1$ Verlinde formula \eqref{eq:Verlinde}.

\subsection{An elementary application --- the free fermion} \label{app:VerRat}

Here, we provide a check of the fermionic Verlinde formula \eqref{eq:FermVerFormulae}, and its derivation, by applying it to the free fermion.  We note that the hypotheses required by the derivation are met in this case:  The supercharacter of the \ns{} module is non-vanishing, while that of the Ramond module vanishes, and the bosonic orbifold of the free fermion is the Virasoro minimal model $\MinMod{3}{4}$ which is rational (the standard Verlinde formula therefore applies with integrals replaced by finite sums).

As in \cref{sec:Fock}, the free fermion \vosa{} is generated by the odd field $\func{b}{z}$ and its conformal structure is defined by
\begin{equation}
\func{T}{z} = \frac{1}{2} \normord{\pd \func{b}{z} \func{b}{z}}.
\end{equation}
This gives $\func{b}{z}$ a conformal weight of $\frac{1}{2}$ and the defining \ope{} and mode anticommutation relations are
\begin{equation}
\func{b}{z} \func{b}{w} \sim \frac{1}{z-w}, \qquad \acomm{b_j}{b_k} = \delta_{j+k=0}.
\end{equation}

As is well known, the free fermion \vosa{} admits a single indecomposable module, the \ns{} vacuum module $\mathsf{NS}$, which is simple, and a single indecomposable twisted module, the Ramond module $\mathsf{R}$, which is also simple.  The space of ground states is one-dimensional for $\mathsf{NS}$, with conformal weight $0$, and two-dimensional for $\mathsf{R}$, with conformal weight $\frac{1}{16}$.  As we consider parity explicitly below, we shall distinguish two \ns{} modules $\mathsf{NS}^{\pm}$ by the parities of their ground states.  The Ramond module $\mathsf{R}$ is isomorphic to its parity-reversed counterpart.

Because the bosonic subalgebra of the free fermion is $\MinMod{3}{4}$, the free fermion may be realised as an order $2$ simple current extension of $\MinMod{3}{4}$.  More precisely, the map that fixes even elements and negates odd elements is an order $2$ automorphism of the free fermion \vosa{} and $\MinMod{3}{4}$ is the corresponding orbifold subalgebra.  This subalgebra possesses three indecomposable modules $\wun$, $\sigma$ and $\epsilon$, all simple, of respective conformal weights $0$, $\frac{1}{16}$ and $\frac{1}{2}$.  Their fusion rules are well known:  $\wun$ is the fusion unit and the remaining rules are
\begin{equation} \label{FR:M34}
\sigma \fuse \sigma = \wun \oplus \epsilon, \qquad \sigma \fuse \epsilon = \sigma, \qquad \epsilon \fuse \epsilon = \wun.
\end{equation}
The last demonstrates that $\epsilon$ is a simple current of order $2$; the corresponding simple current extension of $\MinMod{3}{4}$ recovers the free fermion superalgebra.  

Restricting to $\MinMod{3}{4}$-modules, the free fermion modules decompose as
\begin{equation}
\Res{\mathsf{NS}^{\pm}} = \wun \oplus \epsilon, \qquad \Res{\mathsf{R}} = 2 \, \sigma. 
\end{equation}
We note that the bosonic submodule of $\mathsf{NS}^+$ is $\wun$ and that of $\mathsf{NS}^-$ is $\epsilon$.  Because we are free to regard each $\MinMod{3}{4}$-module as being bosonic (there is no parity in the minimal models), inducing from $\MinMod{3}{4}$ to the free fermion amounts to
\begin{equation}
\Ind{\wun} = \mathsf{NS}^+, \qquad \Ind{\epsilon} = \mathsf{NS}^-, \qquad \Ind{\sigma} = \mathsf{R}.
\end{equation}
We remark that this induction is consistent with fusion orbits on which the simple current fields act freely.

One way to deduce the fusion rules of the free fermion theory is then to utilise the following relation \cite{RidVer14}:
\begin{equation}
\Ind{\Mod{M}} \fuse \Ind{\Mod{N}} = \Ind{\brac{\Mod{M} \fuse \Mod{N}}}.
\end{equation}
This immediately shows that $\mathsf{NS}^+$ is the fusion unit, as expected, that $\mathsf{NS}^-$ reverses the parity of any module (recalling that $\mathsf{R}$ is isomorphic to its parity-reversal), and that
\begin{equation}
\mathsf{R} \fuse \mathsf{R} = \Ind{\sigma} \fuse \Ind{\sigma} = \Ind{\brac{\sigma \fuse \sigma}} = \Ind{\brac{\wun \oplus \epsilon}} = \Ind{\wun} \oplus \Ind{\epsilon} = \mathsf{NS}^+ \oplus \mathsf{NS}^-.
\end{equation}

Our aim is to reproduce these fusion rules using the fermionic Verlinde formula \eqref{eq:FermVerFormulae}, thereby checking the derivation of this formula in \cref{sec:FermVer}.  The free fermion characters and supercharacters are well known:
\begin{equation} \label{eq:FFChars}
\ch{\mathsf{NS}^+} = \sqrt{\frac{\fjth{3}{1;q}}{\func{\eta}{q}}}, \qquad
\sch{\mathsf{NS}^+} = \sqrt{\frac{\fjth{4}{1;q}}{\func{\eta}{q}}}, \qquad
\ch{\mathsf{R}} = \sqrt{\frac{2 \, \fjth{2}{1;q}}{\func{\eta}{q}}}, \qquad
\sch{\mathsf{R}} = 0.
\end{equation}
It is clear that $\ch{\mathsf{NS}^-} = \ch{\mathsf{NS}^+}$ and $\sch{\mathsf{NS}^-} = -\sch{\mathsf{NS}^+}$.  The S-matrix of the free fermion, with respect to the ordered basis $\set{\ch{\mathsf{NS}^+}, \sch{\mathsf{NS}^+}, \ch{\mathsf{R}}}$, is thus
\begin{equation}
\modS =
\begin{pmatrix}
1 & 0 & 0 \\
0 & 0 & \frac{1}{\sqrt{2}} \\
0 & \sqrt{2} & 0
\end{pmatrix}
.
\end{equation}
Comparing with the $\MinMod{3}{4}$ S-matrix, with respect to the ordered basis $\set{\ch{\wun}, \ch{\epsilon}, \ch{\sigma}}$,
\begin{equation}
\OrbmodS = \frac{1}{2}
\begin{pmatrix}
1 & 1 & +\sqrt{2} \\
1 & 1 & -\sqrt{2} \\
+\sqrt{2} & -\sqrt{2} & 0
\end{pmatrix}
,
\end{equation}
we verify \eqref{eq:SFerm=SBos}:
\begin{equation}
\Smat{\mathsf{NS}^+}{\mathsf{NS}^+} = 2 \, \OrbSmat{\wun}{\wun}, \qquad
\Smat{\overline{\mathsf{NS}^+}}{\mathsf{R}} = \OrbSmat{\wun}{\sigma}, \qquad
\Smat{\mathsf{R}}{\overline{\mathsf{NS}^+}} = 2 \, \OrbSmat{\sigma}{\wun}.
\end{equation}
Note that the standard modules of $\MinMod{3}{4}$ and the free fermion are, in both cases, just the simple modules because the relevant module categories are semisimple \cite{RidVer14}.

We can also use the fusion rules \eqref{FR:M34} to verify \eqref{eq:VerCoeff}, which takes the form
\begin{equation}
\begin{aligned}
\fuscoeff{\mathsf{NS}^+}{\mathsf{NS}^+}{\mathsf{NS}^+} &= \orbfuscoeff{\wun}{\wun}{\wun} + \orbfuscoeff{\wun}{\wun}{\epsilon} = 1, &
\fuscoeff{\mathsf{R}}{\mathsf{R}}{\mathsf{NS}^+} &= \orbfuscoeff{\sigma}{\sigma}{\wun} + \orbfuscoeff{\sigma}{\sigma}{\epsilon} = 2, \\
\fuscoeff{\overline{\mathsf{NS}^+}}{\overline{\mathsf{NS}^+}}{\overline{\mathsf{NS}^+}} &= \orbfuscoeff{\wun}{\wun}{\wun} - \orbfuscoeff{\wun}{\wun}{\epsilon} = 1, &
\fuscoeff{\overline{\mathsf{R}}}{\overline{\mathsf{R}}}{\overline{\mathsf{NS}^+}} &= 0,
\end{aligned}
\quad
\fuscoeff{\mathsf{NS}^+}{\mathsf{R}}{\mathsf{R}} = \orbfuscoeff{\wun}{\sigma}{\sigma} = 1.
\end{equation}
Moreover, the Verlinde formulae \eqref{eq:FermVerFormulae} become
\begin{equation}
\begin{aligned}
\fuscoeff{\mathsf{NS}^+}{\mathsf{NS}^+}{\mathsf{NS}^+} &= \frac{\Smat{\mathsf{NS}^+}{\mathsf{NS}^+} \Smat{\mathsf{NS}^+}{\mathsf{NS}^+} \Smat{\mathsf{NS}^+}{\mathsf{NS}^+}^*}{\Smat{\mathsf{NS}^+}{\mathsf{NS}^+}} = 1, \\
\fuscoeff{\mathsf{R}}{\mathsf{R}}{\mathsf{NS}^+} &= \frac{\Smat{\mathsf{R}}{\overline{\mathsf{NS}^+}} \Smat{\mathsf{R}}{\overline{\mathsf{NS}^+}} \Smat{\mathsf{NS}^+}{\mathsf{NS}^+}^*}{\Smat{\mathsf{NS}^+}{\mathsf{NS}^+}} = 2, \\
\fuscoeff{\mathsf{NS}^+}{\mathsf{R}}{\mathsf{R}} &= \frac{1}{2} \frac{\Smat{\mathsf{NS}^+}{\mathsf{NS}^+} \Smat{\mathsf{R}}{\overline{\mathsf{NS}^+}} \Smat{\mathsf{R}^+}{\overline{\mathsf{NS}^+}}^*}{\Smat{\mathsf{NS}^+}{\mathsf{NS}^+}} = 1, \\
\fuscoeff{\overline{\mathsf{NS}^+}}{\overline{\mathsf{NS}^+}}{\overline{\mathsf{NS}^+}} &= 2 \, \frac{\Smat{\overline{\mathsf{NS}^+}}{\mathsf{R}} \Smat{\overline{\mathsf{NS}^+}}{\mathsf{R}} \Smat{\overline{\mathsf{NS}^+}}{\mathsf{R}}^*}{\Smat{\overline{\mathsf{NS}^+}}{\mathsf{R}}} = 1,
\end{aligned}
\end{equation}
because the integration over $\RR$ is replaced by a sum over a single (super)character, that of $\mathsf{NS}^+$ or $\mathsf{R}$.  The counterparts involving $\mathsf{NS}^-$ follow using $\ch{\mathsf{NS}^-} = \ch{\mathsf{NS}^+}$ and $\sch{\mathsf{NS}^-} = -\sch{\mathsf{NS}^+}$.  As the other Verlinde coefficients vanish, these results recover the fusion rules completely.  In particular,
\begin{equation}
\ch{\mathsf{R} \fuse \mathsf{R}} = \ch{\mathsf{R}} \Grfuse \ch{\mathsf{R}} = 2 \ch{\mathsf{NS}^+}, \quad
\sch{\mathsf{R} \fuse \mathsf{R}} = \sch{\mathsf{R}} \Grfuse \sch{\mathsf{R}} = 0 \quad \Ra \quad
\mathsf{R} \fuse \mathsf{R} = \mathsf{NS}^+ \oplus \mathsf{NS}^-,
\end{equation}
by semisimplicity.

\section{Coproduct formulae for fusing twisted modules} \label{app:TwCoprod}

In this appendix, we derive coproduct formulae for fusion products of \emph{twisted} modules.  Such formulae were first deduced by Gaberdiel \cite{GabFus97} as generalisations of his untwisted formulae \cite{GabFus94,GabFus94b}.  While some calculations for ``generic'' $N=1$ and $N=2$ superconformal modules were described, only a bare minimum of information was reported, presumably because of the unwieldy nature of the twisted coproduct formulae.  In particular, the problem of identifying the mathematical structure of these fusion products was not addressed.  Here, we give simplified coproduct formulae that are used in \cref{sec:TwFus} to develop a twisted version of the \NGK{} fusion algorithm \cite{NahQua94,GabInd96}.  This twisted algorithm has been implemented in \textsf{python} for the $N=1$ superconformal algebra; the results reported in \cref{sec:Results} were obtained using this implementation.

Before detailing the derivation of the twisted coproduct formulae, we remark that the aim is to determine the natural action of the generating fields of the \voSa{} on the \opes{} of the fields corresponding to the modules being fused.  Because the algebra fields and the module fields satisfy mutual locality relations, one obtains two different coproduct formulae on the vector space tensor product of the modules.  The fusion product is then defined as the quotient on which these coproduct actions coincide.

Suppose then that the field $\func{\psi_i}{w_i}$, of conformal weight $h_i$, is twisted with respect to the action of a given generator $\func{S}{z}$ of the \voSa{}:
\begin{equation} \label{eq:TwOPE}
\func{S}{z} \func{\psi_i}{w_i} = \sum_{m \in \ZZ - h_i - \eps_i} \func{(S_m \psi_i)}{w_i} \brac{z-w_i}^{-m-h_i}.
\end{equation}
We will refer to the real number $\eps_i$ as the \emph{twist parameter}, with respect to $\func{S}{z}$, of the (twisted) module corresponding to $\func{\psi_i}{w_i}$.  In principle, twist parameters are only defined modulo $\ZZ$.  However, we will find it useful to regard them as real numbers.  When $\eps_i \in \ZZ$, the powers of $z-w_i$ in \eqref{eq:TwOPE} are integers and the corresponding module is said to be untwisted with respect to $\func{S}{z}$.

Because we want to study the action of the modes $S_m$ of $\func{S}{z}$ on the operator products $\func{\psi_1}{w_1} \func{\psi_2}{w_2}$, and because twist parameters are conserved additively under \opes{}, we start with the fusion integral
\begin{equation}
\oint_{\Gamma} \bracket{\phi}{\func{S}{z} \func{\psi_1}{w_1} \func{\psi_2}{w_2}}{\Omega} z^{n+h+\eps-1} \: \frac{\dd z}{2 \pi \ii}.
\end{equation}
Here, $h$ is the conformal weight of $\func{S}{z}$, $\eps = \eps_1 + \eps_2$ is the twist parameter of the \ope{} of $\func{\psi_1}{w_1}$ and $\func{\psi_2}{w_2}$, $\Gamma$ is a contour enclosing $0$, $w_1$ and $w_2$, $\Omega$ is the vacuum, and $\phi$ is an arbitrary spectator state (that may depend on other insertion points that are not enclosed by $\Gamma$).

However, we see that inserting the \ope{} \eqref{eq:TwOPE} leads to a branch cut at $z=w_i$ whenever $\eps_i \notin \ZZ$.  Anticipating that this will be problematic, we note that any branching at $z=w_i$ will be converted to a pole (or a zero) upon multiplying by $\brac{z-w_i}^{-\eps_i}$.  In this way, we arrive at a new proposal for the fusion integral:
\begin{equation} \label{eq:TwistedCoprodInt}
\oint_{\Gamma} \bracket{\phi}{\func{S}{z} \func{\psi_1}{w_1} \func{\psi_2}{w_2}}{\Omega} z^{n+h+\eps-1} \brac{z-w_1}^{-\eps_1} \brac{z-w_2}^{-\eps_2} \: \frac{\dd z}{2 \pi \ii}.
\end{equation}
Of course, we can no longer interpret the result of evaluating this integral as defining a coproduct formula for the modes $S_n$ acting on the coproduct $\func{\psi_1}{w_1} \func{\psi_2}{w_2}$.  Rather, it gives a coproduct for the ($w_1$- and $w_2$-dependent) ``modes'' $\tS_n^{w_1,w_2}$ which are characterised by
\begin{equation}
\sum_{n\in \mathbb{Z}-h-\eps} \tS_n^{w_1,w_2} z^{-n-h} = \func{\tS}{z} \equiv \func{S}{z} z^{\eps} \brac{z-w_1}^{-\eps_1} \brac{z-w_2}^{-\eps_2}.
\end{equation}
Explicitly, the relation between the $S_n$ and the $\tS_n^{w_1,w_2}$ is given
by expanding with $\abs{z} > \abs{w_i}$ (because of the radial ordering
implicit in \eqref{eq:TwistedCoprodInt}):
\begin{subequations} \label{eq:STilde}
\begin{align}
S_n &= \sum_{j_1,j_2=0}^{\infty} \binom{\eps_1}{j_1} \binom{\eps_2}{j_2} \brac{-w_1}^{j_1} \brac{-w_2}^{j_2} \tS_{n-j_1-j_2}^{w_1,w_2}, \\
\tS_{n}^{w_1,w_2} &= \sum_{j_1,j_2=0}^{\infty} \binom{-\eps_1}{j_1} \binom{-\eps_2}{j_2} \brac{-w_1}^{j_1} \brac{-w_2}^{j_2} S_{n-j_1-j_2}.
\end{align}
\end{subequations}

As the integrand of \eqref{eq:TwistedCoprodInt} has no branch cuts, by construction, we may evaluate the fusion integral by computing the residues at $0$, $w_1$ and $w_2$.  Inserting the \ope{} \eqref{eq:TwOPE} with $i=1$, we find that the residue at $z=w_1$ is
\begin{equation} \label{eq:Int1}
\sum_{m=-h-\eps_1+1}^{\infty} \oint_{w_1} z^{n+h+\eps-1} \brac{z-w_1}^{-m-h-\eps_1} \brac{z-w_2}^{-\eps_2} \: \frac{\dd z}{2 \pi \ii} \brac{S_m \otimes \wun},
\end{equation}
where $S_m \otimes \wun$ is shorthand for $\bracket{\phi}{\func{(S_m \psi_1)}{w_1} \func{\psi_2}{w_2}}{\Omega}$.  The residue at $z=w_2$ is computed by applying the mutual locality relation
\begin{equation} \label{eq:Locality}
\func{S}{z} \func{\psi_1}{w_1} = \mu_1 \func{\psi_1}{w_1} \func{S}{z} \qquad \text{(\(\mu_1 \in \CC \setminus \set{0}\))}
\end{equation}
and inserting the \ope{} \eqref{eq:TwOPE} with $i=2$.  The result is
\begin{equation} \label{eq:Int2}
\mu_1 \sum_{m=-h-\eps_2+1}^{\infty} \oint_{w_2} z^{n+h+\eps-1} \brac{z-w_1}^{-m-h-\eps_2} \brac{z-w_2}^{-\eps_1} \: \frac{\dd z}{2 \pi \ii} \brac{\wun \otimes S_m},
\end{equation}
where $\wun \otimes S_m$ is shorthand for $\bracket{\phi}{\func{\psi_1}{w_1} \func{(S_m \psi_2)}{w_2}}{\Omega}$.  We note that the $z=w_2$ result may be obtained from the $z=w_1$ result by swapping $\eps_1$ with $\eps_2$, as well as $w_1$ with $w_2$, and then replacing $\tbrac{S_m \otimes \wun}$ by $\mu_1 \tbrac{\wun \otimes S_m}$:

When $n \ge -h-\eps+1$, there is no residue at $z=0$ and evaluating \eqref{eq:Int1} and \eqref{eq:Int2} gives a coproduct formula:
\begin{multline} \label{eq:TwistedWCont}
\parcoproduct{w_1,w_2}{\tS_n^{w_1,w_2}} = \sum_{m=-h-\eps_1+1}^{\infty} \sum_{j=0}^{m+h+\eps_1-1} \binom{-\eps_2}{j} \binom{n+h+\eps-1}{m+h+\eps_1-1-j} \brac{w_1-w_2}^{-\eps_2-j} w_1^{n-m+\eps_2+j} \brac{S_m \otimes \wun} \\
+ \mu_1 \sum_{m=-h-\eps_2+1}^{\infty} \sum_{j=0}^{m+h+\eps_2-1} \binom{-\eps_1}{j} \binom{n+h+\eps-1}{m+h+\eps_2-1-j} \brac{w_2-w_1}^{-\eps_1-j} w_2^{n-m+\eps_1+j} \brac{\wun \otimes S_m}.
\end{multline}
In this formula, we use the fact that the spectator state $\phi$ is arbitrary to extract $S_m \otimes \wun$ and $\wun \otimes S_m$ acting on the tensor product state $\psi_1 \otimes \psi_2$ that corresponds to the operator product $\func{\psi_1}{w_1} \func{\psi_2}{w_2}$.  We note that the powers of $w_1$ and $w_2$ appearing in this formula are integral, unlike those of $w_1-w_2$ and $w_2-w_1$ (in general).  Moreover, swapping the order of summation lets us truncate the sum over $m$ using the form of the binomial coefficients, assuming still that $n \ge -h-\eps+1$:
\begin{multline} \label{eq:TwistedWCont2}
\parcoproduct{w_1,w_2}{\tS_n^{w_1,w_2}} = \sum_{j=0}^{\infty} \sum_{m=-h-\eps_1+1+j}^{n+\eps_2+j} \binom{-\eps_2}{j} \binom{n+h+\eps-1}{m+h+\eps_1-1-j} \brac{w_1-w_2}^{-\eps_2-j} w_1^{n-m+\eps_2+j} \brac{S_m \otimes \wun} \\
+ \mu_1 \sum_{j=0}^{\infty} \sum_{m=-h-\eps_2+1+j}^{n+\eps_1+j} \binom{-\eps_1}{j} \binom{n+h+\eps-1}{m+h+\eps_2-1-j} \brac{w_2-w_1}^{-\eps_1-j} w_2^{n-m+\eps_1+j} \brac{\wun \otimes S_m}.
\end{multline}

The upshot of this is that the integral powers of $w_1$ and $w_2$ that appear are now manifestly non-negative, so it makes sense to send either $w_1$ or $w_2$ to $0$.  In particular, substituting $w_1 = 0$ and $w_2 = -w$ gives
\begin{align} \label{eq:TwistedWCont3}
\parcoproduct{0,-w}{\tS_n^{0,-w}} &= \sum_{j=0}^{\infty} \binom{-\eps_2}{j} w^{-\eps_2 - j} \brac{S_{n+\eps_2+j} \otimes \wun} \notag \\
&\mspace{100mu} + \mu_1 \sum_{m=-h-\eps_2+1}^{\infty} \sum_{j=0}^{m+h+\eps_2-1} \binom{-\eps_1}{j} \binom{n+h+\eps-1}{m+h+\eps_2-1-j} \brac{-w}^{n-m} \brac{\wun \otimes S_m} \notag \\
&= \sum_{j=0}^{\infty} \binom{-\eps_2}{j} w^{-\eps_2 - j} \brac{S_{n+\eps_2+j} \otimes \wun} + \mu_1 \sum_{m=-h-\eps_2+1}^{\infty} \binom{n+h+\eps_2-1}{m+h+\eps_2-1} \brac{-w}^{n-m} \brac{\wun \otimes S_m}.
\end{align}
Here, we have evaluated the sum over $j$ using the binomial identity
\begin{equation} \label{eq:UsefulBinId}
\sum_{j=0}^n \binom{a}{j} \binom{b}{n-j} = \binom{a+b}{n}.
\end{equation}
We can likewise specialise to $w_1 = w$ and $w_2 = 0$.  The resulting formula (for $\parcoproduct{w,0}{\tS_n^{w,0}}$) may be obtained from \eqref{eq:TwistedWCont3} by swapping $\eps_1$ with $\eps_2$, $w$ with $-w$, and $\brac{S_m \otimes \wun}$ with $\mu_1 \brac{\wun \otimes S_m}$.

We now turn to the case where $n \le -h-\eps$ in which there is a contribution from the residue at $z=0$.  For these $n$, \eqref{eq:TwistedWCont} still gives the sum of the residues at $z=w_1$ and $z=w_2$ (however, \eqref{eq:TwistedWCont2} is only valid if we replace the upper bound on the sums over $m$ by $\infty$).  This time, we can write the contribution from $z=0$ in two forms according as to whether $w_1$ or $w_2$ is assumed to be close to $0$.  Suppose the former, so that we may use the \ope{} $\tfunc{S}{z} \tfunc{\psi_1}{w_1}$ to evaluate the $z=0$ residue as
\begin{multline} \label{eq:ToBeSplit}
\sum_{m\in\ZZ-h-\eps_1} \oint_0 z^{n+h+\eps-1} \brac{z-w_1}^{-m-h-\eps_1} \brac{z-w_2}^{-\eps_2} \: \frac{\dd z}{2 \pi \ii} \brac{S_m \otimes \wun} \\
= \sum_{m\in\ZZ-h-\eps_1} \sum_{j=0}^{-n-h-\eps} \binom{-\eps_2}{j} \binom{-m-h-\eps_1}{-n-h-\eps-j} \brac{-w_1}^{n-m+\eps_2+j} \brac{-w_2}^{-\eps_2-j} \brac{S_m \otimes \wun}.
\end{multline}

We anticipate a partial cancellation of this contribution with that of the residue at $z=w_1$ ---  otherwise, we would not be able to specialise to $w_1 = 0$.  To this end, we split the range of the sum over $m$ into $m \ge -h-\eps_1+1$ and $m \leq -h-\eps_1$.  Because of the identities
\begin{align}
\binom{-m-h-\eps_1}{-n-h-\eps-j} &= \brac{-1}^{-n-h-\eps-j} \binom{m-n-\eps_2-1-j}{-n-h-\eps-j} = \brac{-1}^{-n-h-\eps-j} \binom{m-n-\eps_2-1-j}{m+h+\eps_1-1} \notag \\
&= \brac{-1}^{n-m+\eps_2+j+1} \binom{n+h+\eps-1+j}{m+h+\eps_1-1},
\end{align}
the $m$-sum over the former range may be written in the form
\begin{equation} \label{eq:ToBeCancelled}
-\sum_{m=-h-\eps_1+1}^{\infty} \sum_{j=0}^{-n-h-\eps} \binom{-\eps_2}{j} \binom{n+h+\eps-1+j}{m+h+\eps_1-1} w_1^{n-m+\eps_2+j} \brac{-w_2}^{-\eps_2-j} \brac{S_m \otimes \wun},
\end{equation}
noting that $n-m+\eps_2 \in \ZZ$.  The contribution from $z=w_1$ is given in the first term on the \rhs{} of \eqref{eq:TwistedWCont} which we can write as
\begin{multline}
\sum_{m=-h-\eps_1+1}^{\infty} \sum_{j=0}^{\infty} \binom{-\eps_2}{j} \binom{n+h+\eps-1}{m+h+\eps_1-1-j} \sum_{k=0}^{\infty} \binom{-\eps_2-j}{k} w_1^{n-m+\eps_2+j+k} \brac{-w_2}^{-\eps_2-j-k} \brac{S_m \otimes \wun} \\
= \sum_{m=-h-\eps_1+1}^{\infty} \sum_{j=0}^{\infty} \sum_{k=0}^{\infty} \binom{-\eps_2}{j} \binom{n+h+\eps-1}{m+h+\eps_1-1-j} \binom{-\eps_2-j}{k} w_1^{n-m+\eps_2+j+k} \brac{-w_2}^{-\eps_2-j-k} \brac{S_m \otimes \wun},
\end{multline}
where we have expanded about $w_1 \to 0$.  Writing $\ell = j+k$, converting the $k$-sum to an $\ell$-sum, and then swapping the $j$- and $\ell$- sums, we find that the sum over $j$ may be evaluated using
\begin{equation}
\sum_{j=0}^{\min \set{c,\ell}} \binom{a}{j} \binom{a-j}{\ell-j} \binom{b}{c-j} = \binom{a}{\ell} \sum_{j=0}^{\min \set{c,\ell}} \binom{\ell}{j} \binom{b}{c-j} = \binom{a}{\ell} \binom{b + \ell}{c},
\end{equation}
which itself follows from \eqref{eq:UsefulBinId}.  The result is
\begin{equation} \label{eq:PartiallyCancelled}
\sum_{m=-h-\eps_1+1}^{\infty} \sum_{\ell=0}^{\infty} \binom{-\eps_2}{\ell} \binom{n+h+\eps-1+\ell}{m+h+\eps_1-1} w_1^{n-m+\eps_2+\ell} \brac{-w_2}^{-\eps_2-\ell} \brac{S_m \otimes \wun}
\end{equation}
and we see that the contribution from $z=w_1$ partially cancels \eqref{eq:ToBeCancelled}, as anticipated.

The final result for the coproduct formula is then given by summing the terms in \eqref{eq:ToBeSplit} with $m \le -h-\eps_1$, the second term on the \rhs{} of \eqref{eq:TwistedWCont} and the terms in \eqref{eq:PartiallyCancelled} with $\ell \ge -n-h-\eps+1$.  In each of these terms, the power of $w_1$ is a non-negative integer, so we can consistently set $w_1$ to $0$ and $w_2$ to $-w$.  Combining with \eqref{eq:TwistedWCont3}, valid for $n \ge -h-\eps+1$, and using \eqref{eq:UsefulBinId} judiciously, we arrive at our final form for the twisted coproduct formulae:
\begin{subequations} \label{eq:TwCoprods}
\begin{align}
\parNcoproduct{1}{0,-w}{\tS_n^{0,-w}} &= \sum_{j=0}^{\infty} \binom{-\eps_2}{j} w^{-\eps_2-j} \brac{S_{n+\eps_2+j} \otimes \wun} & &\text{(\(n \ge -h-\eps+1\))} \notag \\
&\mspace{50mu} + \mu_1 \sum_{m=-h-\eps_2+1}^{\infty} \binom{n+h+\eps_2-1}{m+h+\eps_2-1} \brac{-w}^{n-m} \brac{\wun \otimes S_m}, \label{eq:TwCoprods1} \\
\parNcoproduct{1}{0,-w}{\tS_{-n}^{0,-w}} &= \sum_{j=0}^{\infty} \binom{-\eps_2}{j} w^{-\eps_2-j} \brac{S_{-n+\eps_2+j} \otimes \wun} & &\text{(\(n \ge h+\eps\))} \notag \\
&\mspace{50mu} + \mu_1 \sum_{m=-h-\eps_2+1}^{\infty} \binom{m+n-1}{m+h+\eps_2-1} \brac{-1}^{m+h+\eps_2-1} \brac{-w}^{-m-n} \brac{\wun \otimes S_m}. \label{eq:TwCoprods2}
\intertext{If we instead compute the residue at $z=0$ using the \ope{} $\tfunc{S}{z} \tfunc{\psi_2}{w_2}$, valid for $w_2 \to 0$, then the resulting twisted coproduct formulae are}
\parNcoproduct{2}{w,0}{\tS_n^{w,0}} &= \sum_{m=-h-\eps_1+1}^{\infty} \binom{n+h+\eps_1-1}{m+h+\eps_1-1} w^{n-m} \brac{S_m \otimes \wun} & &\text{(\(n \ge -h-\eps+1\))} \notag \\
&\mspace{50mu} + \mu_1 \sum_{j=0}^{\infty} \binom{-\eps_1}{j} \brac{-w}^{-\eps_1-j} \brac{\wun \otimes S_{n+\eps_1+j}}, \label{eq:TwCoprods3} \\
\parNcoproduct{2}{w,0}{\tS_{-n}^{w,0}} &= \sum_{m=-h-\eps_1+1}^{\infty} \binom{m+n-1}{m+h+\eps_1-1} \brac{-1}^{m+h+\eps_1-1} w^{-m-n} \brac{S_m \otimes \wun} & &\text{(\(n \ge h+\eps\))} \notag \\
&\mspace{50mu} + \mu_1 \sum_{j=0}^{\infty} \binom{-\eps_1}{j} \brac{-w}^{-\eps_1-j} \brac{\wun \otimes S_{-n+\eps_1+j}}. \label{eq:TwCoprods4}
\end{align}
\end{subequations}
One can, of course, substitute these formulae into \eqref{eq:STilde} in order to obtain coproduct formulae for the $S_n$.  In practice, we prefer to employ \eqref{eq:TwCoprods} directly and substitute when the explicit mode action is required.  We note that if we set $\eps_1 = \eps_2 = \eps = 0$, then $\tS_n^{0,-w} = \tS_n^{w,0} = S_n$, by \eqref{eq:STilde}, and \eqref{eq:TwCoprods} reduces to the (untwisted) coproduct formulae derived in \cite{GabFus94b} (see also \cite[App.~A]{CanFusI15}).

As in the untwisted case, the twisted coproduct formulae are related by translation:
\begin{equation}
\parNcoproduct{1}{0,-w}{S_{-n}} = \parNcoproduct{2}{w,0}{\ee^{wL_{-1}} S_{-n} \ee^{-wL_{-1}}} = \sum_{m=n}^{\infty} \binom{m-h}{m-n} w^{m-n} \parNcoproduct{2}{w,0}{S_{-m}}.
\end{equation}
Applying \eqref{eq:STilde} twice and \eqref{eq:UsefulBinId}, we obtain a translation formula relating the coproducts of the $\tS_n$:
\begin{align} \label{eq:Translation}
\parNcoproduct{1}{0,-w}{\tS_{-n}^{0,-w}} &= \sum_{j_2=0}^{\infty} \binom{-\eps_2}{j_2} w^{j_2} \parNcoproduct{1}{0,-w}{S_{-n-j_2}} = \sum_{j_2=0}^{\infty} \sum_{m=n+j_2}^{\infty} \binom{-\eps_2}{j_2} \binom{m-h}{m-n-j_2} w^{m-n} \parNcoproduct{2}{w,0}{S_{-m}} \notag \\
&= \sum_{m=n}^{\infty} \sum_{j_2=0}^{m-n} \binom{-\eps_2}{j_2} \binom{m-h}{m-n-j_2} w^{m-n} \sum_{j_1=0}^{\infty} \binom{\eps_1}{j_1} \brac{-w}^{j_1} \parNcoproduct{2}{w,0}{\tS_{-m-j_1}^{w,0}} \notag \\
&= \sum_{j=0}^{\infty} \sum_{k=0}^{\infty} \brac{-1}^j \binom{\eps_1}{j} \binom{n-h-\eps_2+k}{k} w^{j+k} \parNcoproduct{2}{w,0}{\tS_{-n-j-k}^{w,0}}.
\end{align}
This translation formula relates the two different coproducts that have been derived.  The above equalities are therefore non-trivial consequences of imposing mutual locality and amount to defining the fusion product of two twisted modules $\Mod{M}$ and $\Mod{N}$ as the quotient
\begin{equation} \label{eq:DefFusion}
\Mod{M} \fuse \Mod{N} = \frac{\Mod{M} \otimes_{\CC} \Mod{N}}{\left\langle \brac{\parNcoproduct{1}{0,-w}{S_n} - \parNcoproduct{2}{w,0}{\ee^{wL_{-1}} S_n \ee^{-wL_{-1}}}} \brac{\Mod{M} \otimes_{\CC} \Mod{N}} \right\rangle},
\end{equation}
where the submodule that one quotients by is the sum of the images of all modes $S_n$ of all \voSa{} fields $\func{S}{z}$, for all insertion points $w \in \CC \setminus \set{0}$.

\section{Ramond staggered modules} \label{app:RStag}

The logarithmic singularities that give their name to \lcft{} are consequences of a non-diagonalisable hamiltonian \cite{GurLog93}.  At the chiral level, one is therefore led to study \voSa{} modules on which $L_0$ acts non-semisimply.  The simplest such class of modules are the \emph{staggered modules}, originally introduced rather loosely in \cite{RohRed96}, then redefined in \cite{RidSta09} in order to obtain classification results for Virasoro staggered modules.  A general definition for other \voSas{} appeared in \cite{CreLog13}.  In this appendix, we give a brief introduction to the theory of staggered modules for the Ramond algebra.  \ns{} staggered module theory may be found in \cite[App.~B]{CanFusI15}.

We define a \emph{Ramond staggered module} $\Stag{}{}$ to be an extension of a Ramond Kac module by another Ramond Kac module upon which $L_0$ acts non-semisimply:
\begin{equation} \label{es:Stag}
\dses{\Kac{r,s}}{\iota}{\Stag{}{}}{\pi}{\Kac{\rho,\sigma}}.
\end{equation}
Here, $\iota$ and $\pi$ are module homomorphisms.  We remark that it may be useful to replace the Kac modules in this definition by other classes of Ramond modules, \hwms{} in particular.  Most, though not all, of the staggered modules that we discuss in this paper are extensions of Kac modules that are \hw{} --- see \cref{sec:FusNSR} for an example which is not.

We shall customarily affix two sets of indices to the staggered module in \eqref{es:Stag}:  $\Stag{}{} = \Stag{i,j}{k,\ell}$.  These are chosen so that the Kac quotient has labels $\rho = i+k$ and $\sigma = j+\ell$, while the Kac submodule has labels $r=i-k$ and $s=j-\ell$.  The staggered module in \eqref{es:Stag} would then be denoted by
\begin{equation}
\Stag{}{} = \Stag{\frac{1}{2} (\rho + r), \frac{1}{2} (\sigma + s)}{\frac{1}{2} (\rho - r), \frac{1}{2} (\sigma - s)}.
\end{equation}
It is not uncommon to find that the Kac submodule and quotient do not fix the isomorphism class of the staggered module completely.  In this case, additional parameters are needed to pin down the module up to isomorphism and we shall append these parameters, when necessary, using parentheses; for example, $\Stag{i,j}{k,\ell}(\beta)$.  Such parameters were first introduced for Virasoro staggered modules in \cite{GabInd96}, but a general invariant definition does not seem to have appeared until \cite{RidPer07}.  Originally referred to as \emph{logarithmic couplings}, due to their appearance as coefficients of logarithmic terms in correlation functions, the parameters have also been referred to as ``beta-invariants'' \cite{RidSta09} and ``indecomposability parameters'' \cite{VasInd11}.

A basic, but nevertheless powerful, result concerning staggered modules is the following \cite{CreLog13}:  If $w,y \in \Stag{}{}$ form a non-trivial Jordan block for $L_0$, so $(L_0 - h) y = w$ for some $h \in \CC$, and $U$ annihilates $\pi(y)$, for some $U$ in the mode algebra, then $U$ also annihilates $w$.  This has two important consequences.  First, if $y$ projects onto a \hwv{} (or \sv{}) of $\Kac{\rho,\sigma}$, then $w$ is singular.  Second, $\pi(y) \mapsto w$ defines a homomorphism from the submodule of $\Kac{\rho,\sigma}$ generated by $\pi(y)$ into $\Kac{r,s}$.  We often use this fact to rule out certain structures when trying to identify staggered modules in fusion products.

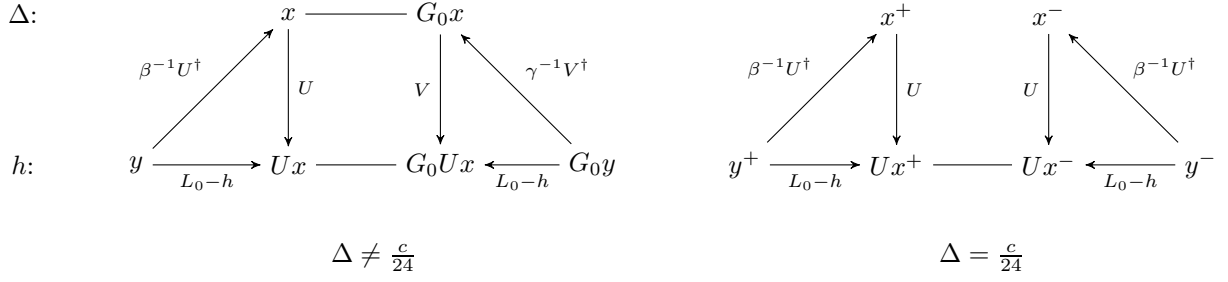
\begin{figure}
\begin{tikzpicture}[>=stealth', node distance=2cm]
\node[] (D) {$\Delta$:};
\node[below of=D] (h) {$h$:};
\node[right=1cm of h] (y) {$y$};
\node[right of=y] (Ux) {$Ux$};
\node[above of=Ux] (x) {$x$};
\node[right of=x] (Gx) {$G_0 x$};
\node[right of=Ux] (GUx) {$G_0 Ux$};
\node[right of=GUx] (Gy) {$G_0 y$};
\draw (x) -- (Gx);
\draw (Ux) -- (GUx);
\draw[->] (x) -- node[right] {$\scriptstyle U$} (Ux);
\draw[->] (Gx) -- node[left] {$\scriptstyle V$} (GUx);
\draw[->] (y) -- node[below] {$\scriptstyle L_0 - h$} (Ux);
\draw[->] (Gy) -- node[below] {$\scriptstyle L_0 - h$} (GUx);
\draw[->] (y) -- node[above left] {$\scriptstyle \beta^{-1} U^{\dag}$} (x);
\draw[->] (Gy) -- node[above right] {$\scriptstyle \gamma^{-1} V^{\dag}$} (Gx);
\node[right=1cm of Ux.center] (tmp) {};
\node[below=0.8cm of tmp] {$\Delta \neq \frac{c}{24}$};
\node[right of=Gy] (y') {$y^+$};
\node[right of=y'] (Ux') {$Ux^+$};
\node[above of=Ux'] (x') {$x^+$};
\node[right of=x'] (Gx') {$x^-$};
\node[right of=Ux'] (GUx') {$Ux^-$};
\node[right of=GUx'] (Gy') {$y^-$};
\draw (Ux') -- (GUx');
\draw[->] (x') -- node[right] {$\scriptstyle U$} (Ux');
\draw[->] (Gx') -- node[left] {$\scriptstyle U$} (GUx');
\draw[->] (y') -- node[below] {$\scriptstyle L_0 - h$} (Ux');
\draw[->] (Gy') -- node[below] {$\scriptstyle L_0 - h$} (GUx');
\draw[->] (y') -- node[above left] {$\scriptstyle \beta^{-1} U^{\dag}$} (x');
\draw[->] (Gy') -- node[above right] {$\scriptstyle \beta^{-1} U^{\dag}$} (Gx');
\node[right=1cm of Ux'.center] (tmp') {};
\node[below=0.8cm of tmp'] {$\Delta = \frac{c}{24}$};
\end{tikzpicture}
\caption{Schematic depictions of the staggered module structures that we have encountered when decomposing fusion products.  The right picture is necessarily different because $G_0 x^+ = G_0 x^- = 0$ when $\Delta = \frac{c}{24}$.} \label{fig:StagMods}
\end{figure}

Let us turn now to the question of logarithmic couplings for Ramond staggered modules.  We shall simplify our considerations by restricting immediately to a set of staggered modules that contains each of the modules that we have encountered in our fusion computations.  This set is characterised by the following properties, referring to \eqref{es:Stag}:
\begin{enumerate}
\item The Kac submodule $\Kac{r,s}$ has a bosonic ground state $x$ of conformal weight $\Delta$ and a \sv{} $Ux$ of conformal weight $h > \Delta \ge -\frac{c}{24}$. \label{it:SV}
\item $U$ cannot be factorised as $U'U''$ such that $U''x$ is a \sv{} of $\Kac{r,s}$ whose conformal weight lies strictly between $\Delta$ and $h$.
\item $Ux$ has a Jordan partner $y \in \Stag{}{}$ satisfying $(L_0 - h) y = \iota(Ux)$.
\item The projection $\pi(y)$ is a ground state of the Kac quotient $\Kac{\rho,\sigma}$.
\end{enumerate}
We depict the structures of the staggered modules from this set in \cref{fig:StagMods}.  Note that if we modify property \ref{it:SV} by replacing $h > \Delta$ by $h = \Delta$, then the corresponding set of staggered modules will have isomorphism classes that are always completely specified by \eqref{es:Stag}.  They therefore require no logarithmic couplings, hence we have excluded them from the present considerations.

Suppose first that $\Delta \neq \frac{c}{24}$, so that $G_0 x \neq 0$ and $G_0 Ux \neq 0$.  We may then define
\begin{equation}
V = \frac{G_0 U G_0}{\Delta - c/24}
\end{equation}
and note that $V G_0 x = G_0 Ux$.  Moreover, $(L_0 - h) G_0 y = G_0 (L_0 - h) y = G_0 Ux$, so we reproduce the left picture in \cref{fig:StagMods} by defining the logarithmic couplings $\beta, \gamma \in \CC$ by
\begin{equation}
U^{\dag} y = \beta x, \qquad V^{\dag} G_0 y = \gamma G_0 x.
\end{equation}
We note that these couplings do not depend upon the choice of $y$.  However, they are not independent:
\begin{gather}
V^{\dag} G_0 y = \frac{G_0 U^{\dag} (L_0 - c/24) y}{\Delta - c/24} = \frac{G_0 U^{\dag} (h - c/24) y + G_0 U^{\dag} Ux}{\Delta - c/24} = \frac{h-c/24}{\Delta - c/24} \beta G_0 x \notag \\
\Ra \qquad \gamma = \frac{h-c/24}{\Delta - c/24} \beta.
\end{gather}
We therefore expect that such a Ramond staggered module requires at most one logarithmic coupling in order to fully specify its isomorphism class.\footnote{A staggered module of this type might actually be completely specified, up to isomorphism, by the exact sequence \eqref{es:Stag}, see \cite{RidLog07,RidSta09}.}

The precise value of a logarithmic coupling depends upon the normalisation chosen for the singular vector $Ux \in \Kac{r,s}$.  For \ns{} staggered modules, we chose \cite{CanFusI15} to normalise by requiring that the coefficient of $G_{-1/2}^{2(h-\Delta)}$ in $U$ be $1$.  For Ramond staggered modules, we choose to normalise by requiring that $U$ be bosonic such that the coefficient of $L_{-1}^{h-\Delta}$ is $1$.  Both choices have the advantage that they do not depend on how one decides to order modes.

The case $\Delta = \frac{c}{24}$ requires a separate treatment because the corresponding Kac submodule is of centre type and is generated by \emph{two} ground states, each of which is annihilated by $G_0$.  We denote the bosonic ground state of $\Kac{r,s}$ by $x^+$ and the fermionic one by $x^-$.  Our first task is to show that $x^-$ may be canonically normalised once we have chosen a normalisation for $x^+$.

To do this, we first compare the submodule $\Kac{r,s}$ with the corresponding Verma module $\RVer{c/24}$, generated by $v$ say.  We see that in these modules, the number of states at each grade, up to that of the \sv{} $Ux^+$, is the same.  It follows that $Uv$ is a \sv{} of $\RVer{c/24}$, hence that each $L_n U$ and $G_k U$, with $n,k>0$, annihilates $v \in \RVer{c/24}$.  Because the modes $L_0 - c/24$, $G_0$, $L_1$ and $G_1$ generate the annihilating ideal of $v \in \RVer{c/24}$, we may therefore write each $L_n U$ and $G_k U$ in the form $U_0 (L_0 - c/24) + V_0 G_0 + U_1 L_1 + V_1 G_1$, for some $U_0$, $V_0$, $U_1$ and $V_1$.  It is now trivial to check that each $L_n U$ and $G_k U$, with $n,k>0$, also annihilates $x^-$.  As $Ux^-$ is an $L_0$-eigenvector, we conclude that it is a \sv{}.  Finally, $G_0 Ux^+$ is easily checked to be singular with the same conformal weight and parity as $Ux^-$.  These two \svs{} are therefore proportional, by the structure of centre type Fock spaces (see \cref{fig:FockStructures}), and so we may normalise $x^-$ so that $G_0 Ux^+ = Ux^-$.

We can now study the staggered module structures when $\Delta = \frac{c}{24}$.  Let $y^+$ and $y^-$ be Jordan partners to $Ux^+$ and $Ux^-$, respectively, as in the right picture of \cref{fig:StagMods}.  Then,
\begin{equation}
(L_0 - h) (y^- - G_0 y^+) = Ux^- - G_0 Ux^+ = 0,
\end{equation}
hence $y^- = G_0 y^+ + u$, where $u \in \Kac{r,s}$ has conformal weight $h$.  Note that $u$ is necessarily annihilated by $U^{\dag}$.  We may again define two logarithmic couplings $\beta^+, \beta^- \in \CC$ by
\begin{equation}
U^{\dag} y^{\pm} = \beta^{\pm} x^{\pm}.
\end{equation}
Once again, these couplings do not depend upon the choice of $y^+$ and $y^-$.  However, this time the couplings appear to be independent unless we make some assumptions, albeit natural ones, about the staggered module.

For example, if the staggered module is invariant under parity-reversal, $\phi \colon \Stag{}{} \ra \Pi \Stag{}{}\xrightarrow{\cong} \Stag{}{}$, then $\phi(x^+) = \alpha x^-$, for some $\alpha \in \CC \setminus \set{0}$, hence
\begin{equation}
(L_0 - h) \phi(y^+) = \phi(Ux^+) = \alpha Ux^- = (L_0 - h) \alpha y^-
\end{equation}
and so $\phi(y^+) - \alpha y^- \in \ker (L_0 - h)$.  In fact, the structure of the staggered module implies now that $\phi(y^+) - \alpha y^-$ belongs to the Kac submodule $\Kac{r,s}$ and is therefore annihilated by $U^{\dag}$.  But then,
\begin{equation}
\alpha \beta^+ x^- = \beta^+ \phi(x^+) = U^{\dag} \phi(y^+) = U^{\dag} \alpha y^- = \alpha \beta^- x^-,
\end{equation}
from which it follows that $\beta^+ = \beta^-$.  Alternatively, if we define an invariant bilinear form $\inner{\cdot}{\cdot}$ on $\Stag{}{}$ by $\inner{x^+}{x^+} = 1$ and $\inner{x^-}{x^-} = \alpha$, for some $\alpha \in \CC \setminus \set{0}$, then
\begin{align}
\alpha \beta^- &= \inner{U^{\dag} y^-}{x^-} = \inner{y^-}{Ux^-} = \inner{y^-}{G_0 Ux^+} = \inner{G_0 y^-}{Ux^+} \notag \\
&= \inner{(L_0 - \tfrac{c}{24}) y^+}{Ux^+} = (h-\tfrac{c}{24}) \inner{y^+}{Ux^+} = (h-\tfrac{c}{24}) \beta^+.
\end{align}
It is not clear to us that either relation must necessarily hold, but it seems reasonable to suppose that the Ramond staggered modules that we construct as fusion products are parity-invariant.  This would, however, require the ratio of $\inner{x^-}{x^-}$ to $\inner{x^+}{x^+}$ to be $h - \frac{c}{24}$, which is a little surprising.

Unfortunately, the only Ramond staggered modules with $\Delta = \frac{c}{24}$ that we have been able to construct are the two $c=0$ examples $\Stag{2,2}{1,0}$ and $\Stag{1,4}{0,2}$, each of which has Kac submodule $\Kac{1,2}$ (see \cref{sec:FusNSR,sec:Edge}).  In both cases, the logarithmic couplings coincide --- they are $\frac{3}{16}$ and $-\frac{3}{16}$, respectively.  While this is consistent with invariance under parity reversal, these examples also have $h-\frac{c}{24} = 1$, so this is also consistent with the existence of an invariant bilinear form with $\inner{x^+}{x^+} = \inner{x^-}{x^-} = 1$.

It would be very revealing to compute the logarithmic couplings of the $c=-\frac{21}{4}$ fusion product $\Kac{2,1} \fuse \Kac{2,4}$, which is expected to be a staggered module with Kac submodule $\Kac{1,4}$ (which is of centre type).  In this case, $h-\frac{c}{24} = 2$, so the norms of the ground states of $\Kac{1,4}$ would have to be different if the couplings coincide.  However, computing this product to depth $2$, required to measure the logarithmic couplings, has remained tantalisingly out of reach with our current implementation of the twisted \NGK{} fusion algorithm.

\flushleft
\singlespacing
%\bibliography{R}
%\bibliographystyle{unsrt}

\end{document}